\documentclass[11pt]{article}
\usepackage[utf8]{inputenc}

\usepackage[doi=false,url=false,isbn=false]{biblatex}
\usepackage{graphicx,graphics,multirow}
\usepackage{subcaption}
\usepackage{amsmath,amssymb,amsthm}
\usepackage{bm}
\usepackage{hyperref}
\usepackage{color}
\usepackage[dvipsnames]{xcolor}
\usepackage[left=2cm,right=2cm,top=2cm,bottom=2cm]{geometry}
\usepackage{MnSymbol}
\usepackage[all]{hypcap}
\usepackage[
        noabbrev,
        capitalise,
        nameinlink,
    ]{cleveref}
\usepackage{hyperref}
\hypersetup{colorlinks, citecolor=MidnightBlue, linkcolor=Blue}

\bibliography{biblio.bib}
\date{}

\newcommand{\Rb}{\mathbb{R}}
\newcommand{\Pb}{\mathbb{P}}

\newcommand{\Ub}{\mathbb{U}}
\newcommand{\Eb}{\mathbb{E}}
\newcommand{\Zb}{\mathbb{Z}}

\newcommand{\Fc}{\mathcal{F}}
\newcommand{\Ac}{\mathcal{A}}
\newcommand{\Bc}{\mathcal{B}}

\newcommand{\Hc}{\mathcal{H}}

\newcommand{\Nc}{\mathcal{N}}

\newcommand{\Uc}{\mathcal{U}}
\newcommand{\Yc}{\mathcal{Y}}
\newcommand{\Dc}{\mathcal{D}}
\newcommand{\Gc}{\mathcal{G}}

\newcommand{\Vc}{\mathcal{V}}

\newcommand{\Xc}{\mathcal{X}}

\newcommand{\xx}{\mathbf{x}}
\newcommand{\yy}{\mathbf{y}}

\DeclareMathOperator{\mmd}{MMD}
\DeclareMathOperator{\var}{Var}
\DeclareMathOperator{\cov}{Cov}

\newcommand{\mds}{\medskip}

\newtheorem{thm}{Theorem}[section]
\newtheorem{lem}{Lemma}
\newtheorem{assum}{Assumption}
\newtheorem{rem}{Remark}

\newtheorem{ex}{Example}

\newtheorem{cor}[thm]{Corollary}
\newtheorem{prop}[thm]{Proposition}

\allowdisplaybreaks

\title{Estimation of time series by Maximum Mean Discrepancy}
\author{Pierre Alquier\thanks{ESSEC Business School, Asia-Pacific campus, 5 Nepal Park, 139408 Singapore, pierre.alquier.stat@gmail.com} \quad Jean-David Fermanian\thanks{ENSAE-CREST, 5 Avenue Le Chatelier, 91120 Palaiseau, France,
jean-david.fermanian@ensae.fr} \quad Benjamin Poignard\thanks{
Keio University, Faculty of Science and Technology, 3-14-1 Hiyoshi, Kohoku-ku, Yokohama, Kanagawa 223-8522, Japan, bpoignard@math.keio.ac.jp. Jointly affiliated at Riken-AIP.}}
\date{\today}

\begin{document}

\maketitle

\begin{abstract}
We define two minimum distance estimators for dependent data by minimizing some approximated Maximum Mean Discrepancy distances 
between the true empirical distribution of observations and their assumed (parametric) model distribution. 
When the latter one is intractable, it is approximated by simulation, allowing to accommodate most dynamic processes with latent variables.
We derive the non-asymptotic and the large sample properties of our estimators in the context of absolutely regular/beta-mixing random elements. Our simulation experiments illustrate the robustness of our procedures to model misspecification, particularly in comparison with alternative standard estimation methods.
\end{abstract}

\begin{flushleft}
\end{flushleft}


\section{Introduction}
\label{sec:intro}


Many modern statistical models  $\{Q_\theta; \theta\in \Theta\subset \Rb^d\}$ are specified algorithmically rather than through tractable likelihoods. This limitation hinders, and in some cases precludes, the use of many classical inference methods. This is particularly common for dynamic latent-variable models arising in machine learning, econometrics, finance, ecology, etc. Examples include discrete choice models, hidden Markov models, agent-based models, ecological simulators, etc.


However, thanks to the algorithmic specification, most of these models can be easily simulated for any given value of $\theta$.
This simple feature has fostered a rich literature devoted to simulation-based inference methods (SBI): simulated likelihood~\cite{pakes1989simulation}, indirect inference~\cite{gourieroux1993indirect}, simulated moments~\cite{mcfadden1989method}, or in a Bayesian setting, synthetic likelihood~\cite{wood2010ecolo} and Approximate Bayesian Computation (ABC)~\cite{tavare1997DNA}. More recently, we can mention neural SBI~\cite{papamakarios2019sequential} and generative adversarial networks (GAN)~\cite{goodfellow2014generative}. A common feature of all these methods is that they avoid explicit computation of likelihoods, but rather involve a comparison of the observed sample $\xx_{1:T}:=(x_t)_{t=1,\ldots,T}$ to simulations from the model $Q_\theta$: the estimator is built so that simulations from the model are "as similar" as possible to the sample. The main difference between these approaches is the criterion used for similarity. Some methods such as simulated moments and (summary-based) ABC involve summary statistics such as moments. On the other hand, to avoid the potential loss of information induced by summary statistics, one can also directly measure a discrepancy between the synthetic and empirical distributions. Here again, multiple choices are possible: integral probability metrics, Kullback–Leibler divergence, etc.


Among the many discrepancies that have recently been proposed for simulation-based inference, kernel-based integral probability metrics -- and in particular the Maximum Mean Discrepancy or MMD -- have emerged as especially attractive~\cite{briol2019statistical,pmlr-v118-cherief-abdellatif20a,dellaporta2022robust,cherief2022finite}. First, the existence of characteristic kernels, such as Gaussian kernels, ensures that SBI with MMD leads to no loss of information, unlike summary-based SBI. Moreover, the Gaussian kernel being also differentiable, optimization of the MMD discrepancy is often feasible. Finally, Gaussian kernels (and other bounded kernels) lead to estimators that are robust under misspecification and to the presence of outliers.


Despite the success of MMD-based inference, essentially all existing work studies i.i.d. observations. While promising applications to stochastic processes are developed in~\cite{briol2019statistical}, the asymptotic theory is established only for i.i.d. observations. The non-asymptotic results in~\cite{cherief2022finite} are in principle valid for $\beta$-mixing processes, but all the examples discussed in the paper are independent. None of these works discusses sampling schemes specifically tailored to time-series models.


In this paper, we make the following contributions: (i) We formulate minimum-MMD estimation for dependent time series through finite-dimensional embeddings of the process. (ii) We introduce and analyse three simulation schemes (independent, pathwise and combined simulation). (iii) We establish non-asymptotic risk bounds, consistency and asymptotic normality under $\beta$-mixing assumptions. (iv) We study the practical behaviour of the resulting estimators on latent-variable and nonlinear time-series models, including robustness under model misspecification and contamination.


The remainder of the paper is organized as follows. Related works are discussed in Section~\ref{subsec:relatedworks}. Section \ref{sec:framework} formally introduces the problem. Section \ref{sec:theory} presents the theoretical properties of our estimators. Section \ref{sec:implementation} describes the implementation details, and numerical experiments are reported in Section \ref{sec:experiments}. All proofs and additional numerical results are deferred to appendices.

\section{Overview of the literature about simulation-based inference methods}
\label{subsec:relatedworks}


In a classical parametric framework, simulation-based inference addresses parameter estimation, hypothesis testing, and confidence interval construction when the likelihood function is intractable, analytical integration is impossible, or standard asymptotics fail. 
Frequentist SBI relies on matching simulated population characteristics to observed data characteristics or replacing intractable expected values with sample averages across repeated simulator runs. Many methods have emerged along these lines. 
Representative examples include indirect inference~\cite{gourieroux1993indirect}, simulated moments~\cite{mcfadden1989method}, synthetic likelihood~\cite{wood2010ecolo}, nonparametric simulated maximum  likelihood~\cite{fermanian2004nonparametric,kristensen2012estimation}. We refer the reader to~\cite{gourieroux1997simulation} for more details and references.
To illustrate, estimators based on indirect inference typically take the form of minimum-distance estimators, where the distance is defined through an auxiliary model fitted to both the observed data and simulated samples.
A usual way to obtain a SBI estimator consists in minimizing some loss 
based on summary statistics considered informative about the model parameters, as in the method of simulated moments.
This is also the idea of synthetic likelihood (SL) inference, where a parametric multivariate normal approximation to the distribution of a vector of summary statistics is constructed.
The latter approach was originally proposed by \cite{wood2010ecolo} to estimate a nonlinear population growth model (the Ricker model). 


From a Bayesian perspective, Approximate Bayesian Computation (ABC), which originated in population genetics (\cite{tavare1997DNA}), shares ideas with SL. ABC approximates the posterior distribution of model parameters by minimizing a discrepancy between synthetic and empirical distributions. While summary statistics are used to quantify this discrepancy, ABC, unlike SL, makes no distributional assumptions about them. Instead, it evaluates their likelihood nonparametrically through an acceptance–rejection criterion. See~\cite{marin2012approximate,gutmann2016bayesian,drovandi2022comparison} for further references on summary-based ABC and synthetic likelihood. To avoid information loss from summary statistics, recent ABC methods use divergences or metrics between the empirical distributions of observed and synthetic data. For instance,~\cite{jiang2018ABC} considered the Kullback-Leibler divergence, while Wasserstein metrics and energy distance were introduced in~\cite{bernton2019abc} and~\cite{nguyen2020energy}, respectively. Notably,~\cite{bernton2019abc} considered a time-series application based on marginal distributions of tuples, closely related to the strategies studied here. More general integral probability metrics, including Wasserstein distance and Maximum Mean Discrepancy, are investigated in~\cite{legramanti2025concentration}.

Modern developments in machine learning have led to new tools for simulation-based inference (SBI). A prominent example is Generative Adversarial Networks (GANs)~\cite{goodfellow2014generative}, which generate images or sounds by mapping standard random vectors (uniform or Gaussian) through neural networks. The Maximum Mean Discrepancy (MMD) was briefly considered for comparing samples~\cite{dziugaite2015training}, but Wasserstein metrics soon became the predominant choice~\cite{arjovsky2017wasserstein}. Although the resulting distribution over images is not explicit, sampling is direct. Neural SBI~\cite{papamakarios2019sequential,hermans2020likelihood} similarly relies on neural networks; see~\cite{cranmer2020frontier,lueckmann2021benchmarking,dalmasso2024likelihood} for comprehensive reviews. Several of these approaches have also been applied to time-series models; see, for example,~\cite{rossi2018indirect,gloeckler2025compositional,warne2022multifidelity}.

More classically, minimum distance estimation (MDE) is a well-established framework dating back to the seminal work of~\cite{wolfowitz1957minimum}. It was subsequently shown to yield highly robust estimators~\cite{parr1980minimum,donoho1988automatic}. While early contributions focused on total variation and Kolmogorov metrics, more recent works have considered alternative discrepancies, including Wasserstein metrics~\cite{bernton2019parameter} and the Maximum Mean Discrepancy (MMD), discussed below.
MDE based on MMD was systematically studied in~\cite{briol2019statistical}, establishing consistency and asymptotic normality. Subsequent non-asymptotic analyses highlighted the estimator's strong robustness to both random and adversarial contamination~\cite{cherief2022finite}. Numerous extensions have since been proposed, including bootstrap- and generalized-Bayes-based methods~\cite{pmlr-v118-cherief-abdellatif20a,dellaporta2022robust,bharti2023optimally}. MMD-based MDE has also been successfully applied to copula models~\cite{alquier2022estimationcopulammd}, hypothesis testing~\cite{schrab2023mmd,fazeli2023semi,bruck2025distribution,key2025composite}, regression~\cite{alquier2024universal}, and missing-data problems~\cite{niu2023discrepancy}. Computational approaches include bootstrap methods~\cite{dellaporta2022robust} and, more commonly, stochastic gradient descent, as studied in~\cite{briol2019statistical,cherief2022finite}.

Most applications of kernel methods, particularly the Maximum Mean Discrepancy (MMD), have focused on i.i.d. observations. However, several kernel constructions have been developed specifically for sequential or path-valued data, notably global alignment kernels~\cite{cuturi2007kernel,cuturi2011fast} and signature kernels~\cite{kiraly2019kernels,cass2024weighted,lee2025signature}; see also~\cite{badiane2022empirical} for alternatives. These methods operate directly on paths, which are infinite-dimensional random elements, introducing additional computational and theoretical challenges. For example, signature kernels can be characterized as solutions to hyperbolic PDEs~\cite{salvi2021signature}, while truncated signatures, which provide finite-dimensional approximations, raise practical concerns~\cite{alden2025signature}.

Instead of comparing complete trajectories, we use a standard characteristic kernel on a finite-dimensional Euclidean space and represent the time series through lagged observations, e.g., $y_t := (x_t,x_{t-1},x_{t-2})$ or $y_t := (x_t,x_{t-p})$, for some $p\geq 0$. This approach, closely related to composite likelihood methods, formulates parameter estimation as a minimum-MMD problem between finite-dimensional distributions. Although it requires selecting an embedding through the lag parameter $p$, it substantially simplifies computation and asymptotic analysis while naturally accommodating simulation-based estimation for models with latent variables. The two approaches are therefore complementary: path-space kernels offer a rich representation of sequential dependence, whereas finite-dimensional lag embeddings provide a simpler and more tractable framework for statistical inference.

Obtaining robust estimators for time series presents new challenges. Classical approaches focus on robust versions of likelihood or estimating equations, leading for example to robust ARMA and GARCH estimators based on bounded score functions, residual autocovariances or transformed moments~\cite{maronna2019robust,hill2015}. These procedures are often tailored to a specific class of models and to particular forms of contamination.
As mentioned above, a complementary line of work studies robustness through general purpose methods as MDE, here through MMD. Since bounded kernels induce bounded influence discrepancies, MMD-based estimators inherit robustness properties under model misspecification, heavy-tailed distributions and contamination. Our work extends these ideas to dependent observations. In particular, we study robustness under both misspecified innovation distributions and contamination models commonly considered in robust time-series analysis, including additive and replacement outliers, as well as the more recent adversarial contamination framework. The numerical experiments compare our estimators to robust procedures specifically designed for ARMA and GARCH models.

Overall, while the literature on simulation-based inference, minimum-distance estimation and MMD is now extensive, the existing theoretical work is largely restricted to independent observations or to path-space discrepancies intended for testing. To the best of our knowledge, there is currently no general theoretical framework for simulation-based MMD estimation of parametric time-series models covering multiple sampling schemes together with non-asymptotic guarantees and asymptotic inference under weak dependence. This is the subject of this paper, which adopts a frequentist approach.

\section{The framework}\label{sec:framework}

The general idea of minimum distance estimators for time series can be summarized as follows: we observe a dataset $\xx_{1:T}:=(x_t)_{t=1,\ldots,T}$ of serially dependent random elements $x_t$, where $x_t$ belongs to a metric space $\Xc$.  
The law of the whole path is denoted by $Q_{0}$.
\textcolor{black}{Our parametric model assumes that the  law of $(x_t)_{t=1,\ldots,T}$ belongs to 
$\{Q_\theta; \theta\in \Theta\subset \Rb^d\}$}
~\footnote{Hereafter, many of our theoretical quantities will implicitly depend on the sample size $T$, that will not appear for notational convenience.}. 
When the model is well-specified, there exists some $\theta_0\in \Theta$ (the so-called ``true'' parameter) such that 
$Q_0=Q_{\theta_0}$, but, more generally, we will consider possibly misspecified models.  
Consider a distance (or divergence) $\Dc$ between distributions on $\Xc^T$ such that there exists a unique parameter
$$   \theta^* := \arg\min_{\theta \in \Theta} \Dc(Q_\theta, Q_{0}),$$
that is called ``the pseudo-true value'' of $\theta$. Obviously, $\theta^*=\theta_0$ for well-specified models. We will no longer consider $\theta_0$, but only $\theta^*$ from now on.
A minimum distance estimator of $\theta^*$ is defined as 
$$    \hat\theta := \arg\min_{\theta \in \Theta} \hat\Dc(Q_\theta,Q_{0}),$$
where $\hat\Dc(Q_\theta,Q_{0})$ is an estimator of the latter distance $\Dc(Q_\theta,Q_0)$.


In the particular case of i.i.d. observations, 
$\hat\Dc(Q_\theta,Q_{0})$ is typically obtained through a distance/divergence between the theoretical law $Q_{1,\theta}$ of a single observation $x_t$ and its empirical distribution
$Q_T:=T^{-1}\sum_{t=1}^T \delta_{x_t}$. 
For instance, when $\Dc$ is the Kullback-Leibler divergence, $\hat\theta$ is the classical 
maximum likelihood estimator of $\theta^*$. 
When $\Dc$ measures the discrepancies between the moments of $Q_{1,\theta}$ and those of $Q_{T}$, this yields the usual method of moments.
Many other distances $\Dc$ can be invoked: Hellinger \cite{lindsay1994efficiency}, Wasserstein \cite{bernton2019parameter}, Maximum Mean Discrepancy \cite{briol2019statistical}, Stein Discrepancy \cite{barp2019minimum}, etc. See other examples in~\cite{basu2011statistical}.


When dealing with dependent data/times series, possibly non stationary, the latter methods have to be revisited. 
Let us illustrate with the maximum likelihood method. With obvious notations, $\hat\Dc(Q_\theta,Q_{0})=-T^{-1}\sum_{t=1}^T \ln f_\theta (x_t)$ in the i.i.d. case and 
\begin{equation}
\hat\Dc(Q_\theta,Q_{0})=-T^{-1}\ln f_\theta(x_1,\ldots,x_T)=-T^{-1}\sum_{t=1}^T \ln f_\theta (x_t|x_{t-1},\ldots,x_1),
\label{cond_dens}
\end{equation}
 with serially dependent data.
The latter expression is analytically intractable for many dynamic models, for instance GARCH or Markov-Switching models. 
In such circumstances, it has been proposed to approximate the distance $\Dc(Q_\theta, Q_{0})$ by simulation, i.e., by using some simulated samples under $Q_\theta$.
For instance, in the case of the method of simulated maximum likelihood, the conditional densities in~(\ref{cond_dens}) are replaced by some averages of so-called ``density generators'' that are obtained by simulation.
This is feasible in Panel Tobit models with individual effects and correlated disturbances: see, e.g., \cite{keane1994computationally}.
For NPSML (\cite{kristensen2012estimation}), the conditional densities in~(\ref{cond_dens}) are replaced with their kernel regression estimates. Hereafter, we assume that the observed process $\xx_{1:T}=(x_t)_{1\leq t\leq T}$ is stationary. 
With i.i.d. data, the law of a single observation $x_t$ is sufficient 
to define the law $Q_\theta$ of the path $\xx_{1:T}$. This is no longer the case with 
stationary dependent data $(x_t)$. In this situation, it is difficult to 
analytically specify $Q_\theta$, and then $\Dc(Q_\theta,Q_0)$. 
Hopefully and \textcolor{black}{for the vast majority of usual statistical models, when $\theta$ is identifiable by the law of $\xx_{1:T}$, this is still the case by the law of some random vector $y_t$ obtained after concatenating some components of $\xx_{1:T}$. This is formalized in the next assumption, that will apply from now on.
\renewcommand{\theassum}{A}
\begin{assum}
There exist some random vectors $y_t$, $t\in \{t_0,\ldots,T'\}$ for some $T'$, such that
\begin{itemize}
    \item the dimension of $y_t$ is fixed (it does not grow with $T$ nor $T'$),
    \item any $y_t$ concatenates a fixed number of components of $\xx_{1:T}$,
    \item the process $(y_t)$ is stationary, and
    \item the law of $y_t$ is sufficient to identify $\theta^*$.
\end{itemize}
\label{hyp_existence_yt}    
\end{assum}
In other words and to be short, by concatenating some components of the original time series $(x_t)$, the underlying parameter becomes identifiable through the marginal law of the new process $(y_t)$. We are not aware of any parametric model (with a fixed number of parameters) for which it is impossible to build such a vector $y_t$.}

Therefore, we build the new process $(y_t)_{t_0\leq t\leq T'}$, such that any $y_t$ is a fixed and deterministic map of the initial process $(x_t)_{1\leq t\leq T}$ and belongs to a new space $\Yc$ (that is different from $\Xc$ in general).
We will focus on $P_0$, the law of $y_t\in \Yc$ instead of the law $Q_0$ of $\xx_{1:T} \in \Xc^T$.
For instance, we could choose $y_t:=(x_t,x_{t-1},x_{t-2})$, or
$y_t:=(x_t,x_{t-p})$ for some $p\geq 0$. 
\textcolor{black}{Hereafter, we denote the concatenation of two arbitrary column vectors $v_1$ and $v_2$ as a new column vector $(v_1,v_2)$.}
More generally, it is usual to set 
$y_t:=(x_t,x_{t-\tau_1},\dots,x_{t-\tau_m})$,
for some positive lags $0<\tau_1<\tau_2<\ldots < \tau_m$ and some given $m$. 
Thus, we have built a new sequence $(y_t)_{t=t_0,\ldots,T'}$ of a priori dependent realizations. 
To lighten notations and without loss of generality, we identify $T'$ and $T$ and we assume $t_0=1$.
The initial model for $(x_t)$ yields a model for $(y_t)$. The corresponding law of $y_t$ is denoted by $P_\theta$.  
By assumption~\ref{hyp_existence_yt}, $\theta^*$ is the unique minimizer of $ \Dc_y(P_\theta, P_{0})$ on $\Theta$, 
for some new distance $\Dc_y$ between probability measures on the space $\Yc$. \textcolor{black}{We will simply denote} $\Dc_y=\Dc$.
The idea of switching from a parametric model $Q_\theta$ for $\xx_{1:T}$ to a model $P_\theta$ for $y_t$, $t\in \{1,\ldots,T\}$, is the basis of composite likelihood methods. See, e.g.,~\cite{varin2011overview}. Therefore, from now on, we focus only on the observations $y_1,\ldots,y_{T}$ and their law, as if the initial process $(x_t)$ were ``forgotten''.
Obviously, we assume $\Dc(P_1,P_2)$ can be easily calculated for two arbitrary discrete distributions $P_1$ and $P_2$ on $\Yc$.
Now, our simulation-based inference method is defined as follows:
for any value of $\theta$, 
let us draw $N$ (independent or dependent) realizations $\tilde y_i$ such that $\tilde y_i\sim P_\theta$, $i\in \{1,\ldots,N\}$. 
Let $\hat P_\theta$ denote the empirical distribution associated with the latter simulated values $\tilde y_i$. 
We then approximate $\Dc(P_\theta, P_0)$ with $\Dc(\hat P_\theta,P_T)$, 
where $P_T$ denotes the empirical law of $(y_t)_{1\leq t \leq T}$.
The corresponding estimator of $\theta^*$ is then the minimizer of $\Dc(\hat P_\theta,P_T)$ over $\Theta$.

Hereafter, the distance between probability measures will be the Maximum Mean Discrepancy, simply called $\mmd$ to be short. Among its comparative advantages, we can cite its generality, its ease of calculation, its robustness: 
see \cite{briol2019statistical,pmlr-v118-cherief-abdellatif20a,cherief2022finite,alquier2022estimationcopulammd,bruck2025distribution}, among others.
To be specific, assume we manipulate random vectors, i.e., $\Yc\subset \Rb^q$.
Let $k:\Yc \times \Yc \mapsto  \Rb^+$ be a Borel measurable kernel on $\Yc$, i.e., a symmetric and positive
definite measurable map. 
Define the squared $\mmd$ distance between two probabilities $P_1$ and $P_2$ by
$$
\mathcal{D}^2(P_1,P_2) = \mathbb{E}_{X,X'\sim P_1} k(X,X') - 2 \mathbb{E}_{X\sim P_1,Y'\sim P_2} k(X,Y') + \mathbb{E}_{Y,Y'\sim P_2} k(Y,Y').$$
From now on, we assume the kernel $k(\cdot,\cdot)$ is characteristic (see~\cite{muandetreview2017}), which ensures that $\mathcal{D}(\cdot,\cdot)$ is a proper distance between distributions.
The $\mmd$ distance between the law $P_0$ of $y_t$ and the law $P_\theta$ of a simulated value $\tilde y_{i}$ is then $ \mathcal{D}(P_\theta,P_0)$.
When the model distribution is analytically tractable, $\mathcal{D}^2(P_\theta,P_0)$ may be approximated by
$$ \mathcal{D}^2(P_\theta,P_T) = \mathbb{E}_{X,X'\sim P_\theta} k(X,X') - \frac{2}{T} \sum_{t=1}^T \mathbb{E}_{X\sim P_\theta} k(X,y_t) + \frac{1}{T^2}\sum_{t,t'=1}^T k(y_t,y_{t'}).$$
Otherwise, simulations are necessary. Replacing the two theoretical laws by their empirical counterparts, this yields a ``feasible'' $\mmd$ criterion
 $$  \Dc^2(\hat P_\theta,P_T) := \frac{1}{N^2} \sum_{i,i'=1}^{N}  k(\tilde y_i,\tilde y_{i'}) -  
 \frac{2}{TN} \sum_{t=1}^{T} \sum_{i=1}^{N} k(\tilde y_i,y_{t})   + \frac{1}{T^2}\sum_{t,t'=1}^{T} k(y_t,y_{t'}),$$
 where $\tilde y_i\sim P_\theta$ for any $i$.
 Let $\tilde \yy_{1:N}:=(\tilde y_1,\ldots,\tilde y_N)$.
Note that the calculation of MMDs between random vectors (our case here) is straightforward \textcolor{black}{compared to competitors} as Wasserstein distances \textcolor{black}{or signature-kernel MMD}, for instance.  
The two estimators of $\theta^*$ we consider are then 
\begin{equation}
\hat\theta_T  :=\arg\min_{\theta \in \Theta}\Dc(P_\theta,P_T),\;\;\text{and}\;\;
\tilde\theta_{N,T}  :=\arg\min_{\theta \in \Theta}\Dc(\hat P_\theta,P_T).
\label{def_estimators}
\end{equation} 
Note that $\hat\theta_T$ is an idealized estimator. Indeed, \textcolor{black}{apart from a few remarkable models discussed in~\cite{briol2019statistical} or \cite{cherief2022finite}, $\Dc(P_\theta,P_T)$ is not available in closed form}. Thus, in practice, we most often rely on sampling from $P_\theta$, yielding $\tilde\theta_{N,T}$. Due to temporal dependencies, there are several ways of generating the simulated $\tilde y_i$. \textcolor{black}{We will consider the three following schemes.}
\begin{enumerate}
    \item[(1)] ISMMD (``independently simulated'' $\mmd$): the realizations $\tilde y_i$, $i\in \{1,\ldots,N\}$, are mutually independent; 
    \item[(2)] PSMMD (``pathwise simulated'' $\mmd$): $(\tilde y_i)_{1\leq i\leq N}$ is a single simulated path of dependent realizations according to the \textcolor{black}{model Data Generating Process (DGP)} given $\theta$;
    \item[(3)] CSMMD (``combined simulated'' $\mmd$): the two latter procedures are combined. Draw $M$ times and independently 
    some simulated paths $\tilde \yy^{(l)}_{1:\bar N}:=(\tilde y^{(l)}_i)_{1\leq i\leq \bar N}$, $l\in \{1,\ldots,M\}$, of dependent realizations according to the model DGP given $\theta$, with $M \bar N =N$; stack all the obtained paths to build a single path $(\tilde y_i)$ of length $N$. In other words, the simulated sample is
    $\tilde \yy_{1:N}:=[\tilde \yy^{(1)}_{1:\bar N}, \tilde \yy^{(2)}_{1:\bar N} ,\ldots, \tilde \yy^{(M)}_{1:\bar N}]$.
    There are different associated schemes as ''fixed $\bar N$, large $M$'', ''large $\bar N$, fixed $M$'', or ''large $\bar N$, large $M$'', possibly with different rates of convergence.
\end{enumerate}
Thus, there are (at least) three corresponding estimators $\tilde \theta_{N,T}$ (as defined in~(\ref{def_estimators})) that will be indexed  by their simulation scheme, yielding $\tilde \theta^{(k)}_{N,T}$, $k\in\{1,2,3\}$.
Such estimators differ only by the way the simulated paths $(\tilde y_i)$ have been generated.
Under similar regularity conditions, they are all consistent and asymptotically normal, even 
if their asymptotic variances generally differ. 

When $y_t=x_t$ and the observations $(x_t)$ are i.i.d., the three latter simulation schemes are identical and we recover the estimators of \cite[Section 2.2]{briol2019statistical}~\footnote{If the observations $(x_t)$ are i.i.d. and $y_t=(x_{t},x_{t-1},\ldots,x_{t-p})$ for some $p\geq 0$, this is no longer the case due to dependencies between successive $y_t$.}. Here, we significantly extend the results of \cite{briol2019statistical} by distinguishing between the processes $(x_t)$ and $(y_t)$, assuming their serial dependencies, managing potential model misspecification and considering several simulation schemes. In particular, ISMMD relies on independent realizations of some random vectors $\tilde y_i$, but the loss $\Dc^2(\hat P_\theta,P_T)$ involves serially dependent observations $y_t$ in general, a situation different from \cite{briol2019statistical}.

\begin{rem}
Note that we need to use the law of the whole process (given $\theta$) to apply PSMMD and CSMMD. Knowing only $P_\theta$ is sufficient to generate the corresponding simulated paths in the latter cases since $\theta$ is identifiable from $P_\theta$ by \textcolor{black}{definition of $(y_t)$}, and knowing $\theta$ implies the model DGP is known. Obviously, under ISMMD, being able to simulate under $P_\theta$ is sufficient to calculate $\tilde\theta_{N,T}^{(1)}$.
Moreover, in every simulation scheme, note that $\Eb\big[ \Dc^2(\hat P_\theta,P_T)\big]$ is not equal to
$\Dc^2(P_\theta,P_0)$, even approximately, because of the serial dependencies in $(y_t)$ and, possibly, in $(\tilde y_i)$. This will constitute a technical difficulty in our proofs. 
\end{rem}

\begin{rem}
An alternative criterion to be minimized could be the average of 
$M$ criteria associated with PSMMD and independently simulated paths of length $\bar N$, i.e.,
$\tilde \theta^{(4)}_{N,T}=\arg\min_\theta M^{-1}\sum_{l=1}^M \Dc_l^2(\hat P_\theta,P_T),$ where
 $$  \Dc_l^2(\hat P_\theta,P_T) := \frac{1}{\bar N^2} \sum_{i,i'=1}^{\bar N}  k(\tilde y^{(l)}_i,\tilde y^{(l)}_{i'}) -  
 \frac{2}{T\bar N} \sum_{t=1}^{T} \sum_{i=1}^{\bar N} k(\tilde y^{(l)}_i,y_{t'})   + \frac{1}{T^2}\sum_{t,t'=1}^{T} k(y_t,y_{t'}).$$
Note that $\tilde \theta^{(4)}_{N,T}$ differ from the CSMMD estimator $\tilde \theta^{(3)}_{N,T}$: some terms $k(\tilde y^{(l)}_i,\tilde y^{(l')}_{i'})$, $l\neq l'$, appear in the latter case, contrary to the former case. We do not consider $\tilde\theta^{(4)}_{N,T}$ in this paper. 
\label{rem_CSMMD_bis}
\end{rem}

\section{Theoretical guarantees}\label{sec:theory}



We still observe a stationary and dependent sequence $(y_t)_{t\geq 1}$, $y_t\sim P_0$.
We restrict ourselves to a (very) large class of parametric dynamic models.
Hereafter, we assume the data generating process \textcolor{black}{of the model} (our ``model assumption'') is induced by a sequence of ``innovations'' $(u_t)_{t\geq 1}$.
\textcolor{black}{
\renewcommand{\theassum}{B}
\begin{assum}
There exists an i.i.d. sequence $(u_t)_{t\in \Zb}$ of random vectors $u_t\in \Rb^m$ and some measurable map $\psi$ such that, when $y_t\sim P_\theta$ for a particular $\theta$, we can write
$$ y_{t} = \psi(\theta;u_t,u_{t-1},\ldots)=:\psi(\theta;\underline{u}_t),\;\; t\geq 1.$$
\label{model_DGP_innovations}
\end{assum}}
The law of $u_t$ obviously does not depend on $\theta$ and may be arbitrarily fixed. Typically, $u_t\sim \Uc(0,1)^{\otimes m}$ or $u_t\sim \Nc(0,1)^{\otimes m}$, the $m$-product of uniform distributions or standard normal distributions, respectively. \textcolor{black}{From now on, \Cref{model_DGP_innovations} applies, in addition to \Cref{hyp_existence_yt}, and we do no longer refer to them.}
\textcolor{black}{Any path $\underline{u}_t$ belongs to a measurable space $\Ub$}.
The law of the (possibly infinite) random vector $\underline{u}_t$ is denoted by $P_\Ub$.

\textcolor{black}{For instance, if $(x_t)$ is a causal univariate ARMA process with Gaussian innovations and an unknown parameter $\theta$, then $(y_t)$ is a causal multivariate ARMA process that can be written as an infinite sum $y_t=\sum_{k\geq 0} A_k u_{t-k}$ with $\sum_{k\geq 0} \|A_k\|<\infty$, for some i.i.d. sequence of standardized Gaussian random vectors $(u_t)$. Here, the matrices $A_k$ are functions of $\theta$.}

\textcolor{black}{As detailed above,} when the model is correctly specified, there exists a parameter $\theta^*$ (also denoted by $\theta_0$) and a sequence of innovations $(u_t)$ such that 
$y_{t} = \psi(\theta^*;\underline{u}_t),$ for every $t$. In the general case of misspecification, this is no longer true, and we will evaluate a ``pseudo-true'' value $\theta^*$ so that $P_0$ is as close as possible to $P_{\theta^*}$.

Since the simulated values $\tilde y_i$ are drawn following $P_\theta$, they are similarly induced by such sequences of innovations.
Therefore, in the ISMMD case (case (1) above) and given the current parameter $\theta$, we have 
$ \tilde y_i = \psi(\theta;u^{(i)}_{T_0},u^{(i)}_{T_0-1},\ldots)=:\psi(\theta;\underline{u}^{(i)}_{T_0})$ for some $T_0$ and any $i\in \{1,\ldots,N\}$, where the sequences of innovations 
$(u^{(i)}_{t})_{t\in \Zb}$ are mutually independent and follow the same law as $(u_{t})_{t\in \Zb}$.
In the PSMMD case (case (2) above), $ \tilde y_i = \psi(\theta;v_i,v_{i-1},\ldots)=:\psi(\theta;\underline{v}_i)$ for a {\it single} sequence $(v_i)_{i\in \Zb}$ that is an independent version of 
$(u_t)_{t\in \Zb}$. 
The CSMMD case (case (3) above) is a combination of both types of innovations.\footnote{\textcolor{black}{In practice, it is obviously not possible to calculate an infinite number of innovations and truncation is sometimes necessary. For instance, in the ISMMD case, set
$ \tilde y_i = \psi(\theta;u^{(i)}_{T_0},u^{(i)}_{T_0-1},\ldots,u^{(i)}_1)$ with $T_0>>1$ and for sequences of innovations $(u^{(i)}_{t})_{1\leq t\leq T_0}$.}}
Hereafter, the sequences $(\tilde y_i)_{i\geq 1}$ implicitly depend on $\theta$ in every case. 
To fix the notations and to cover the three latter situations, any simulated value $\tilde y_i$ will be induced by some (finite or infinite) path $\underline{u}^{(i)}$, without specifying the potential dependencies between such paths, for different indices $i$. 
Obviously, any path $\underline{u}^{(i)}$ belongs to $\Ub$.
Therefore, we denote  
\begin{equation}
\tilde y_i =\psi(\theta;\underline{u}^{(i)}),\; \underline{u}^{(i)} \sim P_\Ub, \; i\in \{1,\ldots,N\},
 \label{not_simul_y}    
\end{equation}
that is our ``model DGP''.
The dependence of $\tilde y_i$ with respect to $\theta$ will be implicit.
For now, recalling~(\ref{def_estimators}), let us study $ \hat{\theta}_T$ first, and then assess how far $ \tilde{\theta}_{N,T}$ can be from $ \hat{\theta}_T$.

\subsection{Non-asymptotic analysis of $\hat{\theta}_T$}

\textcolor{black}{
Let us consider the reproducing kernel Hilbert space (RKHS) $\big(\Hc,\left<\cdot ,\cdot \right>_{\Hc} \big)$ associated with the kernel
$k$. It satisfies the reproducing property $f (y) = \left<f, k(y, ·)\right>_{\Hc}$ for any function $f\in \Hc$ and any $y \in \Yc$.
As in \cite{cherief2022finite}, define} the following measure of serial dependence between the random vectors $y_t$:
$$ \varrho_t := \left|
\mathbb{E} \left< k(y_t,\cdot) - \mathbb{E}[k(y_t,\cdot) ], k(y_0,\cdot) - \mathbb{E}[k(y_0,\cdot)]  \right>_{\Hc}
\right|, \; t>0,  $$
and $\Sigma_t := \sum_{s=1}^t \varrho_s $. Observe that both depend on the distribution of the process $(y_t)$ itself (not only on the law $P_0$ of $y_t$). Let  $\big(\beta(t)\big)_{t\geq 1}$ denote the $\beta$-mixing coefficients of the process $(y_t)$~\footnote{\textcolor{black}{For two $\sigma$-fields $\Ac$ and $\Bc$, their $\beta$-mixing coefficient is $\beta(\Ac,\Bc)=2^{-1}\sup \big\{ \sum_{i\in I} \sum_{j\in J} | \Pb( A_i\cap B_j) - \Pb(A_i)\Pb(B_j)|   \big\}$, the maximum being taken over all finite partitions $(A_i)_{i\in I}$ and $(B_j)_{j\in J}$ of $\Omega$, with the sets $A_i$ in $\Ac$ and the sets $B_j$ in $\Bc$. Set $\beta(t)=\beta\big(\sigma(y_0,y_{-1},\dots),\sigma(y_t,y_{t+1},\dots)\big)$, $t>0$.}}. \cite{cherief2022finite} shows that, for a large class of kernels $k(\cdot,\cdot)$ that includes the Gaussian and the Laplace kernels, we have $\varrho_t \leq \beta(t)$.
We recall this result below.
Recall $y_t$ takes values in $\mathbb{R}^q$ and let $\|\cdot\|$ denote the Euclidean norm in $\mathbb{R}^q$. 
\begin{prop}[Proposition 4.4 in~\cite{cherief2022finite}]
Assume that the kernel $k$ can be written as $k(y,y') = F(\|y-y'\|)$, where $F(x) = \int_{x}^\infty f(u) \,du $ for some nonnegative continuous map $f$, with $\int_0^{+\infty} f(t)\, dt=1$. Then $ \varrho_t \leq \beta(t)$.
\label{prop_rho_beta}
\end{prop}
The proof of Proposition 4.4 in~\cite{cherief2022finite} contains a minor mistake that leads to the slightly worse bound $ \varrho_t \leq 2 \beta(t) $. We thus provide a fixed version of the proof in Section \ref{supp:section_proofs} of the Appendix.

Assume $\sup_y |k(y,y)|\leq 1$.
A direct application of Lemma 7.1 of \cite{cherief2022finite} leads to
\begin{equation}
\label{expectation:badrandpierre}
\mathbb{E}\left[ \mathcal{D}(P_T,P_0) \right] \leq  \sqrt{\frac{1 + 2\Sigma_T}{T}} \cdot
\end{equation}
In particular, if $ \Sigma_\infty := \sum_{s=1}^\infty \varrho_s < \infty $, we have
$
\mathbb{E}\left[ \mathcal{D}(P_T,P_0) \right] \leq  \sqrt{\frac{1 + 2\Sigma_\infty}{T}} \cdot$
More generally, assume only $\Sigma_T = o(T) $ to ensure $
\mathbb{E}\left[ \mathcal{D}(P_T,P_0) \right] =o(1)$.
\textcolor{black}{Deduce the following result, whose proof in given in Appendix~\ref{supp:section_proofs}.}
\begin{prop}
\label{prop:pierre:nonasymptotic}
Assume $\sup_y |k(y,y)|\leq 1$.
We have
$$
\mathbb{E}\left[  \mathcal{D}(P_{\hat{\theta}_T},P_0) \right]
\leq \min_{\theta\in\Theta}  \mathcal{D}(P_\theta,P_0) +  2 \sqrt{\frac{1 + 2\Sigma_T}{T}}\cdot
$$
\end{prop}
\textcolor{black}{
Proposition~\ref{prop:pierre:nonasymptotic} ensures the robustness of $\hat{\theta}_T$ to misspecification. For example, assume that for some $\theta_0\in\Theta$ and $\varepsilon\in(0,1)$, $P_0 = (1-\varepsilon) P_{\theta_0} + \varepsilon Q $, where $Q$ is an arbitrary probability distribution. This means that our model for $P_0$ is essentially correct up to a small deviation. In the i.i.d. setting, this coincides with Huber's contamination model~\cite{huber1992robust}, where a proportion $\varepsilon$ of outliers are sampled from $Q$ instead of $ P_{\theta_0}$. In the context of time series, this includes additive outliers (AO) and replacement outliers (RO), following the terminology used in Chapter 8 in~\cite{maronna2019robust}.
\\
Indeed, note that $ \mathcal{D}(P_{\theta_0},P_0)  = \mathcal{D}(P_{\theta_0},(1-\varepsilon) P_{\theta_0} + \varepsilon Q)
\leq (1-\varepsilon) \mathcal{D}(P_{\theta_0}, P_{\theta_0}) +  \varepsilon  \mathcal{D}(P_{\theta_0},Q) =  \varepsilon  \mathcal{D}(P_{\theta_0},Q) \leq 2\varepsilon$, the final inequality being a consequence of the assumption $\sup_y |k(y,y)|\leq 1$ in Proposition~\ref{prop:pierre:nonasymptotic}~\footnote{Remind that $\mathcal{D}(P_{\theta_0},Q) = \| \mathbb{E}_{y\sim P_{\theta_0}}\Phi(y) - \mathbb{E}_{y\sim Q}\Phi(y) \|_{\mathcal{H}} $ where 
$\Phi$ is the feature map associated with $k$, that is, $k(u,v)=\left<\Phi(u),\Phi(v)\right>_{\mathcal{H}}$ (see~\cite{sriperumbudur2010hilbert}). Thus, $\mathcal{D}(P_{\theta_0},Q)\leq \| \mathbb{E}_{y\sim P_{\theta_0}}\Phi(y) \|_{\mathcal{H}}+\| \mathbb{E}_{y\sim Q}\Phi(y) \|_{\mathcal{H}} \leq \mathbb{E}_{y\sim P_{\theta_0}}  \|\Phi(y) \|_{\mathcal{H}}+\mathbb{E}_{y\sim Q} \| \Phi(y) \|_{\mathcal{H}} = \mathbb{E}_{y\sim P_{\theta_0}} \sqrt{k(y,y)} + \mathbb{E}_{y\sim Q} \sqrt{k(y,y)} \leq 2$.
}.
Moreover, $  \mathcal{D}(P_{\hat{\theta}_T},P_{\theta_0}) \leq \mathcal{D}(P_{\hat{\theta}_T},P_0) +  \mathcal{D}(P_{\theta_0},P_0) $. Substituting in Proposition~\ref{prop:pierre:nonasymptotic}, this gives
$$
\mathbb{E}\left[  \mathcal{D}(P_{\hat{\theta}_T},P_{\theta_0}) \right]
\leq 4\varepsilon +  2 \sqrt{\frac{1 + 2\Sigma_T}{T}}\cdot
$$
Of course, other models of outliers are relevant; see~\cite{diakonikolas2023algorithmic} for a recent survey. A very popular model in modern applications is the adversarial contamination model~\cite{chen2018robust}. In this case, the statistician does not observe $\yy_{1:T}:=(y_t)_{t=1,\ldots,T}$ but instead, he or she observes a contaminated sample $\yy'_{1:T}:=(y'_t)_{t=1,\ldots,T}$ where ${\rm card}\{1\leq t\leq T: y_t \neq y'_t \} \leq k $ for some integer $k\leq T$. There is no assumption on how the $y'_t$ that differ from the $y_t$ are chosen and generated. They can be random or deterministic (as in a short level shift (LS) using the terminology of~\cite{maronna2019robust}). They can even be chosen by an algorithm designed to maximize the loss of the statistician. While $\hat{\theta}_T$ is no longer accessible in this setting, a statistician not aware of the contamination will obtain instead
$$ \hat{\theta}_T' = \arg\min_{\theta\in\Theta} \mathcal{D}(P_\theta,T^{-1}\sum_{t=1}^T \delta_{y_t'} ). $$
Fortunately, it is possible to study minimum distance estimators in such a setting. A general approach can be found in~\cite{donoho1988automatic}, but a simpler approach works here. Indeed, for any $Q$, we have
\begin{eqnarray}
\lefteqn{ |\mathcal{D}(Q,T^{-1}\sum_{t=1}^T \delta_{y_t})
 - \mathcal{D}(Q,T^{-1}\sum_{t=1}^T \delta_{y'_t}) |
\leq
\mathcal{D}(T^{-1}\sum_{t=1}^T \delta_{y_t},T^{-1}\sum_{t=1}^T \delta_{y'_t})} \nonumber
\\
&=& T^{-1} \mathcal{D}(\sum_{\{t:y_t\neq y_t'\}} \delta_{y_t},\sum_{\{t:y_t\neq y_t'\}} \delta_{y'_t})
\leq T^{-1} \sum_{\{t:y_t\neq y_t'\}}  \mathcal{D}( \delta_{y_t},\delta_{y'_t})
\leq T^{-1} 2k,   \label{ineg_MMD_contamin}
\end{eqnarray}
using $\Dc(P_1,P_2)=\sup_{\| f\|_{\Hc}\leq 1} | \int f(y) P_1(dy) - \int f(y) P_2(dy)| $ (see \cite[Definition 2]{gretton2012kernel}, e.g.).
We can apply (\ref{ineg_MMD_contamin}) first to $Q=P_0$, and then to $Q=P_{\hat{\theta}_T}$ to prove
$$
\mathcal{D}(P_{\hat{\theta}'_T},P_0)
\leq T^{-1} 4k + 2 \mathcal{D}(P_{\hat{\theta}_T},P_0).
$$
The details are similar to~\cite[page 208]{cherief2022finite}.
\begin{cor}
In the setting of Proposition~\ref{prop:pierre:nonasymptotic}, for any $k\leq T$ and any random vector $\yy'_{1:T}:=(y'_t)_{t=1,\ldots,T}$ such that  ${\rm card}\{1\leq t\leq T: y_t \neq y'_t \} \leq k $, the contaminated estimator $ \hat{\theta}_T' = \arg\min_{\theta\in\Theta} \mathcal{D}(P_\theta,T^{-1}\sum_{t=1}^T \delta_{y_t'}) $
satisfies
$$
\mathbb{E}\left[  \mathcal{D}(P_{\hat{\theta}'_T},P_0) \right]
\leq
2 \min_{\theta\in\Theta}  \mathcal{D}(P_\theta,P_0) +  4 \sqrt{\frac{1 + 2\Sigma_T}{T}} + \frac{4k}{T} \cdot
$$
\end{cor}
}

Here, we provide a list of commonly encountered stochastic processes $(y_t)_{t\in\mathbb{Z}}$ for which $\Sigma_\infty < \infty $. In Examples~\ref{ex:rho:markov},~\ref{ex:rho:arma},~\ref{ex:rho:garch},~\ref{ex:rho:nonexpo} and~\ref{ex:rho:hidden} below, 
this is proven through $\varrho_t \leq \beta(t) $, when Proposition~\ref{prop_rho_beta} applies. 
However, in Example~\ref{ex:rho:nonbeta}, $\beta(t)\geq 1/4$ while $\varrho_t = \mathcal{O}(2^{-t}) $.
\begin{ex}[Markov chains]
\label{ex:rho:markov}
When $(y_t)$ is an irreducible and aperiodic Markov chain on a finite state space, it is uniformly ergodic. This implies there is some $C>0$ and some $\rho\in [0,1)$ such that, for all $t\geq 1$, $\beta(t) \leq C \rho^{t} $. Beyond the finite case, see the study of the mixing coefficients $\beta(t)$ for a large class of Markov chains in Section 2.4 in~\cite{doukhan1995mixing}.
\end{ex}
\begin{ex}[ARMA process]
\label{ex:rho:arma}
Any stationary ARMA process with i.i.d. noises which admit a density with respect to the Lebesgue measure will satisfy $\beta(t)=\mathcal{O}(\rho^t)$ for some $0<\rho<1$ (Theorem 1 in~\cite{mokkadem1988mixing}).
\end{ex}
\begin{ex}[GARCH and stochastic volatility models]
\label{ex:rho:garch}
Proposition 5 in~\cite{carrasco2002mixing} shows that, under very mild assumptions on the innovations, $\beta(t)$ is exponentially decreasing for a GARCH(1,1) model. They extend this result to many variants of the GARCH(1,1) process (Corollaries 6 to 11). This is also true for general GARCH(p,q) and various extensions again: Proposition 12 and 13, Corollary 14 in~\cite{carrasco2002mixing}. They also cover stochastic volatility models (see their Proposition 15).
\end{ex}
\begin{ex}[An example with non-exponential decay]
\label{ex:rho:nonexpo}
All the results mentioned in Examples~\ref{ex:rho:markov},~\ref{ex:rho:arma} and~\ref{ex:rho:garch} prove $\beta(t)=\mathcal{O}(\rho^t)$ for some $0<\rho<1$ and $t\geq 1$. Exercise 1 of Chapter 9 in~\cite{rio1999theorie} describes the construction of a real-valued, heavy-tailed Markov chain such that $\beta(t) \sim t^{-a}$ for any $a>0$. In particular, when $a>1$, this example satisfies $\varrho_t = \mathcal{O}(t^{-a}) $ and thus $\Sigma_\infty < \infty$.
\end{ex}
\begin{ex}[HMM]
\label{ex:rho:hidden}
Assume that $y_t = F(z_t,\varepsilon_t)$ where the $(\varepsilon_t)_{t\in\mathbb{Z}}$ are i.i.d. and independent from the process $(z_t)$. When $(z_t)$ is a Markov chain, the process $(y_t)$ is refered to as an HMM (Hidden Markov Model). Let $(\beta_z(t))$ denote the $\beta$-mixing coefficients of $(z_t)$. Then $\beta(t) \leq \beta_z(t)$.
\end{ex}
\begin{ex}
\label{ex:rho:nonbeta}
Put
$y_{t+1} = (y_t + \eta_t)/2$ where $y_0\sim\mathcal{U}(0,1)$, the $(\eta_t)$ are i.i.d. and $\eta_t\sim\mathcal{B}e(1/2)$, a Bernoulli distribution. Note that this is actually an AR process, but its noise has no density with respect to the Lebesgue measure, so it does not fit the conditions of Example~\ref{ex:rho:arma}. This is a classical model for which $\beta(t)\geq 1/4$. However, Section 4.3.2 of~\cite{cherief2022finite} shows that $\varrho_t = \mathcal{O}(2^{-t})$, and thus $\Sigma_\infty < \infty$.
\end{ex}

\subsection{Consistency}
Now, let us prove that $\hat\theta_T$ (resp. $\tilde\theta_{N,T}$) is weakly consistent when $T$ (resp. $N$ and $T$) tends to infinity. 

\begin{prop}
\label{prop:pierre:consistency:1}
Assume $\sup_y |k(y,y)|\leq 1$, $\Sigma_T = o(T) $ and the following identifiability conditions:
\begin{enumerate}
\item[(i)] the function $\mathcal{D}(P_\theta,P_0)$ has a unique minimizer $\theta^*\in\Theta$,\;\text{and}
\item[(ii)] for every $r>0$, $\inf_{\{\theta \in \Theta:\|\theta-\theta^*\|\geq r \}} \mathcal{D}(P_\theta,P_0) > \mathcal{D}(P_{\theta^*},P_0) $.
\end{enumerate} 
Then $\hat{\theta}_T \xrightarrow[T\rightarrow\infty]{ \text{ prob.}} \theta^*$.
\end{prop}


We remind that $k(u,v) = \left<\Phi(u),\Phi(v) \right>_{\mathcal{H}}$ for some function $\Phi:\Rb^q\rightarrow \Hc$. 
Thus, for two probability measures $P_1$ and $P_2$,
\begin{equation}
    \| \int \Phi(x)\, (P_1-P_2)(dx) \|_\Hc^2 = \int < \Phi(x),\Phi(y)>_{\Hc}\, (P_1-P_2)(dx)\,\, (P_1-P_2)(dy) = \Dc(P_1,P_2)^2.
    \label{dec_MMD}
\end{equation}

\begin{prop}
\label{prop:pierre:consistency:2}
Under the assumptions of Proposition~\ref{prop:pierre:consistency:1}, assume additionally that
\begin{enumerate}
\item[(iii)] there is a real number $\pi>0$ large enough so that the covering numbers $\Nc(\alpha,\Theta,\|\cdot\|) =\mathcal{O}(\alpha^{-\pi})$ when $\alpha$ tends to zero~\footnote{The covering number 
$\Nc(\alpha,\Theta,\|\cdot\|)$ is the minimal number of 
$\|\cdot\|$-balls of
radius $\alpha$ needed to cover the set $\Theta$.};
\item[(iv)] there exists $\gamma>0$ such that, for any sequence of innovations $\underbar{u}\in \Ub$ and for any $(\theta,\theta')\in\Theta^2$,
$$
\| \Phi\big(\psi(\theta,\underbar{u}) \big) - \Phi\big(\psi(\theta',\underbar{u}) \big) \|_{\mathcal{H}}
\leq m(\underbar{u}) \|\theta-\theta'\|^{\gamma},
$$
for some measurable map $m(\cdot)$, with $\mathbb{E}\big[m(\underbar{u})\big] < +\infty $;
\item[(v)] 
we sample a trajectory $(\tilde{y}_i)_{i\geq 1}$ and, whatever the parameter $\theta\in \Theta$, the sampled sequences $(\tilde y_i)$, $\tilde y_i\sim P_\theta$, have $\beta$-mixing coefficients $\big(\beta(n;\theta)\big)_{n\geq 1}$ that decrease fast enough:  $\sup_{\theta\in \Theta}\beta(n;\theta)= o(n^{-(2\pi+1)}) $ for $n\rightarrow\infty$ where $\pi$ is as in (iii).
\end{enumerate}
Then $\tilde{\theta}_{N,T} \xrightarrow[T,N\rightarrow\infty]{ \text{prob.} } \theta^*$.
\end{prop}
Note that \textcolor{black}{\textit{(iii)}} is satisfied for $\pi=d$ when $\Theta \subset\mathbb{R}^d$ is relatively compact.
Moreover, \textit{(v)} is satisfied when we sample i.i.d. pseudo-observations $(\tilde{y}_i)_{i=1,\dots,N}$ (the ISMMD scheme).
Indeed, $\beta(n;\theta) = 0$ for any $n\geq 1$ in the latter case. Moreover, for CSMMD, this is still the case since 
$\beta(n;\theta) = 0$ for any $n> \bar N$.
Finally, observe that \textit{(iv)} can we rewritten more explicitly as
\begin{equation}
k\big(\psi(\theta,\underbar{u}) ,\psi(\theta,\underbar{u}) \big) - 2k\big(\psi(\theta,\underbar{u}),\psi(\theta',\underbar{u}) \big) + k\big(\psi(\theta',\underbar{u}) ,\psi(\theta',\underbar{u}) \big)  
\leq m^2(\underbar{u}) \|\theta-\theta'\|^{2 \gamma}.
\label{CS_iv}
\end{equation}
\textcolor{black}{In particular, \Cref{CS_iv} is satisfied 
when the real map $(y,y')\mapsto \|\nabla k(y,y') \|$ is bounded by a constant $K$ (as for the Gaussian kernel), and  
the maps $\theta\mapsto \psi(\theta,\underbar{u})$, $\underbar{u}\in \Ub$ satisfy the H\"older-type condition 
$$ \| \psi(\theta,\underbar{u}) - \psi(\theta',\underbar{u}) \|\leq \frac{m^2(\underbar{u})}{2K} \| \theta - \theta' \|^{2\gamma}. $$
}


\subsection{Asymptotic Normality}

Recall we observe a strictly stationary dependent sequence $(y_t)_{t\geq 1}$ in $\Rb^q$. For any $\theta$, we are able to simulate sequences $(\tilde y_i)_{1\leq i \leq N}$ where $\tilde y_i\sim P_\theta$, as described in~\eqref{not_simul_y}.
The simulated vectors $\tilde y_i$ are independent in the ISMMD case but not under the simulation schemes PSMMD or CSMMD, as explained above.
\textcolor{black}{We now state the asymptotic normality of $\hat\theta_T$ and $\tilde\theta_{N,T}$.}

\subsubsection{Asymptotic normality of $\hat\theta_T$}
\label{AN_hattheta_T}

Let the map $I:\Theta \mapsto \Rb^{d\times d}$ be
{\small{\begin{eqnarray*}
\lefteqn{ I(\theta) := 
2\int \Big\{\nabla_{\theta} \psi^\top(\theta,u) \nabla^2_{1,1} k\big(\psi(\theta,u),\psi(\theta,u')\big)  \nabla_{\theta^\top} \psi(\theta,u) 
    + \nabla_{\theta} \psi^\top(\theta,u) \nabla^2_{1,2} k\big(\psi(\theta,u),\psi(\theta,u')\big)  \nabla_{\theta^\top} \psi(\theta,u')  }\\
    &+& \sum_{l=1}^q \partial_{l} k\big(\psi(\theta,u),\psi(\theta,u')\big)  \nabla^2_{\theta,\theta^\top} \psi_l(\theta,u) \Big\}
    \, P_\Ub(du)\, P_\Ub(du'),\hspace{6cm}
\end{eqnarray*}}}
where $\nabla^2_{1,1} k(y,y')$ denotes the second order differential of the kernel with respect to its first argument, evaluated at $(y,y')$. 
Similarly, $\nabla^2_{1,2} k(y,y')$ is the crossed second order differential of $k$ with respect to its first and second arguments respectively, at the same point. 
Moreover, define the family of maps
$\Fc:=\{ f_\theta: \Rb^q \mapsto \Rb^{d\times d};\theta\in \Theta\}$, where
$$ f_\theta(y) :=2 \int 
\Big\{ \nabla_{\theta} \psi^\top(\theta,u) \nabla^2_{1,1} k\big(\psi(\theta,u),y\big)\nabla_{\theta^\top} \psi(\theta,u) + \sum_{l=1}^q \partial_{l} k\big(\psi(\theta,u),y\big)\nabla^2_{\theta,\theta^\top} \psi_l(\theta,u)\Big\}\, P_\Ub (du),$$
and set 
$  V_1(\theta) := I(\theta) - \int f_\theta (y) \, P_0(dy).$

\begin{prop}
\label{AN_theta_T}
In addition to $\hat\theta_T - \theta^*=o_P(1)$, assume 
\begin{itemize}
\item[(a)] the process $(y_t)$ is absolutely regular/beta-mixing, 
$\sum_{t\geq 1} \beta(t)^{\delta/(2+\delta)}<\infty$ for some $\delta>0$  and
$\beta(t)=O(t^{-r_\beta})$, $r_\beta\in (0,1]$; moreover, $\Eb[\| v_t \|^{2+\delta}]<\infty$ where 
$$ v_t:=\int \nabla_\theta \psi^\top(\theta^*,u) \nabla_1 k\big(\psi(\theta^*,u),y_{t}\big)\, P_\Ub (du) ;$$
\item[(b)] the kernel $k$ is two times continuously differentiable on $\Rb^{2q}$, $\theta^*$ belongs to the interior of $\Theta$, $\theta \mapsto V_1(\theta)$ is continuous at $\theta^*$
and $V_1(\theta^*)$ is invertible;
\item[(c)] there exists some measurable map $H:\Rb^q\mapsto \Rb$ such that, for every $\theta$ and $\theta'$ in a neighborhood of $\theta^*$ and every $y\in \Rb^q$, we have
\begin{equation*}
\big\| f_\theta(y) - f_{\theta'}(y) \big\| \leq  H(y) \| \theta - \theta'\|^\pi,\; \text{with} \; \Eb[ H(y_t) ]<\infty,
\label{cond_LLN_Ustat_1}
\end{equation*}
 for some $\pi>0$; finally, $\Eb\big[ \| f_{\theta^*}(y_t) \| \big]<\infty$.
\end{itemize}
Then,
$\sqrt{T}\big\{\hat{\theta}_{T} - \theta^*\big\} \xrightarrow[T\rightarrow\infty]{ \text{law} } \Nc\big(0, V_1(\theta^*)^{-1}\,V_0(\theta^*)\, V_1(\theta^*)^{-1}\big),\; \text{with}$
\begin{eqnarray*}
    \lefteqn{ V_0(\theta^*):=
4  \int  \nabla_\theta \psi^\top(\theta^*,u) \var \Big( \nabla_1 k\big(y_{1},\psi(\theta^*,u)\big)   \Big)
 \nabla_{\theta^\top} \psi(\theta^*,u)\, P_\Ub (du) }\\
&+&   8 \sum_{t=2}^\infty\int   \nabla_{\theta} \psi^\top(\theta^*,u') \cov \Big( \nabla_1 k\big(y_{1},\psi(\theta^*,u)\big),
\nabla_1 k\big(y_{t},\psi(\theta^*,u')\big)   \Big)
\nabla_{\theta^\top} \psi(\theta^*,u)\, P_\Ub (du)\, \Pb_\Ub(du').
\end{eqnarray*}
\end{prop}
 In the particular case of a correctly specified model, i.e., when $P_0=P_{\theta^*}$, we simply have
 $$ V_1(\theta^*)=2\int \nabla_\theta \psi^\top(\theta^*,u) \nabla^2_{1,2} k\big(\psi(\theta^*,u),\psi(\theta^*,u')\big)   \nabla_{\theta^\top} \psi(\theta^*,u')\,P_\Ub(du)\, P_\Ub(du').$$
If, in addition, the observations $(x_t)$ are i.i.d. and $y_t=x_t$, it is easy to check that the asymptotic variance in~\Cref{AN_theta_T} is indeed the same as in \cite[Th. 2, point 4]{briol2019statistical}.
\textcolor{black}{Note that the condition $\Eb[\| v_t \|^{2+\delta}]<\infty$ in {\it (a)} is satisfied with a Gaussian kernel and when 
$\Eb\big[ \|  \nabla_\theta \psi^\top(\theta^*,\underline{u_t}) \| \big]<\infty$.}
\textcolor{black}{
\begin{rem}
The H\"older condition of order $\pi$ labelled {\it (c)} may be easily satisfied when similar conditions apply to the kernel, to the maps $\theta\mapsto \psi(\theta,u)$, and to their derivatives (up to second order), under some convenient conditions of integrability.
To illustrate, consider the Gaussian kernel $k(y,y') := \exp\big(-\|y-y'\|^2/(2\sigma^2)\big)$. Note that its derivatives are uniformly bounded. 
Assume $\| \nabla^j_\theta \psi(\theta,u) - \nabla^j_\theta \psi(\theta',u) \|\leq H_j(u)\|\theta - \theta'\|^{\pi_j},$ for some positive constants $\pi_j$, $j\in \{0,1,2\}$, and  
    \begin{eqnarray*}
        \lefteqn{\sup_{\theta \in \Theta} \int \big\{ H_1(u) \} 
        \| \nabla_\theta \psi(\theta,u)\| +
    H_0(u) \| \nabla_\theta \psi(\theta,u)\|^2 }\\
    &+& H_0(u) H_1(u) \| \nabla_\theta \psi(\theta,u)\| 
    + H_0(u) \| \nabla^2_{\theta,\theta^\top} \psi(\theta,u)\|
    + H_2(u) \big\}\, P_{\Ub}(du)<\infty.
    \end{eqnarray*}
Then, {\it (c)} is satisfied with a constant map $y\mapsto H(y)$ and $\pi=\min(\pi_0,\pi_1,\pi_2)$.
\label{rem_CS_AN_hattheta_T}    
\end{rem}
}

\begin{rem}
\label{remark_ULLN_thetaT}
In some cases, our Assumption (c) may be seen as restrictive. Nonetheless, it can be significantly weakened. Indeed, the uniform consistency of $\int f_\theta(y) \, P_T(dy)$ on a neighborhood of $\theta^*$ may be obtained thanks to a ULLN for strongly stationary and $\beta$-mixing time series. 
Typically, {\it (c)} could be replaced by some conditions on the covering numbers associated with $\Fc$ and for a convenient semi-metric between such maps. For instance, see Theorem 3.4 in~\cite{yu1994rates}, Theorem 1 in~\cite{doukhan1995invariance} or, more recently,~\cite{kuersteiner2019invariance}.
\end{rem}

\subsubsection{Asymptotic normality of $\tilde\theta_{N,T}$}
\label{AN_hattheta_NT}

Before stating our regularity assumptions, additional notations are required.
For any $u\in \Ub$, 
introduce the $q$-dimensional vector 
$\tilde y(u):=\psi(\theta^*;u)$, the $d\times q$ matrix $z(u) := \nabla_\theta \psi^\top(\theta^*;u)$, the $q(1+d)$-dimensional vector 
$\zeta(u):=\Big(\tilde y(u),\text{Vec}\big(z(u)\big)\Big)=\Big(\psi(\theta^*;u),\text{Vec}\big(\nabla_\theta \psi^\top(\theta^*;u\big)\Big)$,
and the $q(1+d+d^2)$-dimensional vector 
$$\xi(u):=\Big(\psi(\theta^*,u),\text{Vec} \big(\nabla_{\theta} \psi^\top(\theta^*,u)\big),\text{Vec}\big(
\nabla^2_{\theta,\theta^\top} \psi_1(\theta^*,u)\big),\ldots,\text{Vec}\big(
\nabla^2_{\theta,\theta^\top} \psi_q(\theta^*,u) \big)\Big).$$ 
Hereafter, the argument $u$ of the latter maps will be left aside when the context is obvious.
If $u=\underline{u}^{(i)}$, one of our simulated elements, then we simply set
$$ \tilde y_i:=\tilde y(\underline{u}^{(i)}),\; z_i:=z(\underline{u}^{(i)}), \; 
\zeta_i:= \zeta (\underline{u}^{(i)}),\; \text{and}\; \xi_i:=\xi(\underline{u}^{(i)}),$$
to be short. In other words, the random vectors $\tilde y_i$, $z_i$, $\zeta_i$ and $\xi_i$ are here simulated in the particular case $\theta=\theta^*$.
Let $P_\zeta$ denote the law of the random vectors $\zeta_i$.



For arbitrary vectors 
$\zeta=\big(\tilde y,\text{Vec}(z)\big)$ and $\zeta'=\big(\tilde y',\text{Vec}(z')\big)$ in $\Rb^{q(1+d)}$, define the symmetric map
$ h(\zeta,\zeta'):= z\nabla_1 k(\tilde y,\tilde y')  + z'\nabla_1 k(\tilde y',\tilde y) \in \Rb^{d}.$
\textcolor{black}{Obviously, $\nabla_1 k(\tilde y,\tilde y')$ denotes the gradient of the kernel $k$ with respect to its first argument and evaluated at $(\tilde y,\tilde y')$.}
Moreover, set 
$$h_{(1)}(\zeta'):= \Eb_{\zeta\sim P_\zeta}[h(\zeta,\zeta')]-\Eb_{(\bar\zeta_1,\bar\zeta_2) \sim P_\zeta \otimes P_\zeta}[h(\bar\zeta_1,\bar\zeta_2)] .$$ 


Additionally, for any $y\in \Rb^q$, set the $d$-dimensional vectors $g(\zeta,y):= 2z\nabla_1 k(\tilde y,y)$, 
$$g_{(1)}(\zeta):= \Eb_{Y\sim P_0}[g(\zeta,Y)]- \Eb[g(\zeta_i,y_t)],\; \text{and} \; g_{(2)}(y):= \Eb_{\zeta\sim P_{\zeta}}[g(\zeta,y)]- \Eb[g(\zeta_i,y_t)].$$
For any $l\in \{1,\ldots,d \}$, define the map $H_l:\Rb^{2q(1+d)} \times \Rb^{2q}$ to $\Rb$ by
\begin{eqnarray*}
\lefteqn{  H_l(\zeta,\zeta',y,y'):=
\Big\{ g_l(\zeta,y) - \Eb_{Y\sim P_0}\big[g_l(\zeta,Y)\big] - \Eb_{\zeta\sim P_{\zeta}}\big[g_l(\zeta,y)\big]+ \Eb\big[g_l(\zeta_i,y_t)\big]\Big\}   }\\
&\times & \Big\{ g_l(\zeta',y') - \Eb_{Y\sim P_0}\big[g_l(\zeta',Y)\big] - \Eb_{\zeta\sim P_{\zeta}}\big[g_l(\zeta,y')\big]+ \Eb\big[g_l(\zeta_i,y_t)\big]\Big\} . \hspace{3cm}  
\end{eqnarray*}

\setcounter{assum}{0}
\begin{assum}
There exists a positive number $\delta_1$ such that 
$$  \max\Big( \int | H_l(\zeta,\zeta',y,y')|^{1+\delta_1} \, dP_{\zeta}(\zeta)\, dP_{\zeta}(\zeta') ;  \int | H_l(\zeta,\zeta',y,y')|^{1+\delta_1} \, 
d P_{(\zeta_{i_1},\zeta_{i_2})}(\zeta,\zeta') \Big) \leq M(y,y')<\infty ,$$
for any couple $(y,y')\in \Rb^{2q}$, any $l\in \{1,\ldots,d \}$, and any couple of indices $(i_1,i_2)$, $i_1<i_2$, with
$$ \sup_{t,t'}\Eb\Big[  M^{1/(1+\delta_1)}(y_{t},y_{t'}) \Big]=: M^* <\infty .$$
In addition, for some $\delta_2>0$, 
$$  \max\Big( \int | H_l(\zeta,\zeta',y,y')|^{1+\delta_2} \, dP_0(y)\, dP_0(y') ;  \int | H_l(\zeta,\zeta',y,y')|^{1+\delta_2} \, 
d P_{(y_{t_1},y_{t_2})}(y,y') \Big) \leq \bar M(\zeta,\zeta')<\infty ,$$
for any couple $(\zeta,\zeta')\in \Rb^{2q(1+d)}$ and any couple of indices $(t_1,t_2)$, $t_1<t_2$, and
$$ \sup_{i,j}\Eb\Big[  \bar M^{1/(1+\delta_2)}(\zeta_{i},\zeta_{j}) \Big]=:\bar M^* <\infty .$$
\label{AN_theta_NT_technik}
\end{assum}

\begin{assum}
The process $(y_t)$ is absolutely regular, with $\Eb\big[ \| g_{(2)}(y_t)\|^{2+\delta}\big]<\infty$
and $\sum_{t\geq 1} \beta(t)^{\delta/(2+\delta)}<\infty$ for some $\delta>0$.
Moreover, $\beta(t)=O(t^{-(2+\delta_2')/\delta_2'})$ for some $\delta_2' \in \big(0,\min(\delta,\delta_2)\big)$, where $\delta_2$ is defined in \Cref{AN_theta_NT_technik}.
\label{AN_theta_NT_beta}
\end{assum}

\begin{assum}
The $\Rb^{q(1+d+d^2)}$-valued process $(\xi_i)_{i\geq 1}$ is absolutely regular. 
Its mixing coefficients $\big(\beta_\xi(n)\big)_{n\geq 1}$ satisfy 
$\beta_\xi(n)=O(n^{-\kappa})$ for some $\kappa>0$.
\label{AN_theta_NT_u}
\end{assum}

\begin{assum}
The process $(\zeta_i)$ is absolutely regular. 
Its sequence of beta mixing coefficients $\big(\beta_\zeta(n)\big)_{n\geq 1}$ satisfies
$\sum_{n\geq 1} \beta_\zeta(n)^{\delta_\zeta/(2+\delta_\zeta)}<\infty$ for some $\delta_\zeta>0$; moreover,
\begin{equation}
\Eb\big[ \|2h_{(1)}(\zeta_i)-g_{(1)}(\zeta_i)\|^{2+\delta_\zeta}\big]<\infty,\; \text{and}\;
\sup_{ i<j} \Eb\big[ \|h(\zeta_{i},\zeta_j)\|^{2+\delta_\zeta}  \big] < \infty.
\label{cond_moment_h_g1}
\end{equation}
Finally,
$\beta_\zeta(n)=O(n^{-(2+\delta_1')/\delta_1'})$ for some $\delta_1' \in \big(0,\min(\delta_\zeta,\delta_1)\big)$, where $\delta_1$ is defined in \Cref{AN_theta_NT_technik}. 
\label{AN_theta_NT_beta_zeta}
\end{assum} 

Obviously, \Cref{AN_theta_NT_technik} is satisfied when the map $(u,y) \mapsto \nabla_\theta 
\psi^\top(\theta^*;u) \nabla_1 k\big(\psi(\theta^*;u),y\big) $ is a.s. bounded on $\text{Supp}(P_\Ub) \times \big\{ \text{Supp}(y_t)\cup \text{Supp}(P_0) \big\}\subset \Ub\times \Rb^q$.
If, moreover, $(u,u') \mapsto \nabla_\theta 
\psi^\top(\theta^*;u) \nabla_1 k\big(\psi(\theta^*;u),\psi(\theta^*;u')\big) $ is a.s. bounded on $\text{Supp}(P_\Ub)^2$, then~(\ref{cond_moment_h_g1}) is satisfied too.
\textcolor{black}{Note that, in the ISMMD case, the sequences $(\zeta_i)$ and $(\xi_i)$ are i.i.d. As a consequence, the conditions on mixing coefficients in \Cref{AN_theta_NT_u} and \Cref{AN_theta_NT_beta_zeta} are trivially satisfied.
}


In addition, define the family of maps $\Gc:=\{ \bar g_\theta : \Ub\times \Rb^q \rightarrow \Rb^{q\times q} ; \theta \in \Theta  \}$ by
$$ \frac{\bar g_\theta(u,y)}{2}:=\nabla_{\theta} \psi^\top(\theta,u) \nabla^2_{1,1} k\big(\psi(\theta,u),y\big)\nabla_{\theta^\top} \psi(\theta,u) + \sum_{l=1}^q \partial_{l} k\big(\psi(\theta,u),y\big)\nabla^2_{\theta,\theta^\top} \psi_l(\theta,u).$$

Moreover, set $\Hc:=\{ \bar h_\theta :  \Ub^2 \rightarrow \Rb^{q\times q} ; \theta \in \Theta  \}$, where
$ \bar h_\theta(u,u'):=\tilde h_\theta(u,u')+\tilde h_\theta(u',u)$, with
\begin{eqnarray*}
    \lefteqn{ \tilde h_\theta(u,u'):=  
    \nabla_{\theta} \psi^\top(\theta,u) \nabla^2_{1,1} k\big(\psi(\theta,u),\psi(\theta,u')\big)  \nabla_{\theta^\top} \psi(\theta,u) }\\
    &+& \nabla_{\theta} \psi^\top(\theta,u) \nabla^2_{1,2} k\big(\psi(\theta,u),\psi(\theta,u')\big)  \nabla_{\theta^\top} \psi(\theta,u')  
    + \sum_{l=1}^q \partial_{l} k\big(\psi(\theta,u),\psi(\theta,u')\big)  \nabla^2_{\theta,\theta^\top} \psi_l(\theta,u).    
\end{eqnarray*}
Note that $\bar g_{\theta^*}(u,y)$ (resp. $\bar h_{\theta^*}(u,u')$) is a measurable map of $\big(\xi(u),y\big)$ (resp. $\big(\xi(u),\xi(u')\big)$).

\begin{assum}
\label{cond_Holder_AN_hattheta_NT}
There exist some measurable maps $\bar H(\cdot,\cdot)$ and $\bar G(\cdot,\cdot)$ such that, for every $\theta$ and $\theta'$ in a neighborhood of $\theta^*$ and every $(u,v)\in \Ub^2$, we have
\begin{equation*}
\big\|\bar h_\theta(u,v) - \bar h_{\theta'}(u,v) \big\| \leq 
\bar H(u,v) \| \theta - \theta'\|^\varsigma,\;\;    
\sup_{1\leq i<i'< \infty}\Eb[\bar H(\underline{u}^{(i)},\underline{u}^{(i')}) ]<\infty,
\label{cond_LLN_Ustat}
\end{equation*}
for some $\varsigma>0$. Moreover, for some $\tilde\varsigma>0$ and any $y\in \Rb^q$, 
$$  \big\|\bar g_\theta(u,y) - \bar g_{\theta'}(u,y) \big\| \leq 
\bar G(u,y) \| \theta - \theta'\|^{\tilde\varsigma},\;\; \Eb[ \bar G(\underline{u}^{(i)} ,y_t)  ]<\infty.$$
Finally, for some $\nu \in (0,1]$ and some $r>2$, we have 
\begin{equation}
    \sup_{i<i'}\Eb\big[ \| \bar h_{\theta^*}(\underline{u}^{(i)},\underline{u}^{(i')}) \|^{1+\nu} \big]<\infty,\;\; \text{and} \;\; \Eb\big[ \| \bar g_{\theta^*}(\underline{u}^{(i)},y_t)\|^r   \big] <\infty.
    \label{cond_LLN_barh}
\end{equation}
\label{lipschitz_ULLN}
\end{assum}
Define 
$\tilde \Sigma:=\int \bar h_{\theta^*}(u,u')\, dP_\Ub(u)\, dP_\Ub(u') -2\Eb\big[ \bar g_{\theta^*}(\underline{u}^{(i)},y_t) \big],$
\begin{eqnarray*}
\lefteqn{  \Sigma_1 := \Eb\big[ (2h_{(1)}- g_{(1)})(\zeta_1) (2h_{(1)}- g_{(1)})^\top(\zeta_1)\big] }\\
&+&
\sum_{i>1} \Big\{ \Eb\big[ (2h_{(1)}- g_{(1)})(\zeta_1) (2h_{(1)}- g_{(1)})^\top(\zeta_i)\big] + \Eb\big[ (2h_{(1)}- g_{(1)})(\zeta_i) (2h_{(1)}- g_{(1)})^\top(\zeta_1)\big] \Big\},\;\text{and} 
\end{eqnarray*}  
$$  \Sigma_2 := \Eb\big[g_{(2)}(y_1) g_{(2)}^\top(y_1)\big] +
\sum_{t>1} \Big\{ \Eb\big[ g_{(2)}(y_1) g_{(2)}^\top(y_t)\big] + \Eb\big[ g_{(2)}(y_t) g_{(2)}^\top(y_1)\big] \Big\}.$$

\begin{prop}
\label{AN_theta_NT}
In addition to \Cref{AN_theta_NT_technik}-\Cref{lipschitz_ULLN}, 
suppose $\tilde \theta_{N,T}- \theta^*=o_P(1)$,
$\theta^*$ belongs to the interior of $\Theta$, and $N$ and $T$ tend to the infinity.
\begin{enumerate}
    \item[(1)] If $N/(N+T)\rightarrow \lambda$ for some $\lambda \in (0,1)$, then
$$\sqrt{N+T}\Big\{\tilde{\theta}_{N,T} - \theta^*\big\} \xrightarrow[N,T\rightarrow\infty]{ \text{law} } \Nc\Big(0, \tilde\Sigma^{-1}\Sigma \tilde\Sigma^{-1}\big),\;\; \text{where} \; \; \Sigma:= \Sigma_1/\lambda + \Sigma_2/(1-\lambda);$$ 
\item[(2)] if $N/T\rightarrow 0$, then
$\sqrt{N}\Big\{\tilde{\theta}_{N,T} - \theta^*\big\} \xrightarrow[N,T\rightarrow\infty]{ \text{law} } \Nc\Big(0, \tilde\Sigma^{-1}\Sigma_1 \tilde\Sigma^{-1}\big);$
\item[(3)] if $N/T\rightarrow \infty$, then
$\sqrt{T}\Big\{\tilde{\theta}_{N,T} - \theta^*\big\} \xrightarrow[N,T\rightarrow\infty]{ \text{law} } \Nc\Big(0, \tilde\Sigma^{-1}\Sigma_2 \tilde\Sigma^{-1}\big).$
\end{enumerate}
\end{prop}

Since the laws of the processes $(y_t)$ and $(\tilde y_i)$ may be different (except when the true underlying model is well-specified, i.e., $P_0=P_{\theta^*}$), their associated mixing coefficients are generally different.
Moreover, since $\zeta_i$ is a measurable map of $\xi_i$,  $\beta_\zeta(n)\leq \beta_\xi(n)$. 

\textcolor{black}{As in \Cref{rem_CS_AN_hattheta_T}, note that \Cref{cond_Holder_AN_hattheta_NT}
may be easily checked when similar H\"older-type conditions apply to the kernel, to the maps $\theta\mapsto \psi(\theta,u)$, and to their derivatives (up to second order), under some convenient conditions of integrability.}

\begin{rem}
As in \Cref{remark_ULLN_thetaT}, our H\"older-type \Cref{lipschitz_ULLN} could be replaced by alternative conditions of regularity.
For example, Theorem 3.1 in~\cite{arconesyu1994ustatmixing} yields a ULLN for the U-process indexed by the kernels $\bar h_\theta$, $\theta\in \Theta$, but under the rather demanding assumption that the maps in $\Hc$ are uniformly bounded. 
\end{rem}

\section{Implementation}\label{sec:implementation}

In this section, 
we discuss a practical method for solving Program (\ref{def_estimators}) to obtain the ISMMD $\tilde{\theta}^{(1)}_{N,T}$ and PSMMD $\tilde{\theta}^{(2)}_{N,T}$ estimators.
Moreover, \textcolor{black}{without a lack of generality}, we restrict ourselves to observations of the form $y_t = (x_{t},x_{t-1},\ldots,x_{t-p})$ for a given $p\geq 0$. 
The simulated vectors $\tilde{y}_i$ will be constructed in the same manner. 
To simplify notations, we will drop the $N,T$ indexing in $\tilde{\theta}^{(k)}_{N,T}$, $k\in \{1,2\}$. 
We propose to employ a gradient descent-based procedure to obtain a numerical solution of (\ref{def_estimators}). 
The iterations will depend on the  generation method of the simulated realizations $\tilde{y}_i$. 

Let us consider ISMMD, i.e., the estimator $\tilde{\theta}^{(1)}$.
First, we generate paths of innovations $\underbar{u}^{(i)}$ for $i\in \{1,\ldots , N\}$.
Let $\tilde{\theta}^{(1)}_l$ be the current estimator of $\theta^*$ at the $l$-th iteration of the algorithm, given an initial parameter value $\tilde{\theta}^{(1)}_0$. For a step size vector $\eta_l$, the gradient descent algorithm updates the estimator $\tilde{\theta}^{(1)}_{l}$ at step $l+1$ as follows:
\begin{itemize}
    \item[(i)] Compute the vector $\tilde{y}_i = (x^{(i)}_{T_0},x^{(i)}_{T_0-1},\ldots,x^{(i)}_{T_0-p})$ based on $\tilde{\theta}^{(1)}_l$, some fixed $T_0>>1$ and the innovations $\underbar{u}^{(i)}$ for $i\in \{1,\ldots,N\}$, using~\eqref{not_simul_y}.
    \item[(ii)] Update $\tilde{\theta}^{(1)}_l$ as $\tilde{\theta}^{(1)}_{l+1}=\tilde{\theta}^{(1)}_l-\eta_l \circ \nabla_{\theta}\Dc^2(\hat P_{\tilde{\theta}^{(1)}_l},P_T)$, with $\circ$ the element-by-element product.
\end{itemize}
Alternatively, it is tempting to re-sample the innovations at each step of the algorithm. In other words, we can also consider the following strategy, that we will refer to as ISMMD-SGD: for a given initial parameter value $\tilde{\theta}^{(1)}_0$ and a step size sequence $(\eta_l)$, update $\tilde{\theta}^{(1)}_{l}$ at step $l+1$ by the following procedure:
\begin{itemize}
    \item[(i)] Generate the innovations $\underbar{u}^{(i),l}$ for $i\in \{1,\ldots, N\}$. For any $i$, compute the vector $\tilde{y}_i = (x^{(i)}_{T_0},x^{(i)}_{T_0-1},\ldots,x^{(i)}_{T_0-p})$ based on $\tilde{\theta}^{(1)}_l$ and the innovations $\underbar{u}^{(i),l}$.
    \item[(ii)] Update $\tilde{\theta}^{(1)}_l$ as $\tilde{\theta}^{(1)}_{l+1}=\tilde{\theta}^{(1)}_l-\eta_l \circ \nabla_{\theta}\Dc^2(\hat P_{\tilde{\theta}^{(1)}_l},P_T)$.
\end{itemize}
The first version is designed to target ISMMD, or equivalently $\tilde{\theta}^{(1)}$.
The second version is a stochastic gradient algorithm and is therefore denoted by ISMMD-SGD; it targets the ``ideal'' estimator $\hat{\theta}_T$.
This is actually the algorithm that was implemented in~\cite{briol2019statistical,cherief2022finite} or \cite{alquier2022estimationcopulammd}. Under suitable convexity assumptions on the MMD criterion, it can be shown that ISMMD-SGD actually converges to $\hat{\theta}_T$: see Section 5.3 in~\cite{cherief2022finite} for example. 
Our simulation experiments in Subsection \ref{supp:subsec:SGD} of the Appendix suggest that ISMMD-SGD performs better in practice than ISMMD. Therefore, all experiments presented hereafter are conducted using ISMMD-SGD. To simplify the notation, unless stated otherwise, the estimator obtained by ISMMD refers to ISMMD-SGD.

We now turn to the PSMMD case, where we aim to obtain $\tilde{\theta}^{(2)}$. Let $\tilde{\theta}^{(2)}_l$ denote the current estimator of $\theta^*$ at the $l$-th iteration of the algorithm. 
Following the stochastic gradient version of the ISMMD procedure, for a given initial parameter value $\tilde{\theta}^{(2)}_0$ and a step size $\eta_l$, the gradient descent procedure updates $\tilde{\theta}^{(1)}_{l}$ at step $l+1$ as follows:
\begin{itemize}
    \item[(i)] Generate a path of innovations $\underline{u}^{(l)}$. Using~\eqref{not_simul_y}, compute the associated path $(x_{1-p},\ldots,x_{N})$ given $\tilde{\theta}^{(2)}_l$. 
    Define $\tilde{y}_i = (x_{i},x_{i-1},\ldots,x_{i-p})$ for $i\in \{1,\ldots,N\}$.
    \item[(ii)] Update $\tilde{\theta}^{(2)}_l$ as $\tilde{\theta}^{(2)}_{l+1}=\tilde{\theta}^{(2)}_l-\eta_l \circ \nabla_{\theta}\Dc^2(\hat P_{\tilde{\theta}^{(2)}_l},P_T)$.
\end{itemize}
This version, denoted PSMMD-SGD, also relies on stochastic gradients. As in the ISMMD method, our experiments reported in Subsection \ref{supp:subsec:SGD} of the Appendix indicate that PSMMD-SGD outperforms a variant of PSMMD in which the innovations are generated only once. Therefore, unless stated otherwise, all experimental results referring to PSMMD correspond to PSMMD-SGD.
The CSMMD case can be handled in an analogous manner.
In any case, 
$\Dc^2(\hat P_{\tilde{\theta}_l},P_T) :=  \sum_{i,i'=1}^{N}  k(\tilde y_i,\tilde y_{i'})/N^2 -  
 2 \sum_{t'=1}^{T} \sum_{i=1}^{N} k(\tilde y_i,y_{t'})/(NT)   + \sum_{t,t'=1}^{T} k(y_t,y_{t'})/T^2$, with $\tilde{\theta}_l=\tilde{\theta}^{(1)}_l$ or $\tilde{\theta}_l=\tilde{\theta}^{(2)}_l$. 
 
 Obviously, an explicit calculation of the derivative $\nabla_{\theta}\Dc^2(\hat P_{\tilde{\theta}_l},P_T)$ is most of the time impossible. Therefore, we employ numerical differentiation to estimate the latter derivative. Furthermore, the parameter $\eta_l$ must be suitably calibrated. In the same spirit as adaptive gradient descent, we use adaptive weights to update each coordinate of $\tilde{\theta}_l$, and construct the $j$-th component of $\eta_l$ as $\eta_{l,j}=R/(\sqrt{\sum^l_{m=1}\{\nabla_{\theta_j}\Dc^2(\hat P_{\tilde{\theta}_m},P_T)\}^2+\epsilon})$, for some $\epsilon>0$. We will set $R=0.025$ and $\epsilon=10^{-6}$. 
 Other forms of updating formulas are obviously possible: see, e.g., \cite{duchi2011}. Finally, we will use the Gaussian kernel $k(y,y') := \exp\big(-\|y-y'\|^2/(2\sigma^2)\big)$ throughout our experiments. To calibrate $\sigma$, following \cite{gretton2012kernel}, we rely on the median heuristic applied to the sample of observations $y_t$, $t\in \{1,\ldots,T\}$. \textcolor{black}{Subsection \ref{supp:subsec:kernel_comp} provides numerical experiments on the sensitivity of the estimation accuracy with respect to using the Gaussian and the Laplace kernels. These experiments suggest that the Gaussian kernel provides more precise estimators.}
 The code for implementation and the numerical experiments detailed hereafter are publicly available in the following Github repository: \url{https://github.com/Benjamin-Poignard/MMD_time_series}. 

\section{Numerical experiments}\label{sec:experiments}

The performance of our estimation methods is assessed through numerical experiments. We successively consider settings in which the true underlying data generating process (DGP) follows a stochastic volatility (SV), GARCH(1,1), ARMA(1,1), nonlinear MA(1), or Ricker model. We generate $100$ independent batches of observations $(x_t)$. 
For each batch, once a path $(y_t)_{1\leq t\leq T}$, with $y_t = (x_{t},x_{t-1},\ldots,x_{t-p})$, is generated from the true DGP and paired with the simulated path $(\tilde{y}_i)_{1\leq i \leq N}$, we solve (\ref{def_estimators}) by the gradient descent algorithm previously described. This yields the estimators $\tilde{\theta}^{(k)}_{N,T}$, $k\in \{1,2\}$, for a given $p$. \textcolor{black}{These correspond to the ISMMD-SGD and PSMMD-SGD estimators of \Cref{sec:implementation}}. \textcolor{black}{For ISMMD, we set $T_0=100$. For both ISMMD and PSMMD, we set $N = 1000$}. At each iteration of the algorithm, innovations are drawn from a standard normal distribution. 
\textcolor{black}{Then, under correct model specification, we will evaluate the estimation accuracy by the $\ell_2$ distance $\|\tilde{\theta}^{(k)}_{N,T}-\theta^*\|_2$, where $\theta^*=\theta_0$ is the true parameter. This distance will be averaged over the $100$ batches. For each model setting, we will also consider estimation under misspecification (which will be called ``Case 2'')
\textcolor{black}{, where the true underlying DGP does not belong to the family of model DGPs (but is not ``too far away'' from them).}
\textcolor{black}{We will only consider misspecified models whose parameters can be ``mapped'' to those of the true DGP (see below).}
In this case, \textcolor{black}{we can still calculate} the $\ell_2$ error $\|\tilde{\theta}^{(k)}_{N,T}-\theta^*\|_2$, using the same value $\theta^*$ as before; the resulting simulation outcomes should be interpreted as a robustness assessment of the estimation procedure with respect to perturbations in the DGP.} All experiments are performed for a single \textcolor{black}{value $\theta^*$}. Section \ref{supp:sec:additional_experiments} of the Appendix contains the same experiments but implemented \textcolor{black}{with alternative values of $\theta^*$}. 

\subsection{SV model}\label{subsec:SV}

\begin{figure}[htbp]
\centering
  \begin{minipage}[b]{0.45\textwidth}
    \centering
    \includegraphics[width=0.92\linewidth]{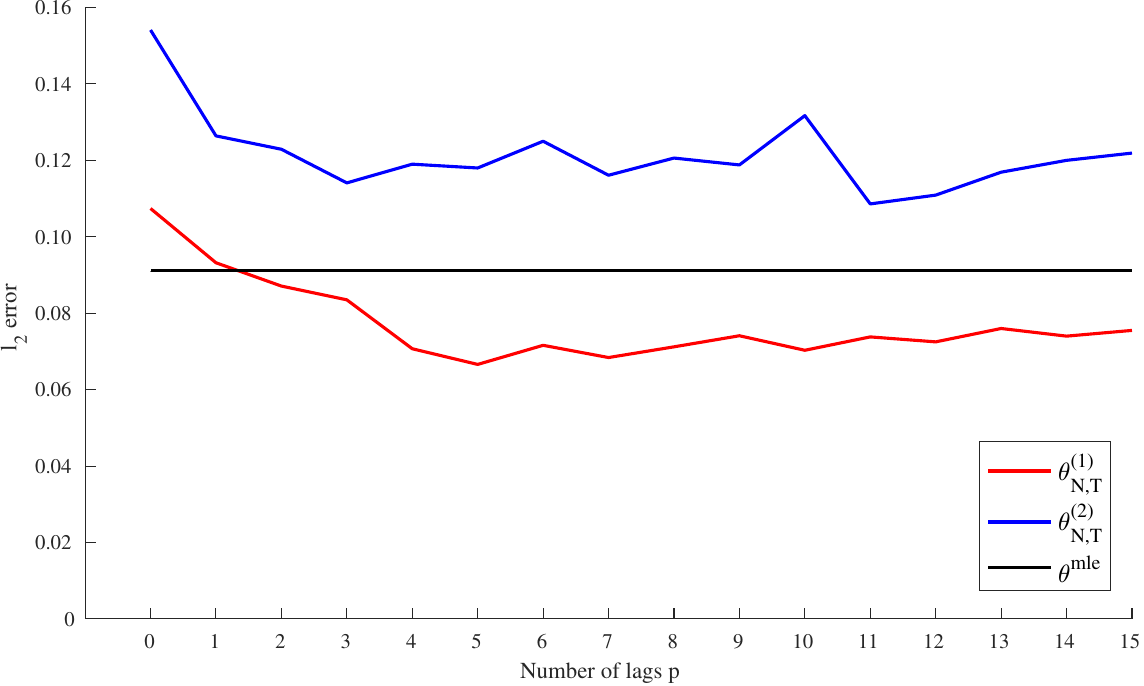}
    \subcaption{Case 1: $\eta_t \sim \mathcal{N}(0,1)$. $T=300$. \\ Estimation under correct specification.}\label{figure_SV_G_b9_300}
  \end{minipage}  
  \begin{minipage}[b]{0.45\textwidth}
    \centering
    \includegraphics[width=0.92\linewidth]{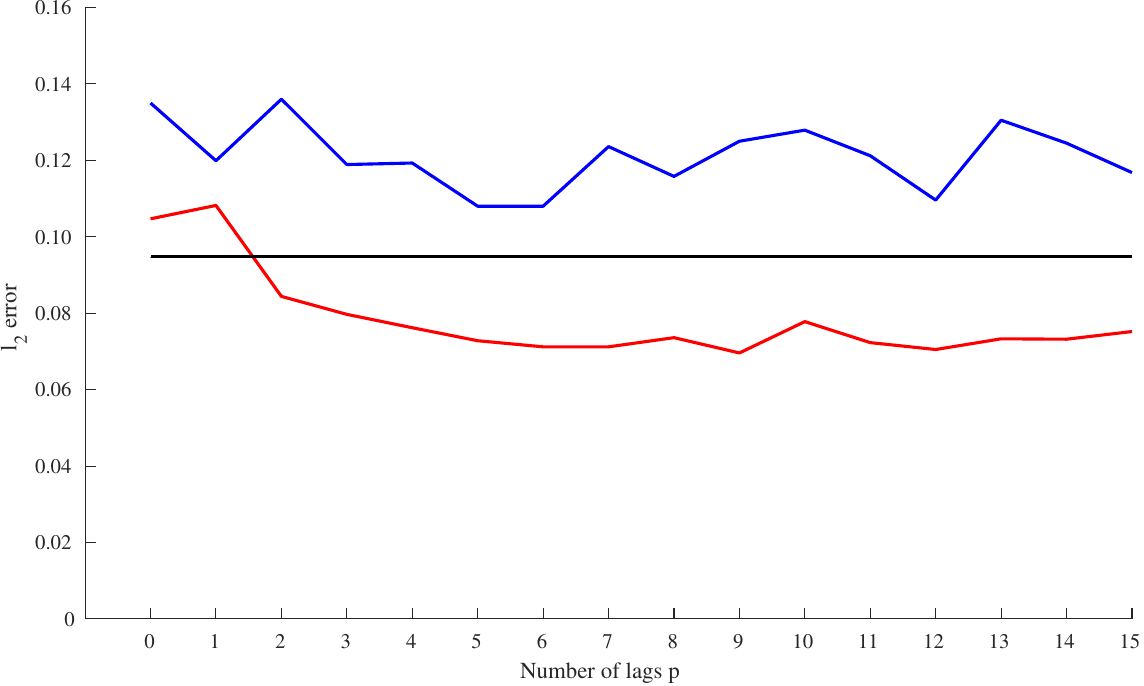}
    \subcaption{Case 2:  $\eta_t = t(3)/\sqrt{3}$. $T=300$. \\ Estimation under misspecification.}\label{figure_SV_S_b9_300}
  \end{minipage}

  \vspace*{0.8cm}
  
  \begin{minipage}[b]{0.45\textwidth}
    \centering
    \includegraphics[width=0.92\linewidth]{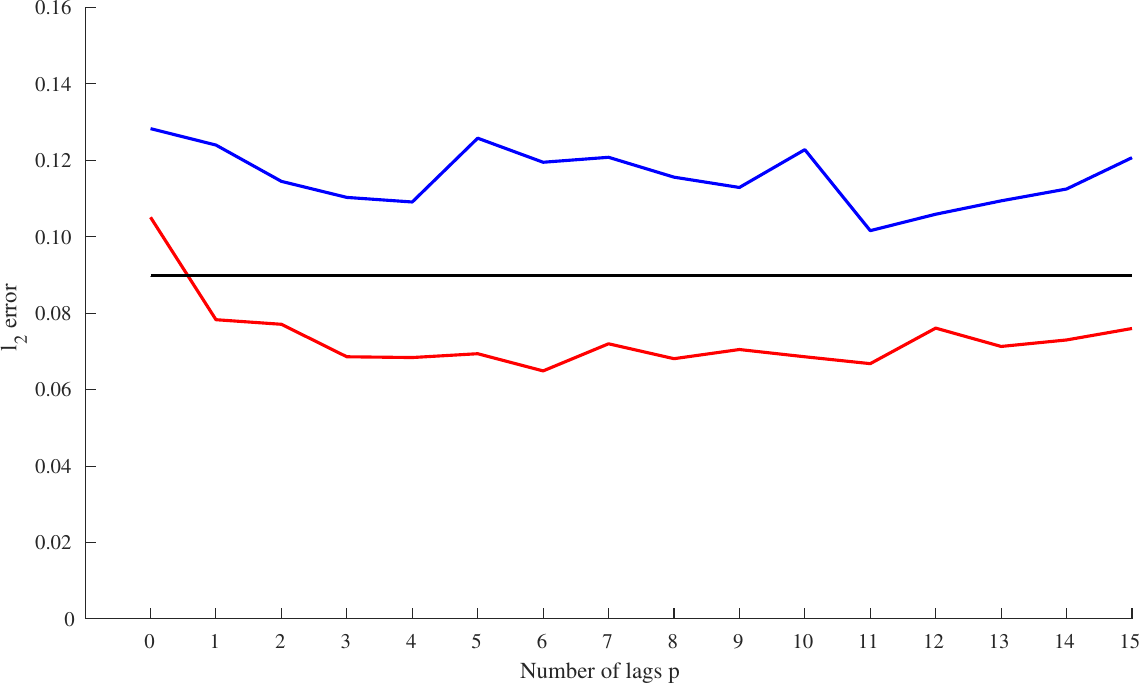}
    \subcaption{Case 1: $\eta_t \sim \mathcal{N}(0,1)$. $T=1000$. \\ Estimation under correct specification.}\label{figure_SV_G_b9_1000}
  \end{minipage}
  \begin{minipage}[b]{0.45\textwidth}
    \centering
    \includegraphics[width=0.92\linewidth]{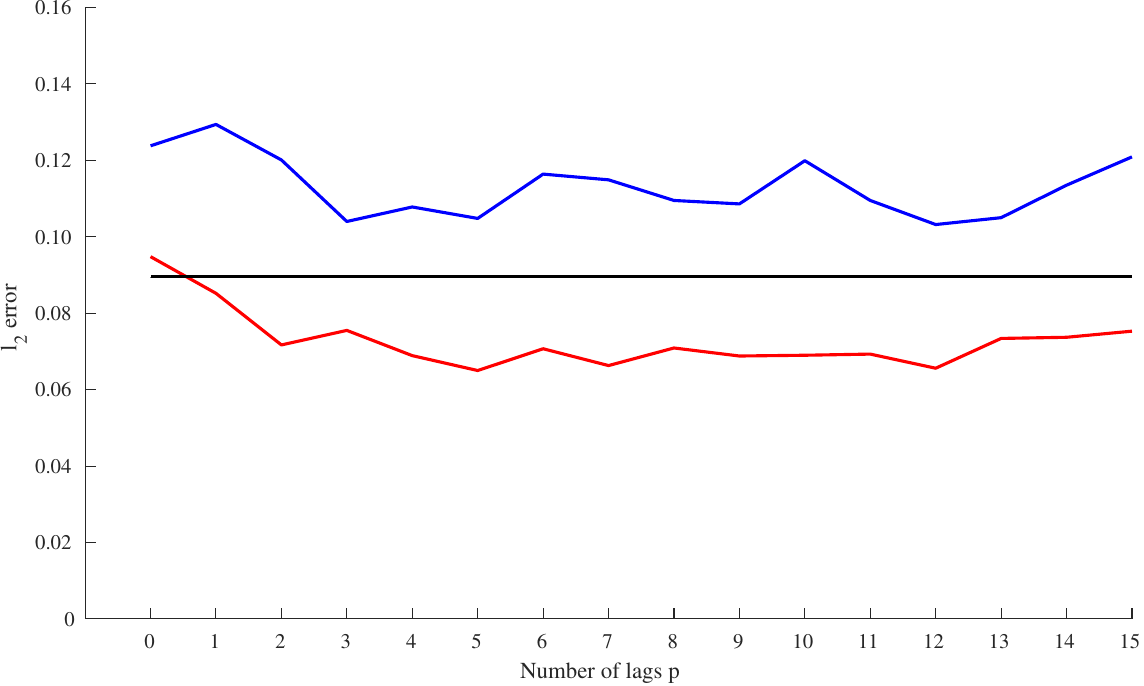}
    \subcaption{Case 2:  $\eta_t = t(3)/\sqrt{3}$. $T=1000$. \\ Estimation under misspecification.}\label{figure_SV_S_b9_1000}
  \end{minipage}
  \caption{SV model - $\theta^*=(0.9,0.1,0.2) $. Each point shows the average of 100 batches.}
\end{figure}

The time series is generated according to a stochastic volatility (SV) process:
\begin{equation*}
\forall t=1,\ldots,T,\;\;
x_t = \sigma_x \exp(0.5 \, h_t) v_t, \;\; \text{with}\;\;
h_t = \phi h_{t-1} + \sigma_{\eta} \eta_t,
\end{equation*}
where $(v_t)$ are i.i.d. centered Gaussian innovations with unit variance and $(\eta_t)$ is an i.i.d. centered sequence of innovations. We will consider two cases: in case 1, $\eta_t \sim \mathcal{N}(0,1)$; in case 2, $\eta_t = t(3)/\sqrt{3}$ where $t(3)$ is a centered Student distribution with $3$ degrees of freedom. Therefore, the sequence  $(u_t)$ of innovations is defined by $u_t=(v_t,\eta_t) $, for any $t$, the two processes $(v_t)$ and $(\eta_t)$ being independent. The parameter vector is $\theta=(\phi,\sigma_{\eta},\sigma_x) $ and we set the true parameter as $\theta^*=(0.9,0.1,0.2) $. We select the sample sizes $T=300$ and $T=1000$. 
For the sake of comparison, we consider the simulation-based MLE estimator $\hat{\theta}^{\text{mle}}$. The later maximizes a log likelihood function $\mathbb{L}(\theta)$ which is evaluated using the particle filter technique as detailed in Section 3.1 in \cite{pitt2014}. More precisely, at each time $t\in \{1,2,\ldots,T\}$, we consider the following steps: we have built $h^{(i)}_{t-1}$, for $i\in \{1,\ldots,K\}$; then,
\begin{itemize}
    \item for each given particle $i\in \{1,\ldots,K\}$, set $\tilde{h}^{(i)}_t=\phi h^{(i)}_{t-1}+\sigma_{\eta} \eta^{(i)}_t$, $\eta^{(i)}_t \sim \mathcal{N}(0,1)$; then compute $w^{(i)}_t=\exp\big(-x^2_t \exp(-\tilde{h}^{(i)}_t)/(2\sigma^2_x)\big)/ \sqrt{2\pi\sigma^2_x\exp(\tilde{h}^{(i)}_t)}$ and $W^{(i)}_t = w^{(i)}_t/\sum^K_{k=1}w^{(k)}_t$.
    \item for each $i\in \{1,\ldots,K\}$, sample $a_t^{(i)}\in\{1,\ldots,K\}$ with $\mathbb{P}(a_t^{(i)}=k)=W^{(k)}_t$, and set $h^{(i)}_t=\tilde{h}^{a^{(i)}_t}_t$.
\end{itemize}
The log likelihood is then defined by $\mathbb{L}(\theta):=\sum^T_{t=1}\log(K^{-1}\sum^K_{i=1}w^{(i)}_t)$. The iterative procedure is initialized by $h^{(i)}_0 \sim \Nc\big(0,\sigma^2_{\eta}/(1-\phi^2)\big)$ for any $i$.
We set $K = 5000$. The $\ell_2$ distances under correct specification and misspecification are displayed in Figures \ref{figure_SV_G_b9_300}-\ref{figure_SV_S_b9_1000}. 
\textcolor{black}{Overall, the ISMMD-based procedure outperforms the simulated likelihood-based approach, contrary to PSMMD. Apparently, ISMMD is better able to capture the latent dynamics for the SV model.}

\subsection{GARCH model}\label{subsec:GARCH}

The second model is a GARCH(1,1) process, where the true DGP is as follows: 
\begin{equation*}
\forall t=1,\ldots,T, \;\; x_t = \sqrt{h_t}u_t, \;\; \text{with} \;\;
h_t = \omega + \beta h_{t-1} + \alpha x^2_{t-1},
\end{equation*}
where the innovations of the sequence $(u_t)$ are i.i.d. and centered variables with unit variance. We still specify two distribution cases: $u_t \sim \mathcal{N}(0,1)$ (case 1) and $u_t= t(3)/\sqrt{3}$ (case 2). The parameter is $\theta=(\omega,\beta,\alpha) $ and the true parameter is set as $\theta^*=(0.05,0.92,0.05) $.
The sample sizes are $T=300$ and $T=1000$. For each batch, we also compute the Gaussian quasi maximum likelihood (QML) estimator $\hat{\theta}^{\text{mle}}$ of the GARCH model. \textcolor{black}{In addition to the Gaussian QML, we will compute the Method of Negligibly Weighted Moment (MNWM) estimator proposed by \cite{hill2015}, which is a robust estimator. The MNWM procedure replaces the Gaussian
score equations by moment equations using transformed standardized innovations. Define $u_t=x_t/\sqrt{h_t}$ and let $s_t=\partial \log(h_t)/ \partial\theta$ be the score vector. The MNWM method replaces the usual score $(u_t^2-1)s_t$ by $m_t(\theta)=\big\{\psi(u_t,c)^2-\frac{1}{T}\sum_{j=1}^T\psi(u_j,c)^2\big\}s_t$, where $\psi(\cdot,c)$ is a bounded transformation depending on a threshold $c$. The MNWM estimator is obtained by minimizing the quadratic norm of the sample moments, that is $\hat{\theta}=\arg\min_{\theta}\left\| T^{-1} \sum_{t=1}^Tm_t(\theta)\right\|^2$. We consider three transformations $\psi(\cdot,c)$:
\begin{itemize}
    \item $\psi(u,c)=u\,\mathbf{1}(|u|\leq c)$, that is a simple trimming transform;
    \item $\psi(u,c)=u\big(1-u^2/c^2\big)^2\,\mathbf{1}(|u|\leq c)$, the Tukey bisquare transform;
    \item $\psi(u,c)=u\exp\big(-|u|/c\big)\,\mathbf{1}(|u|\leq c)$, i.e. an exponential transformation.
\end{itemize}
The resulting estimators will be denoted by $\hat{\theta}^{\text{trim}}$, $\hat{\theta}^{\text{tukey}}$ and $\hat{\theta}^{\text{exp}}$, respectively.
We employ the procedure detailed by \cite{hill2015} to specify $c$. For a given $\theta$, define the descending order statistics of the absolute standardized
residuals by $|u|_{(1)}\geq|u|_{(2)}\geq \cdots \geq|u|_{(T)}$. Let
$k_T=\min\big\{T-1,\,\max\big(1,\big\lfloor\frac{0.025T}{\log T}\big\rfloor\big)\big\}$
and use the parameter-dependent threshold $c_T(\theta)=|u|_{(k_T)}$. Then, we set $c=c_T(\theta)$ in the transformation $\psi(\cdot,c)$. 
}

\begin{figure}[htbp]
\centering
  \begin{minipage}[b]{0.45\textwidth}
    \centering
    \includegraphics[width=0.92\linewidth]{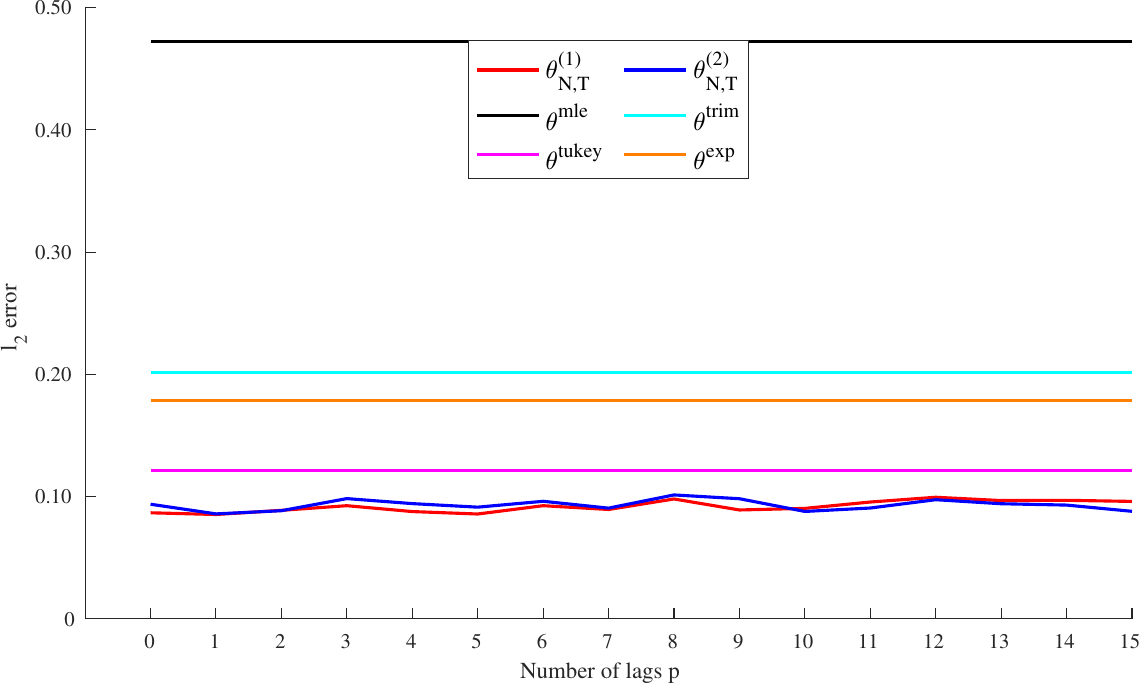}
        \subcaption{Case 1: $u_t \sim \mathcal{N}(0,1)$. $T=300$. \\ Estimation under correct specification.}\label{figures_GARCH_G_b92_c05_300}
  \end{minipage}
  \begin{minipage}[b]{0.45\textwidth}
    \centering
    \includegraphics[width=0.92\linewidth]{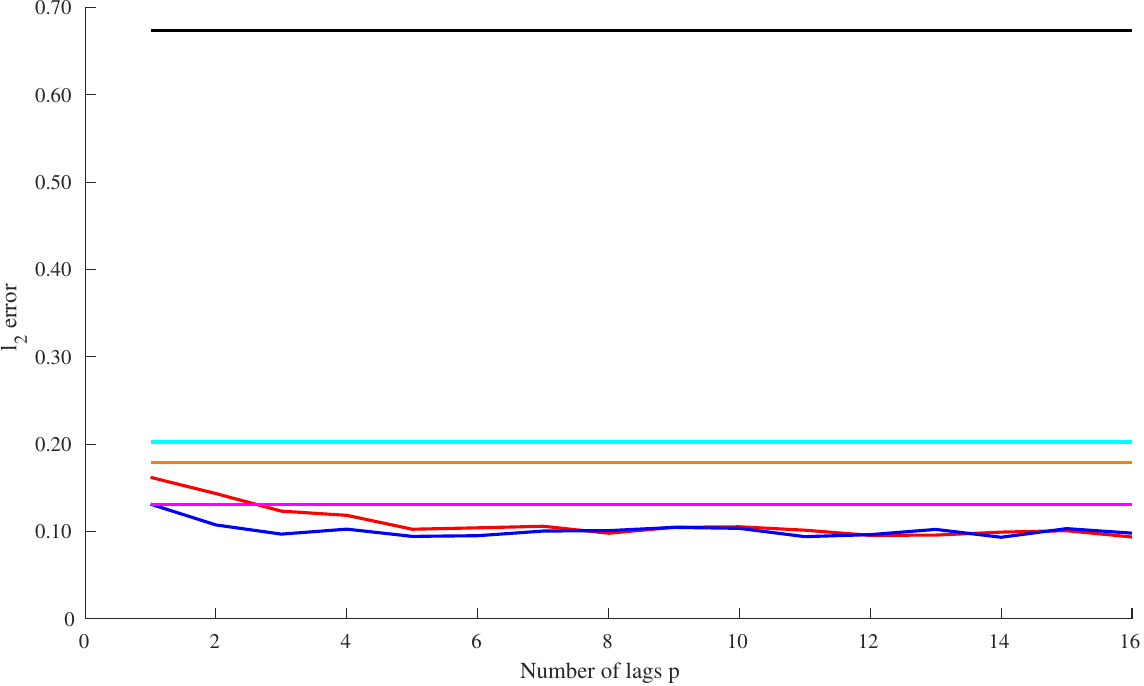}
        \subcaption{Case 2: $u_t = t(3)/\sqrt{3}$. $T=300$. \\ Estimation under misspecification.}\label{figures_GARCH_S_b92_c05_300}
  \end{minipage}

  \vspace*{0.8cm}
  
  \begin{minipage}[b]{0.45\textwidth}
    \centering
    \includegraphics[width=0.92\linewidth]{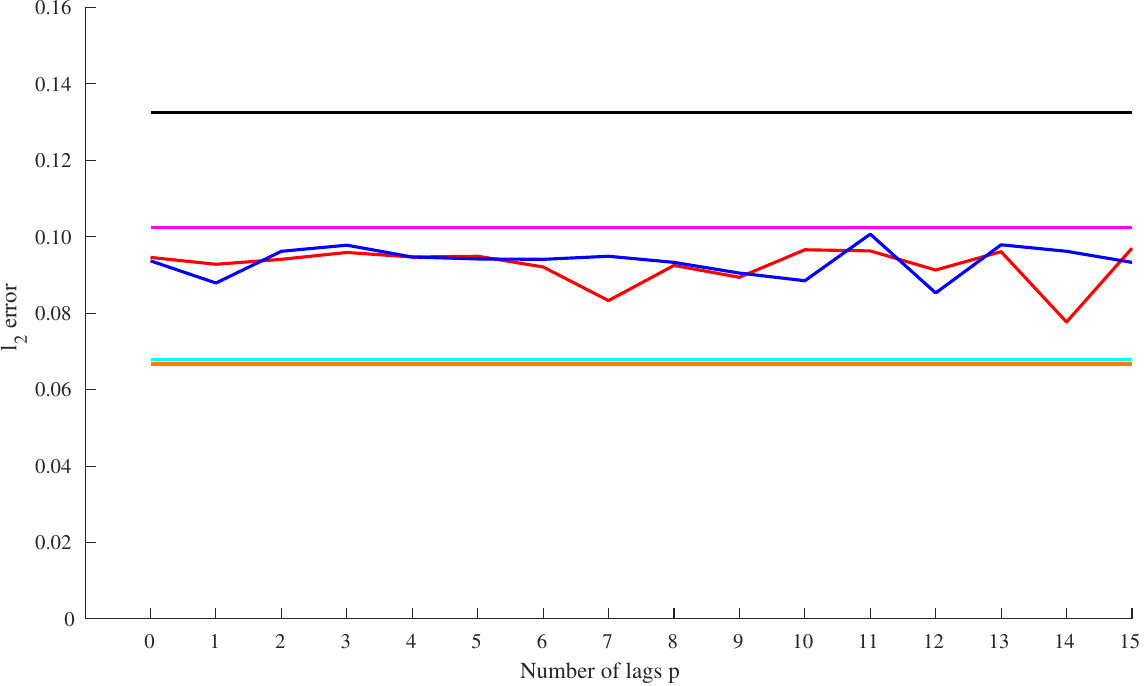}
        \subcaption{Case 1: $u_t \sim \mathcal{N}(0,1)$. $T=1000$. \\ Estimation under correct specification.}\label{figures_GARCH_G_b92_c05_1000}
  \end{minipage}
  \begin{minipage}[b]{0.45\textwidth}
    \centering
    \includegraphics[width=0.92\linewidth]{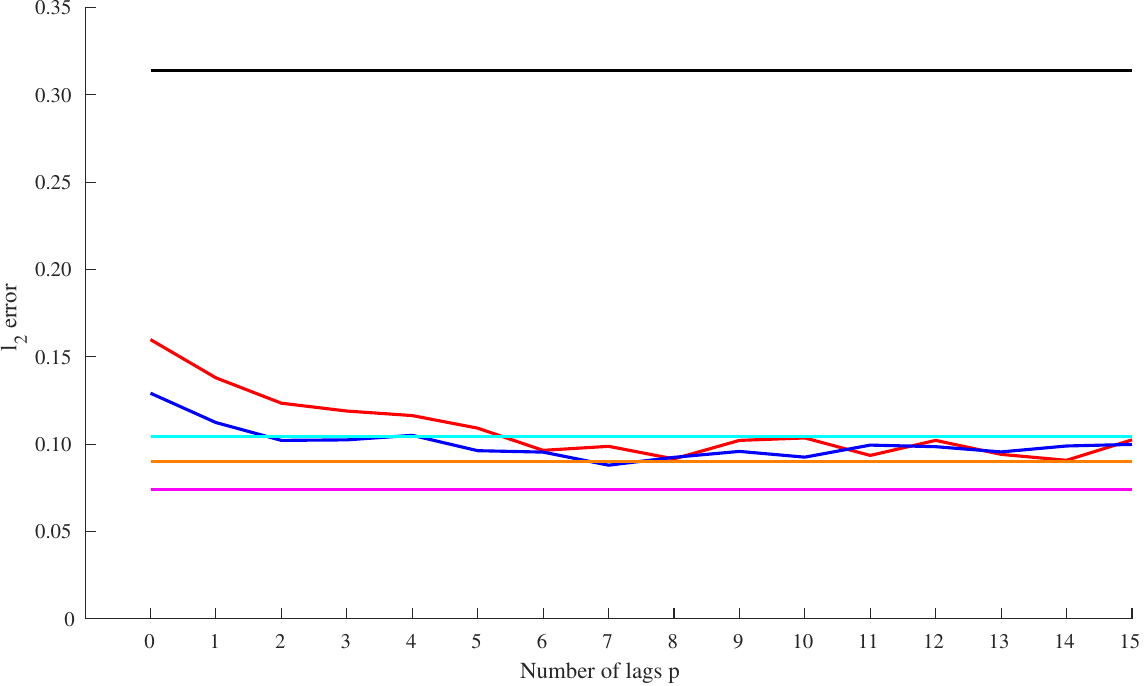}
        \subcaption{Case 2: $u_t = t(3)/\sqrt{3}$. $T=1000$. \\ Estimation under misspecification.}\label{figures_GARCH_S_b92_c05_1000}
  \end{minipage}
  \caption{GARCH model - $\theta^*=(0.05,0.92,0.05) $. Each point shows the average of 100 batches.}
\end{figure}

\textcolor{black}{The $\ell_2$ distances between the pseudo-true and the estimated parameters are displayed in Figures \ref{figures_GARCH_G_b92_c05_300}-\ref{figures_GARCH_S_b92_c05_1000}.
In any case (correct/mis-specification, different sample sizes), the QML estimator is clearly outperformed by the robust and the MMD methods. Interestingly,
the comparative advantage of MMD-based over QML-based estimators is significantly greater under misspecification, as depicted in Figures \ref{figures_GARCH_S_b92_c05_300}-\ref{figures_GARCH_S_b92_c05_1000}. The MMD also outperforms the MNWM method, for any given transformation $\psi(\cdot,\cdot)$, in the small sample regime. In the large sample regime and under misspecification, only the Tukey method provides better performances. 
This illustrates the robustness of our method when the estimated model slightly deviate from the true DGP. 
\textcolor{black}{We emphasise that our MMD-based methods are universal, whereas MNWM was designed specifically for GARCH models.}
Both ISMMD and PSMMD perform similarly, although in the misspecified case, ISMMD is slightly outperformed by PSMMD for $p\leq 10, T=1000$.}
    
\subsection{ARMA model}\label{subsec:ARMA}

\begin{figure}[htbp]
\centering
  \begin{minipage}[b]{0.45\textwidth}
    \centering
    \includegraphics[width=0.92\linewidth]{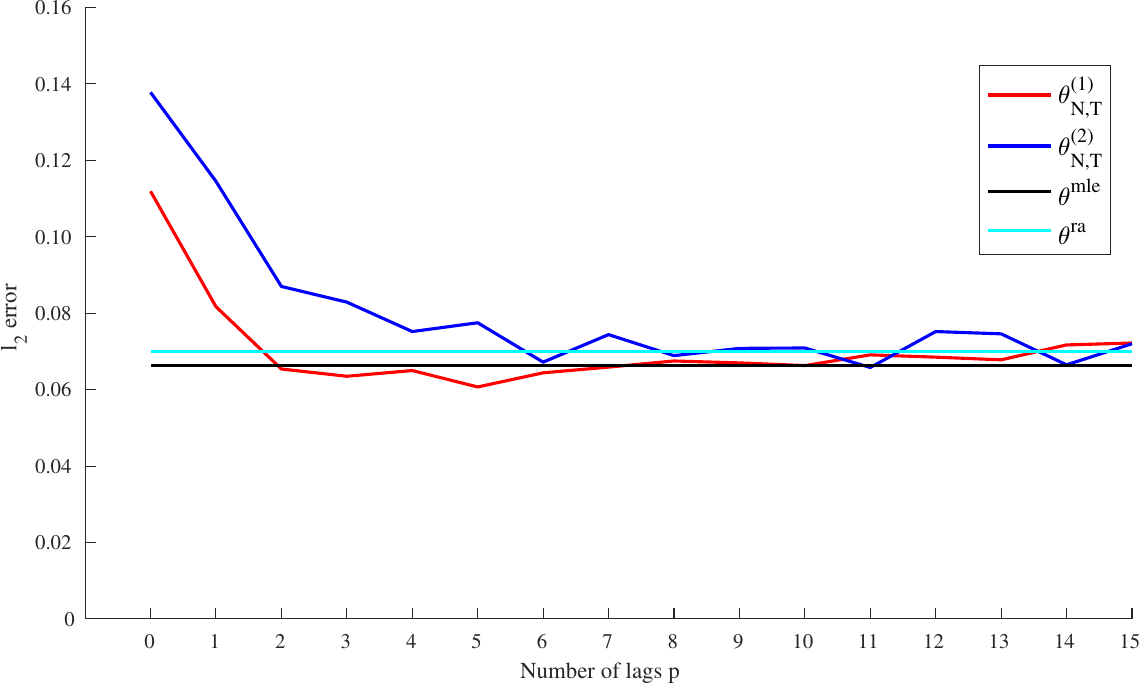}
    \subcaption{Case 1: $u_t \sim \mathcal{N}(0,1)$. $T=300$. \\ Estimation under correct specification.}\label{figures_ARMA_G_p8_t15_300}
  \end{minipage}
  \begin{minipage}[b]{0.45\textwidth}
    \centering
    \includegraphics[width=0.92\linewidth]{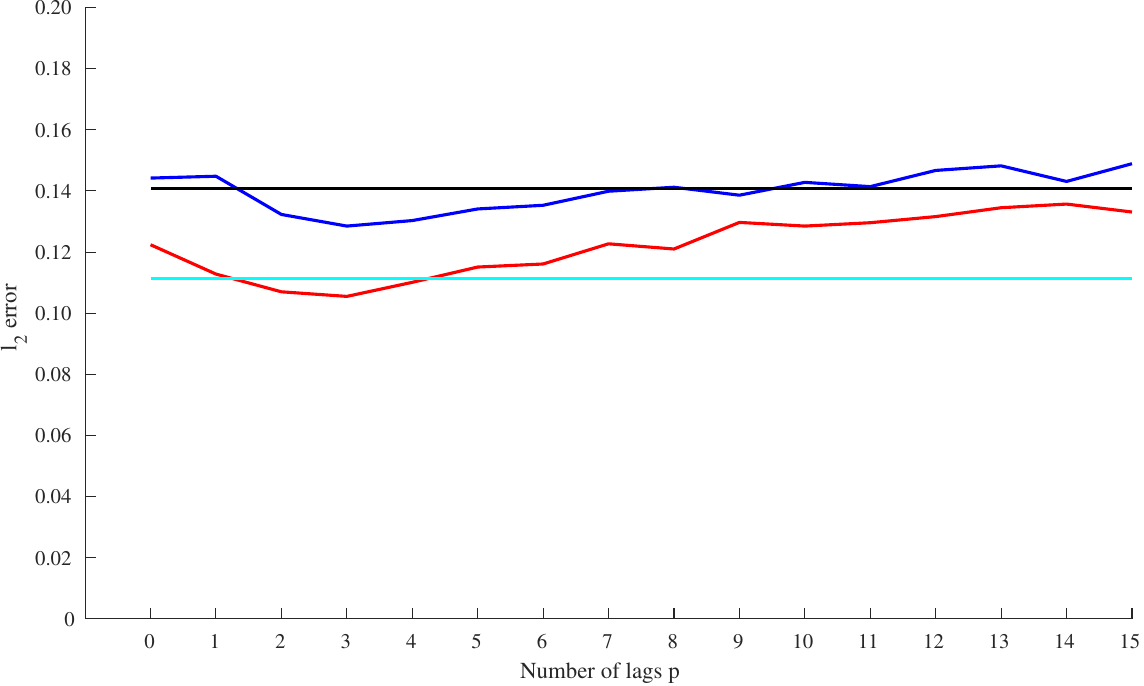}
    \subcaption{Case 2: $u_t \sim (1-\epsilon)\mathcal{N}(0,1)+\epsilon \mathcal{N}(0,\kappa)$. $T=300$. \\ Estimation under contamination.}\label{figures_ARMA_IO_p8_t15_300}
  \end{minipage}

  \vspace*{0.8cm}
  
  \begin{minipage}[b]{0.45\textwidth}
    \centering
    \includegraphics[width=0.92\linewidth]{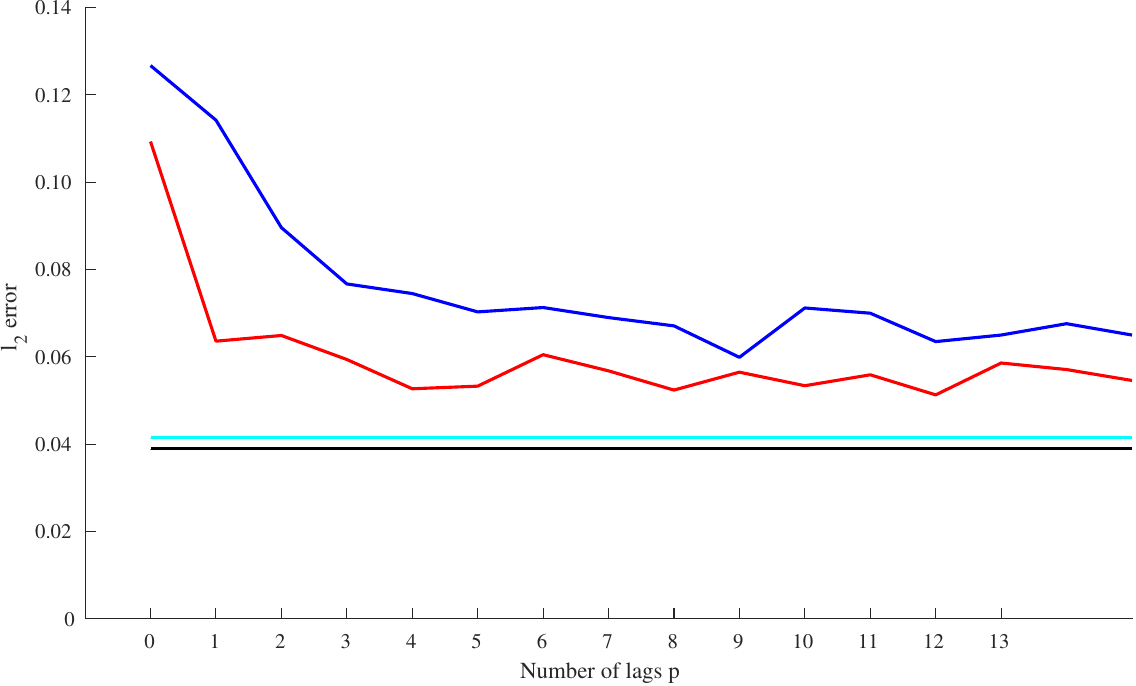}
    \subcaption{Case 1: $u_t \sim \mathcal{N}(0,1)$. $T=1000$. \\ Estimation under correct specification.}
\label{figures_ARMA_G_p8_t15_1000}  \end{minipage}
  \begin{minipage}[b]{0.45\textwidth}
    \centering
    \includegraphics[width=0.92\linewidth]{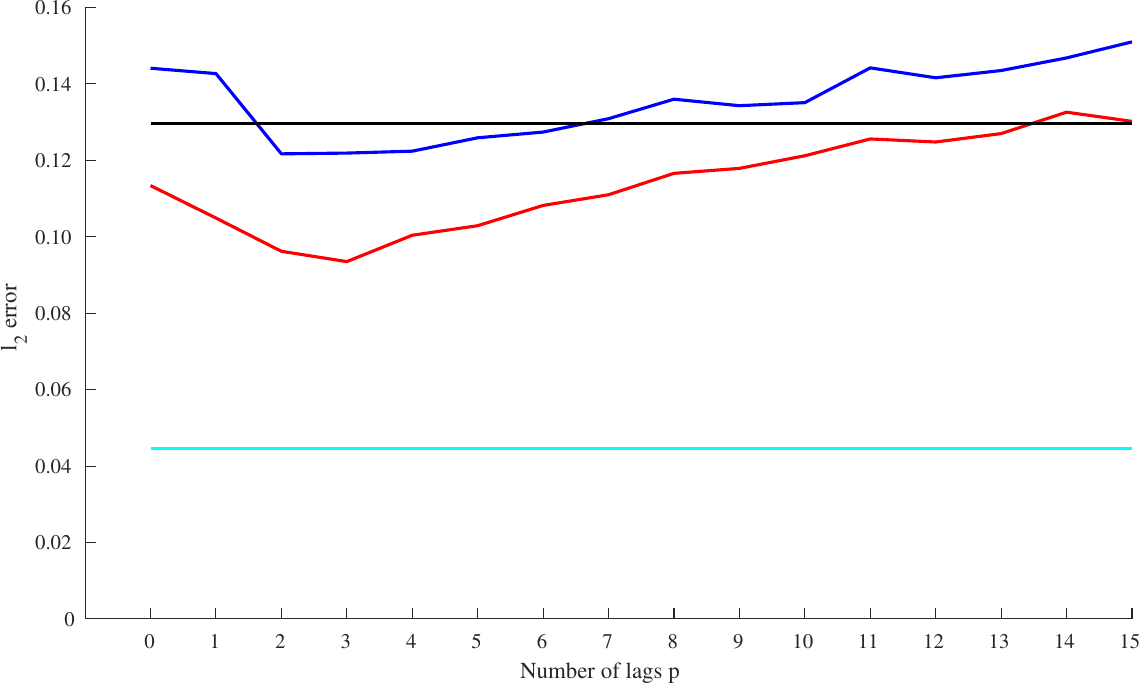}
    \subcaption{Case 2: $u_t \sim (1-\epsilon)\mathcal{N}(0,1)+\epsilon \mathcal{N}(0,\kappa)$. $T=1000$. \\ Estimation under contamination.}\label{figures_ARMA_IO_p8_t15_1000}
  \end{minipage}
  \caption{ARMA model - $\theta^*=(0.8,0.15,0.05) $. Each point shows the average of 100 batches.}
\end{figure}

Let us consider a classical ARMA(1,1) process: $x_t  = \phi x_{t-1} + v_t + \psi v_{t-1}$,
for any $t$, where $v_t = \sqrt{\sigma^2_{u}}u_t$, with $(u_t)$ an i.i.d. sequence. 
\textcolor{black}{We consider two cases for $u_t$: $u_t \sim \mathcal{N}(0,1)$ (case 1) and $u_t \sim (1-\epsilon)\mathcal{N}(0,1)+\epsilon \mathcal{N}(0,\kappa)$ (case 2),  where $0<\epsilon<1$ is the contamination probability and $\kappa>1$ is the outlier variance multiplier. Case 2 builds upon Equation (1.8) in \cite{bustos1986}. Throughout our experiments, we will set $\epsilon=0.1$ and $\kappa=25$. The parameter is $\theta=(\phi,\psi,\sigma^2_u) $ and the true parameter value is set as $\theta^* = (0.8,0.15,0.05) $.
The sample sizes are $T=300$ and $T=1000$. For each batch, in addition to the MMD-based estimator, we also compute the Gaussian QML estimator $\hat{\theta}^{\text{mle}}$ and the robust Residual Autocovariances (RA) estimator $\hat{\theta}^{\text{ra}}$ of \cite{bustos1986} of the ARMA model. The RA estimator replaces the classical residual autocovariances that appear in the estimating equations by bounded robust counterparts. For a given $\theta$, the residuals are computed recursively as $r_t=x_t-\phi x_{t-1}-\psi r_{t-1}$, $t=2,\ldots,T$, with the initial residual set equal to zero. Following Section 4.1 of \cite{bustos1986}, a robust estimator of the innovation scale is $\hat{\sigma}(\theta)=\text{median}_{2 \leq t \leq T}|r_t|/0.6745$. For each lag $\ell\geq 1$, we define the robust residual autocovariance by $\hat{\gamma}_{\ell}(\theta)=\frac{1}{T-1-\ell}\sum_{t=2+\ell}^{T}\eta\big(r_t(\theta)/\hat{\sigma}(\theta),r_{t-\ell}(\theta)/\hat{\sigma}(\theta)\big)$,
where $\eta(\cdot,\cdot)$ is a bounded bivariate transformation. Following \cite{bustos1986}, Equation (2.9'), we employ the Mallows transformation $\eta(u,v)=\rho(u)\rho(v)$, where $\rho$ is the Tukey bisquare function $\rho(u)=u\big(1-u^2/c^2\big)^2\mathbf{1}(|u|\leq c)$. Still following \cite{bustos1986}, we set the latter function with $c=5.58$, which corresponds to approximately 95\% Gaussian efficiency.
The RA estimating equations are based on the inverse autoregressive and moving-average polynomials. 
The estimating equations become $g_{\phi}(\theta)=\sum_{h=0}^{\infty}\phi^h\hat{\gamma}_{h+1}(\theta)=0$, and $g_{\psi}(\theta)=\sum_{h=0}^{\infty}(-\psi)^h\hat{\gamma}_{h+1}(\theta)=0$ (see \cite[Equation (2.8)]{bustos1986}). Now, as the coefficients decay geometrically under stationarity and
invertibility, we truncate the infinite sums. To simplify the computations, we set the truncation with $L=\min\{200,T-3\}$. Then, the estimator is obtained by minimizing the squared norm of the truncated estimating equations, that is $\hat{\theta}^{\text{ra}}=\underset{\theta}{\arg\min}\Big\{\Big(\sum_{h=0}^{L-1}\phi^h\hat{\gamma}_{h+1}(\theta)\Big)^2+\Big(\sum_{h=0}^{L-1}(-\psi)^h\hat{\gamma}_{h+1}(\theta)\Big)^2\Big\}$.
Figures \ref{figures_ARMA_G_p8_t15_300}-\ref{figures_ARMA_IO_p8_t15_1000} display the $\ell_2$ distances between $\theta^*$ and the estimated parameters. In the small sample size regime, the MMD-based estimators perform better than the QML method for a wide range of $p$ values, particularly in case 2. Interestingly, for some values of $p$, ISMMD also outperforms the RA method, which was designed for this latter case. In the large sample regime and in case 2, MMD still outperforms the QML method. However, in this setting, the RA procedure performs best. 
This is expected because the RA procedure was designed to cover the true DGP in case 2.
Globally, ISMMD tends to perform better than PSMMD.}

\subsection{Non-linear MA model}\label{subsec:NL_MA}

\begin{figure}[htbp]
\centering
  \begin{minipage}[b]{0.45\textwidth}
    \centering
    \includegraphics[width=0.92\linewidth]{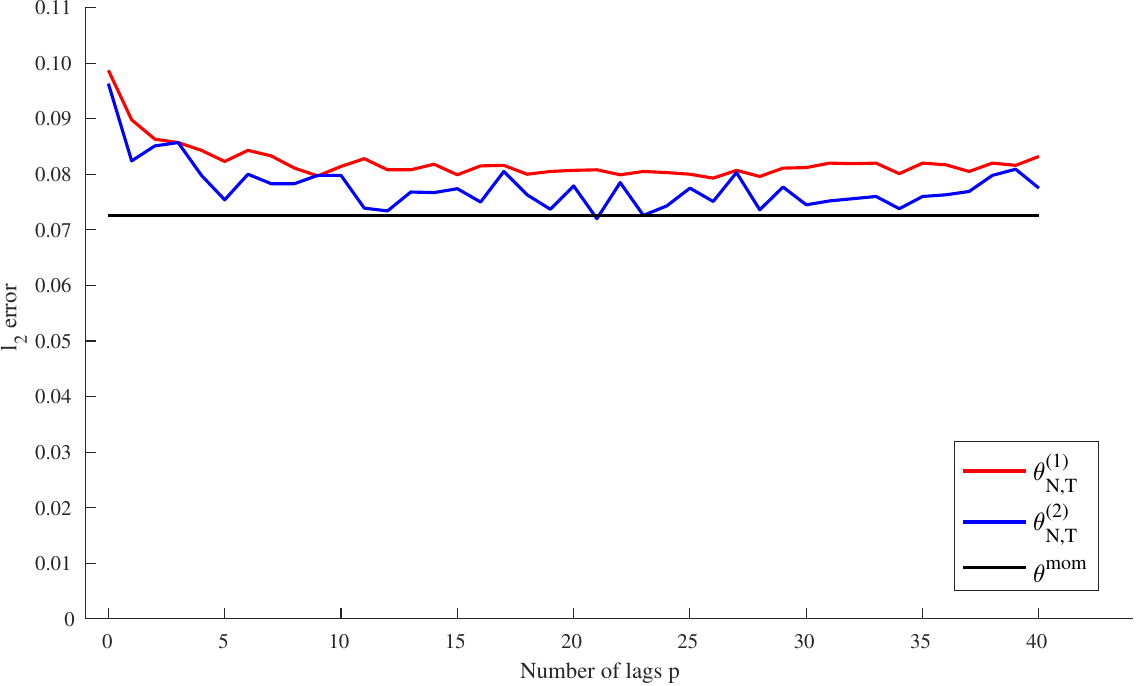}
    \subcaption{Case 1: $u_t \sim \mathcal{N}(0,1)$. $T=300$. \\ Estimation under correct specification.}
  \end{minipage}\label{figures_NLS_G_a9_300}
  \begin{minipage}[b]{0.45\textwidth}
    \centering
    \includegraphics[width=0.92\linewidth]{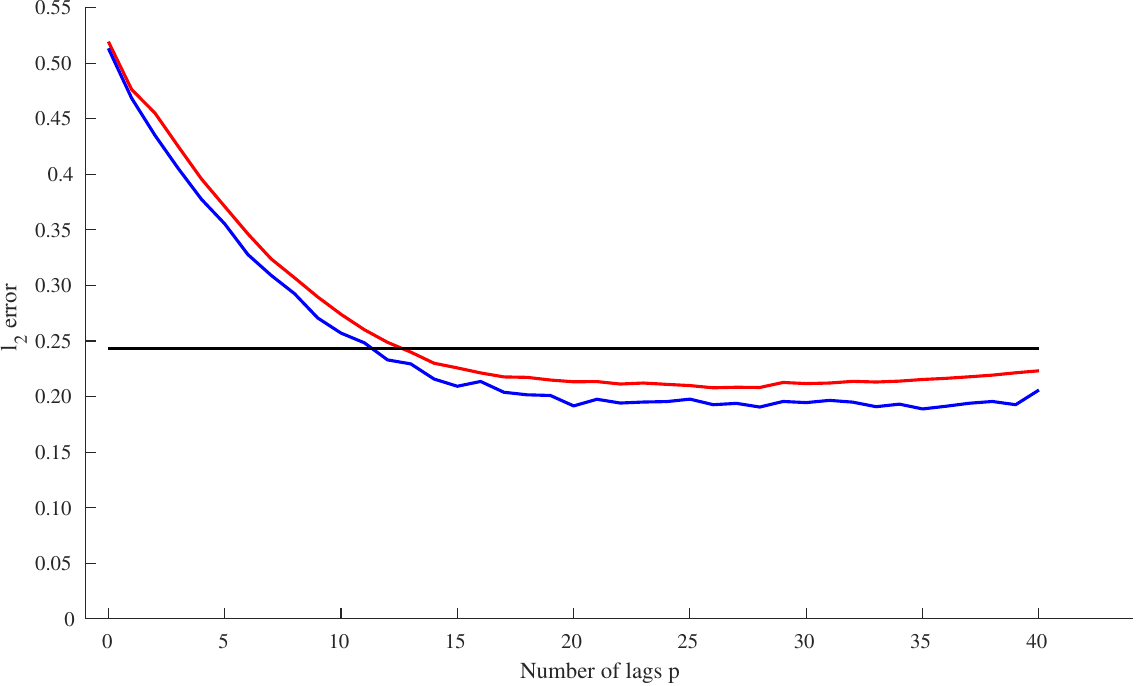}
    \subcaption{Case 2: $u_t = t(3)/\sqrt{3}$. $T=300$. \\ Estimation under misspecification.}\label{figures_NLS_S_a9_300}
  \end{minipage}

  \vspace*{0.8cm}
  
  \begin{minipage}[b]{0.45\textwidth}
    \centering
    \includegraphics[width=0.92\linewidth]{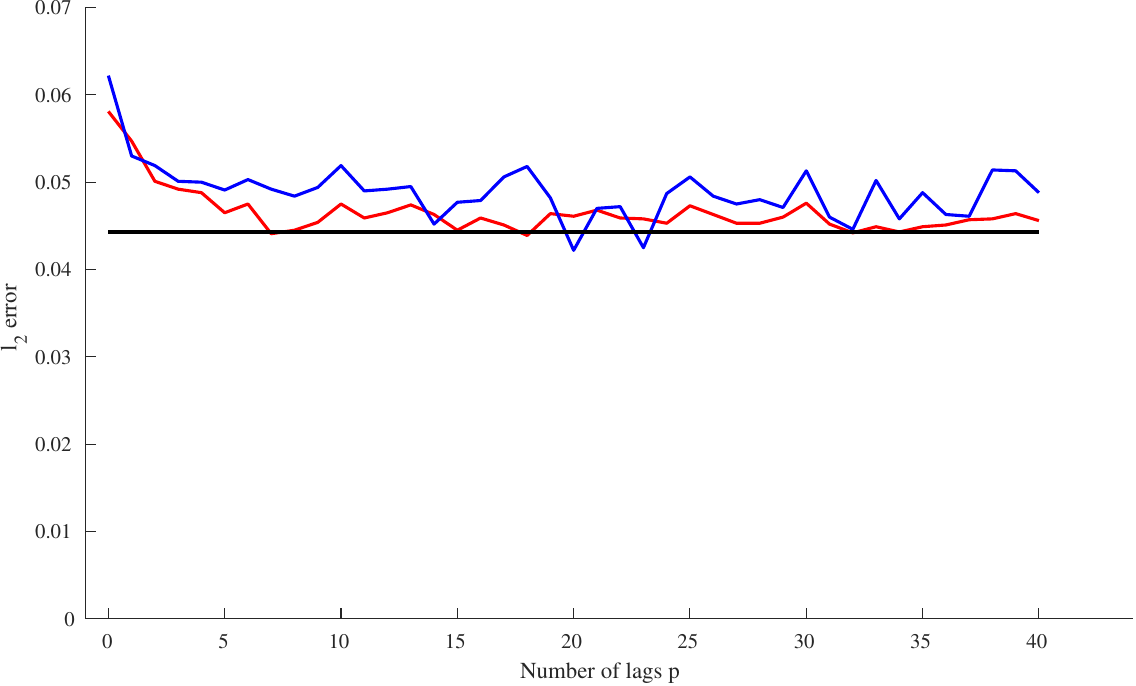}
    \subcaption{Case 1: $u_t \sim \mathcal{N}(0,1)$. $T=1000$. \\ Estimation under correct specification.}\label{figures_NLS_G_a9_1000}
  \end{minipage}
  \begin{minipage}[b]{0.45\textwidth}
    \centering
    \includegraphics[width=0.92\linewidth]{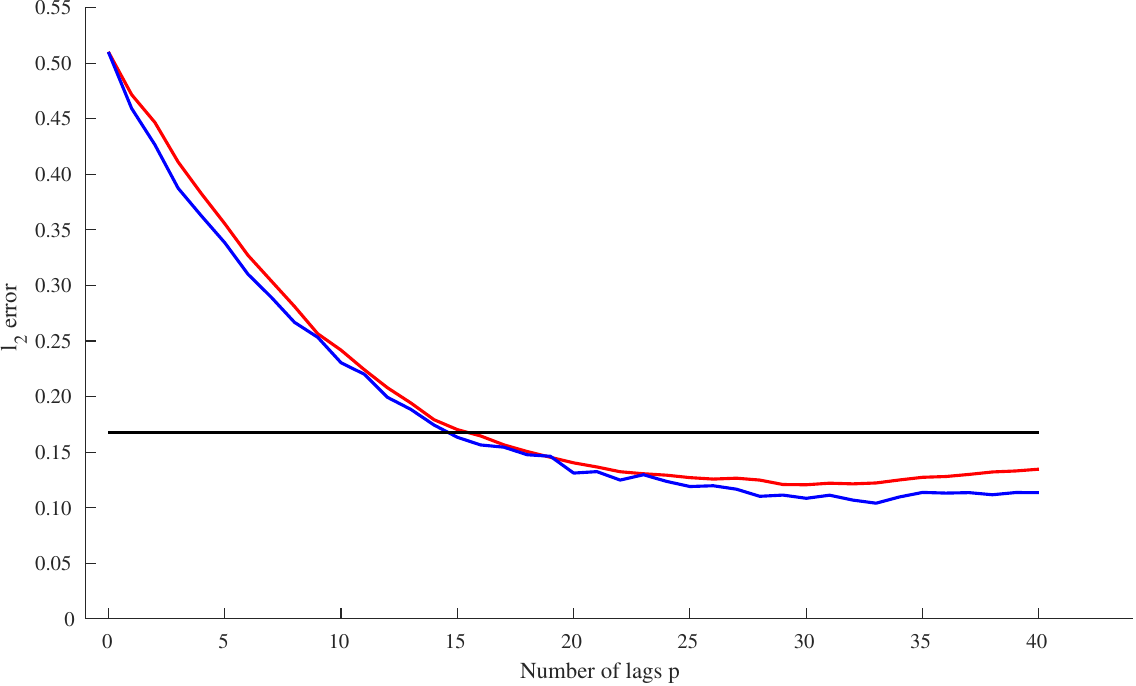}
    \subcaption{Case 2: $u_t = t(3)/\sqrt{3}$. $T=1000$. \\ Estimation under misspecification.}\label{figures_NLS_S_a9_1000}
  \end{minipage}
  \caption{Non-linear MA model - $\theta^*=0.9$. Each point shows the average of 100 batches}
\end{figure}

We propose to estimate the following non-linear MA(1) model: for any $t \in \{ 1,\ldots, T\}$, 
$ x_t = u_t + \psi u^2_{t-1}$, where $(u_t)$ is a non-observable innovation process. Every $u_t$ is 
centered with unit variance. Again, we consider the cases $u_t \sim \mathcal{N}(0,1)$ (case 1) and $u_t= t(3)/\sqrt{3}$ (case 2). Here, $\theta=\psi$. The true parameter is set as $\theta^*=0.9$. The sample sizes are set as $T=300$ and $T=1000$. We will compute ISMMD and PSMMD for lags up to $p=40$. Indeed, the non-linear MA model requires more lags for a suitable estimation than the previously studied models. In addition, we will estimate the non-linear MA model parameter by the moment estimator $\hat{\theta}^{\text{mom}}$, which is simply the sample average of the observations. In the correct specification case and the large sample regime (Figure \ref{figures_NLS_G_a9_1000}), the moment estimator and the $\tilde{\theta}_{N,T}^{(k)}$, $k \in \{ 1,2\}$, are very close when $10 \leq p \leq 30$. The former method slightly outperforms the MMD-based method in the low sample regime, except for $\tilde{\theta}_{N,T}^{(2)}$ when $p=21$. The results are altered in the misspecified case: $\tilde{\theta}_{N,T}^{(k)},k\in\{1,2\}$, are outperformed when $p \leq 12$ and $T=300$, but perform better than $\hat{\theta}^{\text{mom}}$ when $p\geq 13$, \textcolor{black}{emphasizing the robustness of the MMD procedure in the presence of \textcolor{black}{heavy-tailed innovations}}. Moreover, Figures \ref{figures_NLS_S_a9_300} and \ref{figures_NLS_S_a9_1000} suggest the existence of an optimal set of $p$ values between $20$ and $25$.

\subsection{Ricker model}\label{subsec:Ricker}

\begin{figure}[htbp]
\centering
  \begin{minipage}[b]{0.45\textwidth}
    \centering
    \includegraphics[width=0.92\linewidth]{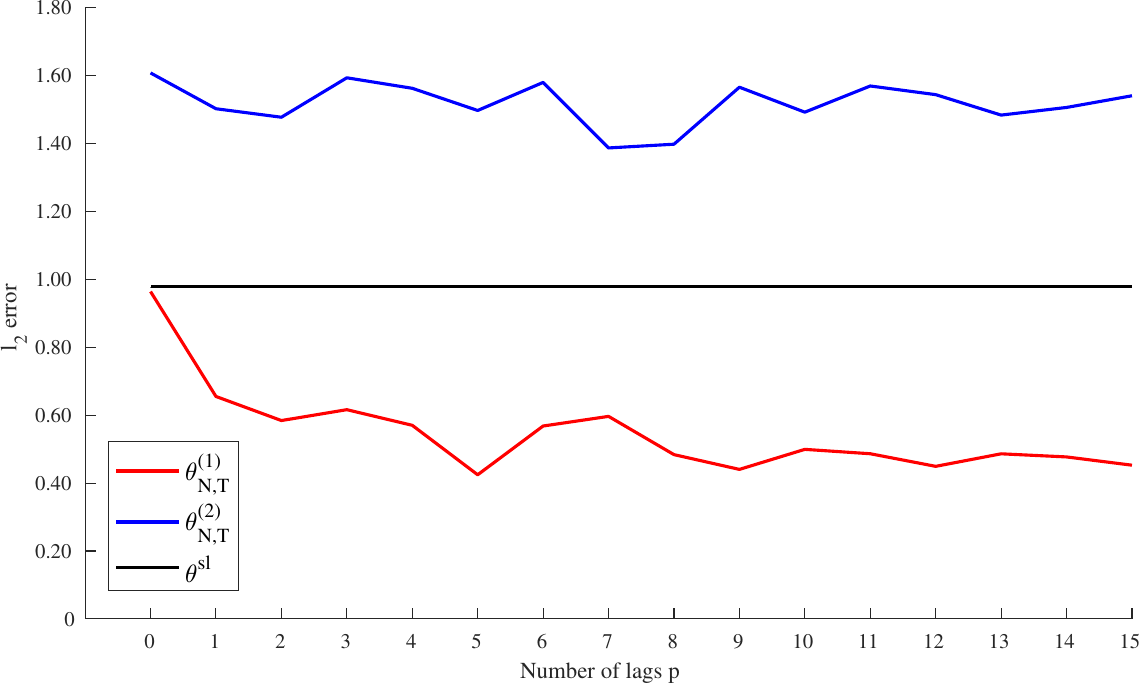}
    \subcaption{Case 1: $u_t \sim \mathcal{N}(0,1)$. $T=300$. \\ Estimation under correct specification.} \label{figure_Ricker_G_l7_p7_300}
  \end{minipage}
   \begin{minipage}[b]{0.45\textwidth}
    \centering
    \includegraphics[width=0.92\linewidth]{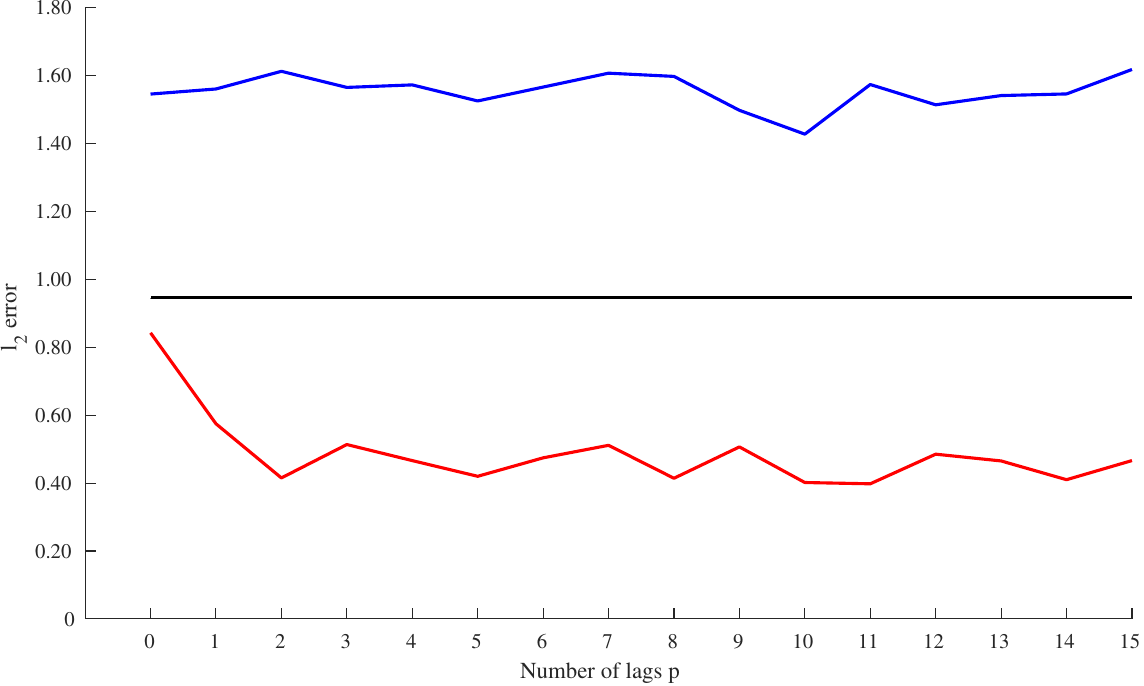}
    \subcaption{Case 2: $u_t = t(3)/\sqrt{3}$. $T=300$. \\ Estimation under misspecification.}\label{figures_Ricker_S_l7_p7_300}
  \end{minipage}

  \vspace*{0.8cm}
  
  \begin{minipage}[b]{0.45\textwidth}
    \centering
    \includegraphics[width=0.92\linewidth]{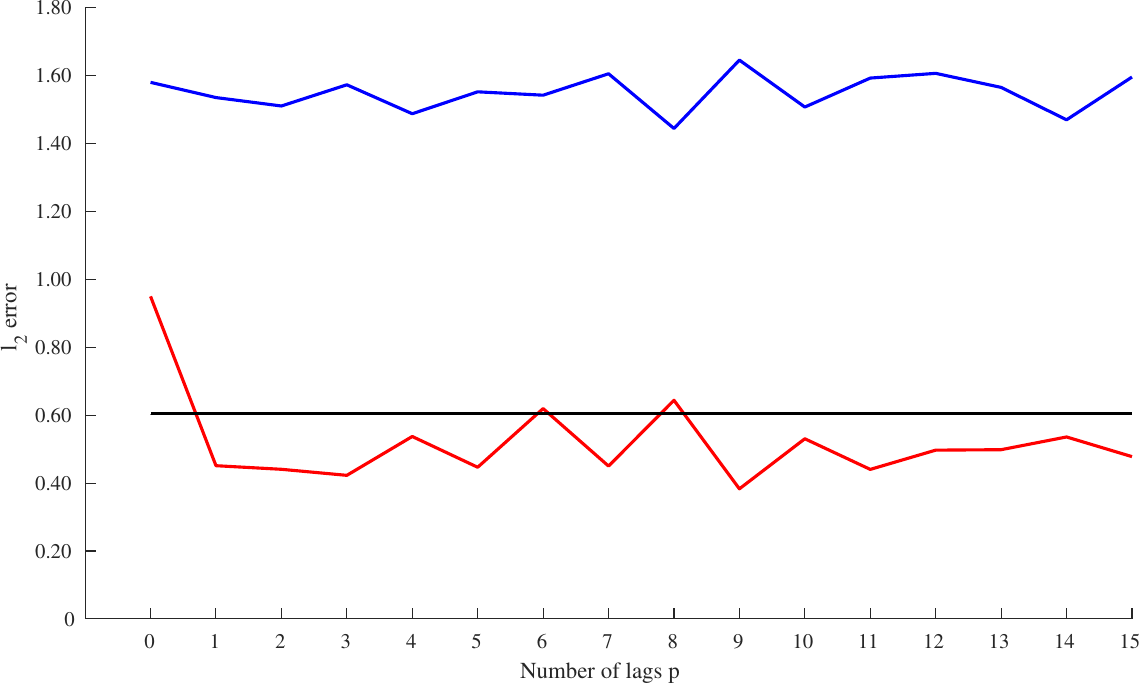}
    \subcaption{Case 1: $u_t \sim \mathcal{N}(0,1)$. $T=1000$. \\ Estimation under correct specification.}  \label{figure_ricker_model_1000}
  \end{minipage}
  \begin{minipage}[b]{0.45\textwidth}
    \centering
    \includegraphics[width=0.92\linewidth]{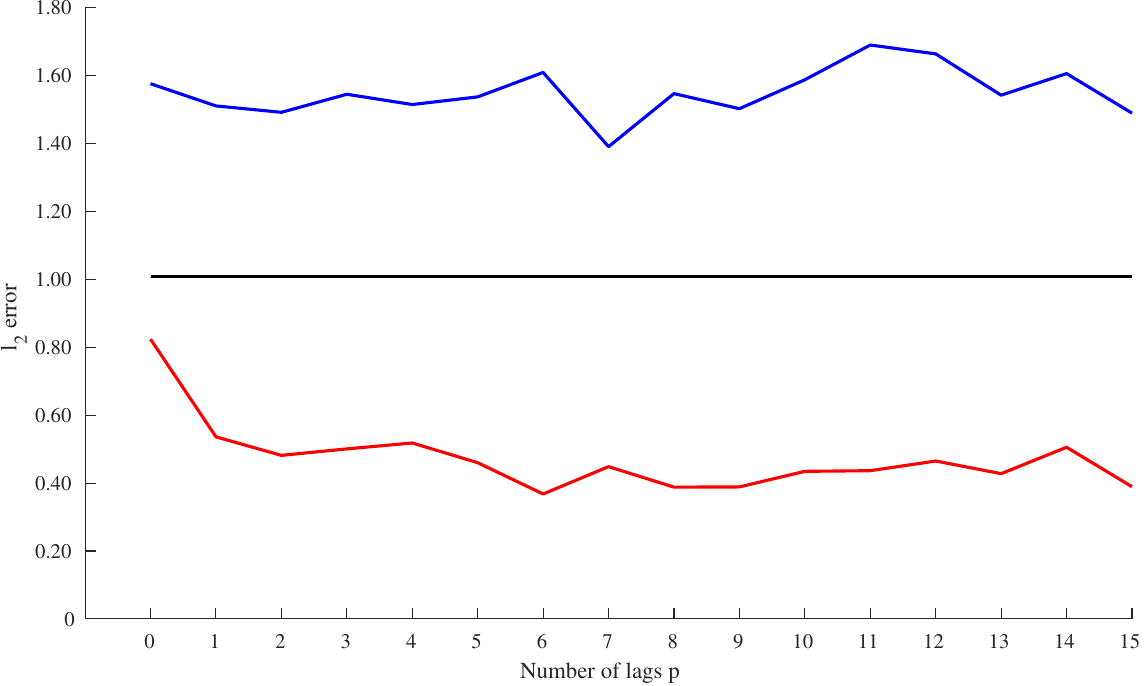}
    \subcaption{Case 2: $u_t = t(3)/\sqrt{3}$. $T=1000$. \\ Estimation under misspecification.}\label{figures_Ricker_S_l7_p7_1000}
  \end{minipage}
  \caption{Ricker model - $\theta^* = (\log(7),0.05,7) $. Each point shows the average of 100 batches}
\end{figure}

In the Ricker model, a latent stochastic process drives the observations. More precisely, it is a non-linear process where the observed time series $(x_t)$ is generated as $x_t \sim \mathcal{P}(\phi\,N_t)$ with
\begin{equation*}
\log(N_t) = \log(r)+\log(N_{t-1})-N_{t-1} + \sigma_{u} u_t,
\end{equation*}
for every $t$, with the initial condition $N_0=1$. Here, $\mathcal{P}(\phi\,N_t)$ denotes a Poisson distribution with intensity $\phi\,N_t$, for some $\phi>0$. The parameter vector is $\theta=(\log(r),\sigma_{u},\phi) $. The innovations $(u_t)$ are i.i.d.; we still consider the cases $u_t \sim \mathcal{N}(0,1)$ (case 1) and $u_t= t(3)/\sqrt{3}$ (case 2). The true parameter is $\theta^* = (\log(7),0.05,7) $ and the sample sizes are $T\in \{300,1000\}$.
The standard maximum likelihood method cannot be applied because $(N_t)$ is a latent process, which renders the likelihood intractable. To address this issue, \cite{wood2010ecolo} proposed the synthetic likelihood (SL) method based on summary statistics, which are assumed to follow a Gaussian distribution. Following this approach, we compare our method with the SL estimator $\hat{\theta}^{\text{sl}}$. The latter is obtained as a minimizer of the negative Gaussian log-likelihood of the latter summary statistics. To evaluate their likelihood for a given parameter $\theta$, we sample $R = 1000$ datasets $(x^{(r)}_t), r = 1,\ldots,R$, from the Ricker model, under the Gaussian assumption on $(u_t)$, and transform these $R$ synthetic samples into $R$ $d$-dimensional vectors of summary statistics $S_1=S(x^{(1)}_1,\ldots,x^{(1)}_T),\ldots,S_R=S(x^{(R)}_1,\ldots,x^{(R)}_T)$. 
Using this synthetic sample of summary statistics, we empirically estimate their mean vector and their variance–covariance matrix, which allows us to compute the Gaussian log-likelihood evaluated at the observed summary statistics. We use the following summary statistics: the mean, the variance, the mean first difference, the variance of first difference, the autocorrelations at lag 1 and lag 2 of the synthetic data, the autocorrelation at lag 1 of the differenced synthetic data and the mean of the squared first difference, so that $d=8$. 
Figures~\ref{figure_Ricker_G_l7_p7_300}-\ref{figures_Ricker_S_l7_p7_1000} show that the ISMMD estimator $\tilde{\theta}_{N,T}^{(1)}$ performs very well, compared to the PSMMD estimator $\tilde{\theta}_{N,T}^{(2)}$ and the SL estimator $\hat{\theta}^{\text{sl}}$. 

\section{Optimal lag selection}\label{sec:lag_selection}
The specification of $y_t = (x_{t},x_{t-1},\ldots,x_{t-p})$ requires choosing \textcolor{black}{a lag $p\geq p_0$, where $p_0$ is the smallest lag for which the parameter $\theta^*$ is identified by the law of the corresponding $y_t$.}
The results presented in the previous section were established for a range of values of $p$. 
Nevertheless, in practice, one must select a ``convenient'' value for $p$ from which an MMD-based estimator $\tilde{\theta}^{(k)}_{N,T}$ is deduced, for some $k\in\{1,2,3\}$. 
Ideally, the selected $p$ should provide an estimator that should be as close as possible to the pseudo-true parameter $\theta^*$. 
Since the true model is unknown in practice, we propose to employ the MMD distance as an information criterion for the selection of $p$ in a data-driven way. 
More precisely, we consider the following strategy:
\begin{itemize}
    \item[(i)] Split the sample of observations as follows: for any $p$, the training (resp. test) set consists of values of  $(y_t)$ observed before (resp. after) some arbitrary chosen date $t_0$. The corresponding empirical distributions are denoted by $P^{\text{train}}_T$ and $P^{\text{test}}_T$, respectively. 
    \item[(ii)] Solve the criterion (\ref{def_estimators}) based on $P^{\text{train}}_T$ for $p\in \{0,1,2,\ldots,p_{\max}\}$, providing 
    $\tilde{\theta}^{(k)}_{N,T}$ for each $p$.
    \item[(iii)] For a given $p$, generate the synthetic process $(\tilde{y}_i)_{1 \leq i \leq N_0}$ given $\tilde{\theta}^{(k)}_{N,T}$. The empirical distribution of $(\tilde{y}_i)_{1 \leq i \leq N_0}$ is $\hat P_{\tilde{\theta}^{(k)}_{N,T}}$, as denoted before. Then, recalling the test set and for a given $p$, compute the MMD-based information criterion $\mathcal{D}_p(\hat{P}_{\tilde{\theta}^{(k)}_{N,T}},P^{\text{test}}_T)$.
    \item[(iv)] Select the optimal lag $p$ as $\hat{p}:=\underset{0 \leq p \leq p_{\max}}{\arg\min}\;\mathcal{D}_p(\hat{P}_{\tilde{\theta}^{(k)}_{N,T}},P^{\text{test}}_T)$.
\end{itemize}
\textcolor{black}{To stress that the MMD distances now depend on $p$, we have used the notation $\mathcal{D}_p$ above.}
Hereafter, the date $t_0$ is chosen in such a way that the training sample corresponds to the first $75\%$ of the entire sample and the test sample to the last $25\%$. In our experiments, we will set $N_0=10000$ and all the considered models will \textcolor{black}{be misspecified exactly as in Section~\ref{sec:experiments}}.

\medskip

\begin{figure}[htbp]
\centering
  
  \begin{minipage}[b]{0.45\textwidth}
    \centering
    \includegraphics[width=0.92\linewidth]{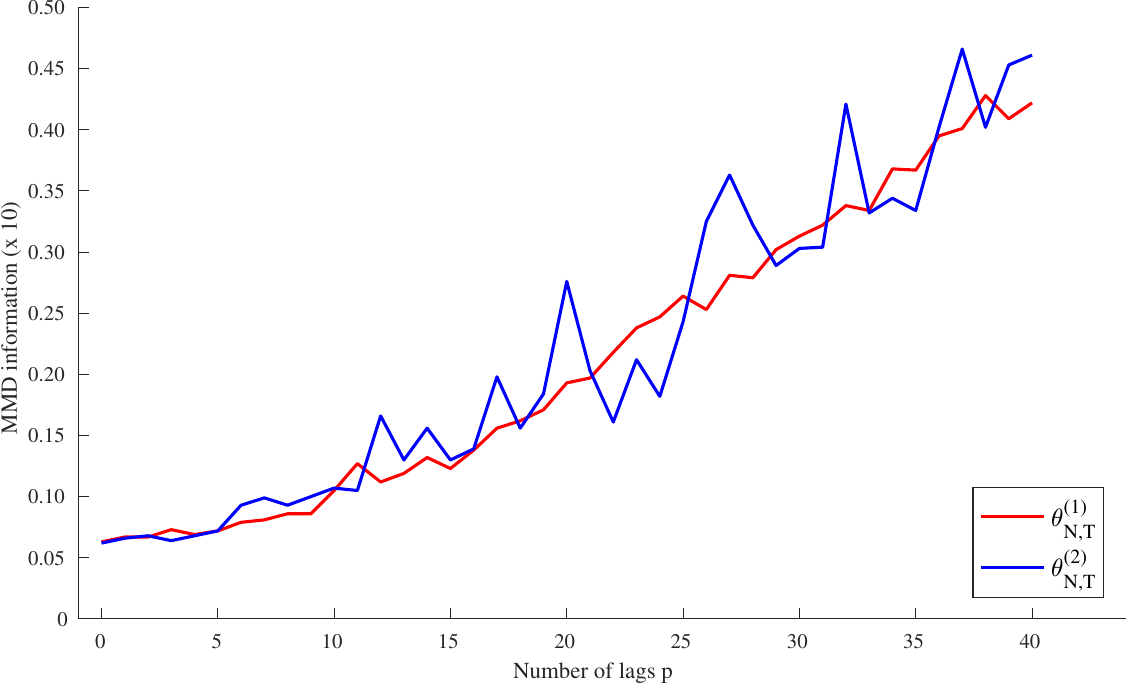}
    \subcaption{MMD criterion. ARMA model.}\label{MMD_validation_ARMA}
  \end{minipage}
  \begin{minipage}[b]{0.45\textwidth}
    \centering
    \includegraphics[width=0.92\linewidth]{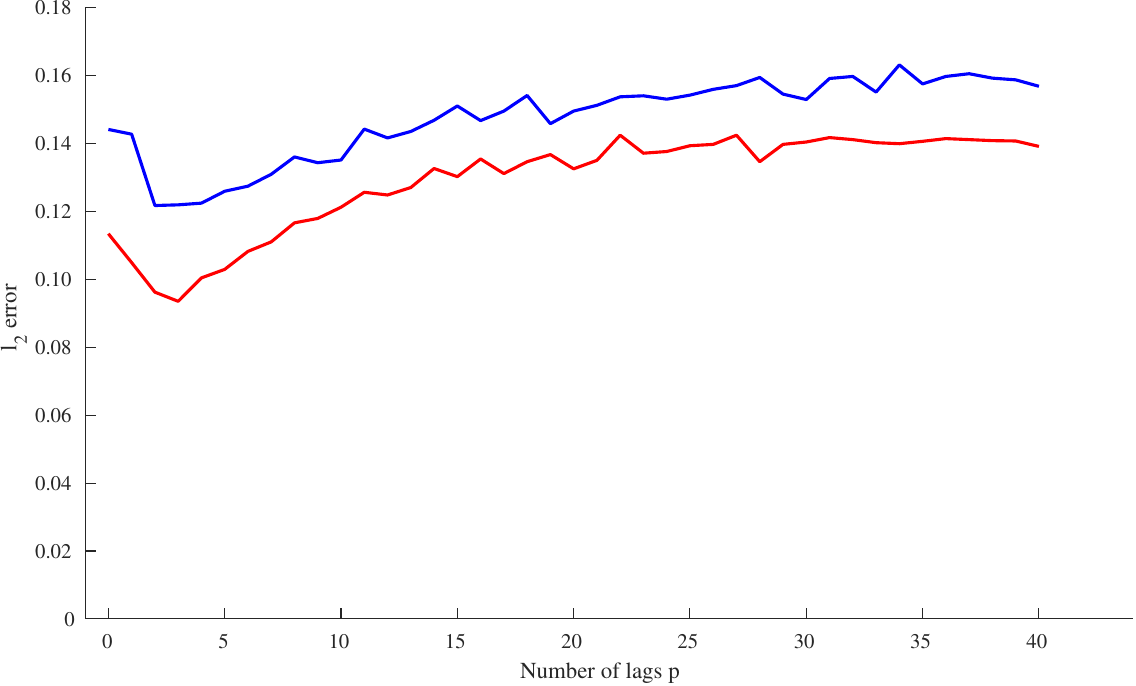}
    \subcaption{$\ell_2$-error. ARMA model.}\label{ARMA_error}
  \end{minipage}

\vspace*{0.8cm}

  \begin{minipage}[b]{0.45\textwidth}
    \centering
    \includegraphics[width=0.92\linewidth]{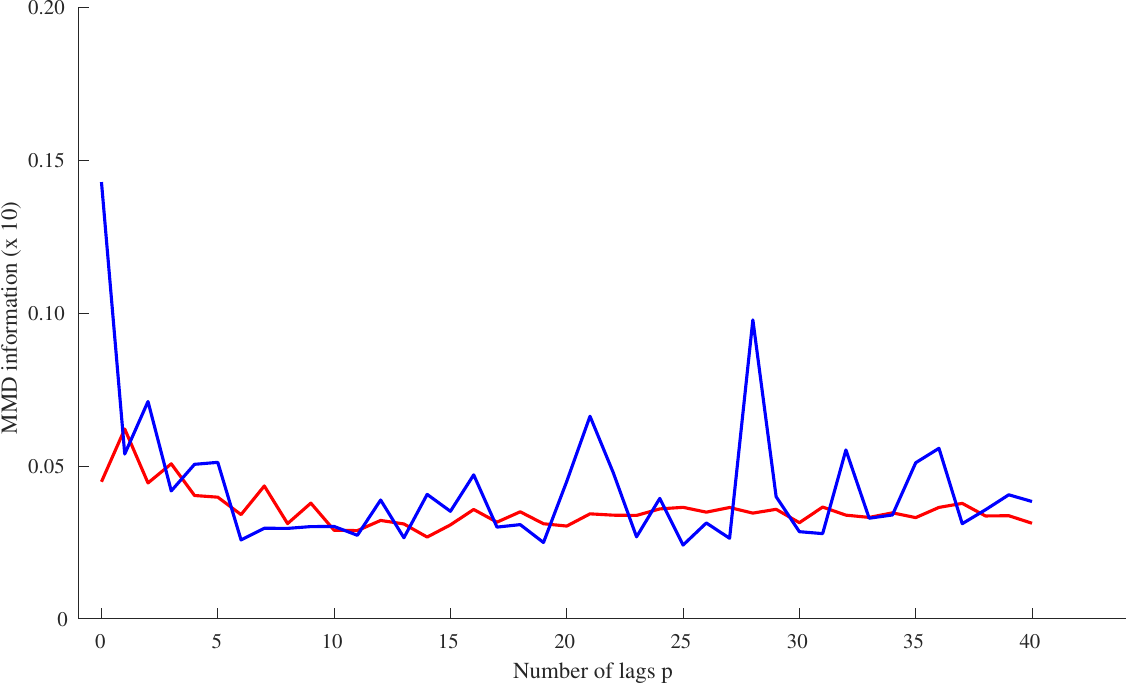}
    \subcaption{MMD criterion. GARCH model.}\label{MMD_validation_GARCH}
  \end{minipage}
  \begin{minipage}[b]{0.45\textwidth}
    \centering
    \includegraphics[width=0.92\linewidth]{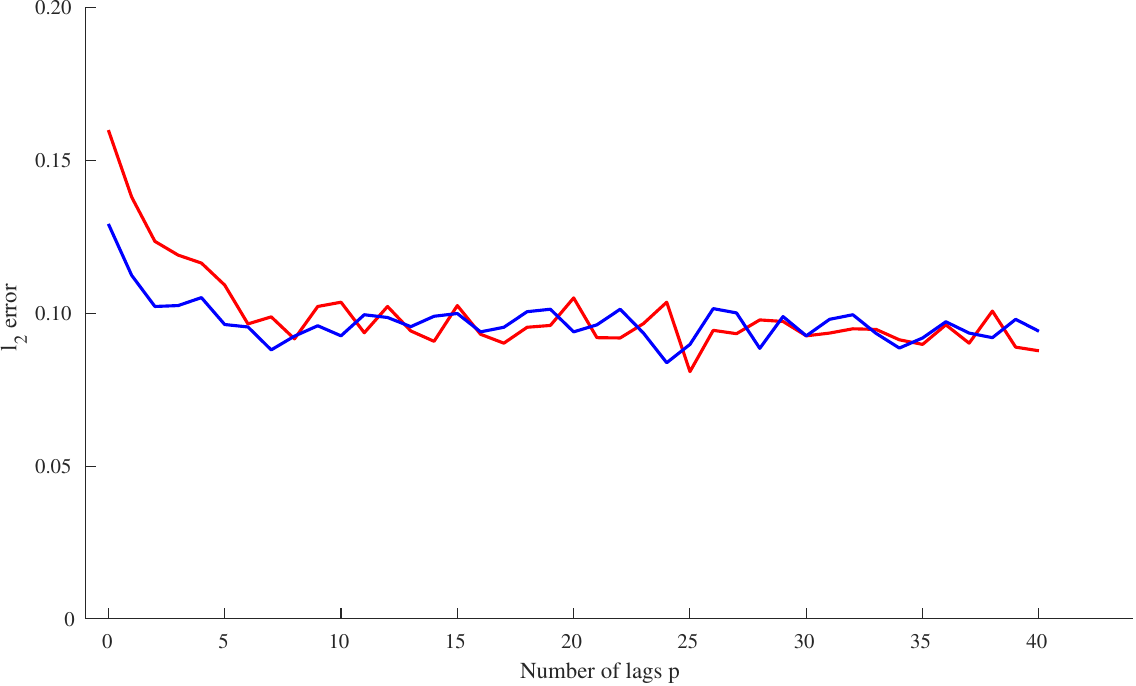}
    \subcaption{$\ell_2$-error. GARCH model.}\label{GARCH_error}
  \end{minipage}

\vspace*{0.8cm}

  \begin{minipage}[b]{0.45\textwidth}
    \centering
    \includegraphics[width=0.92\linewidth]{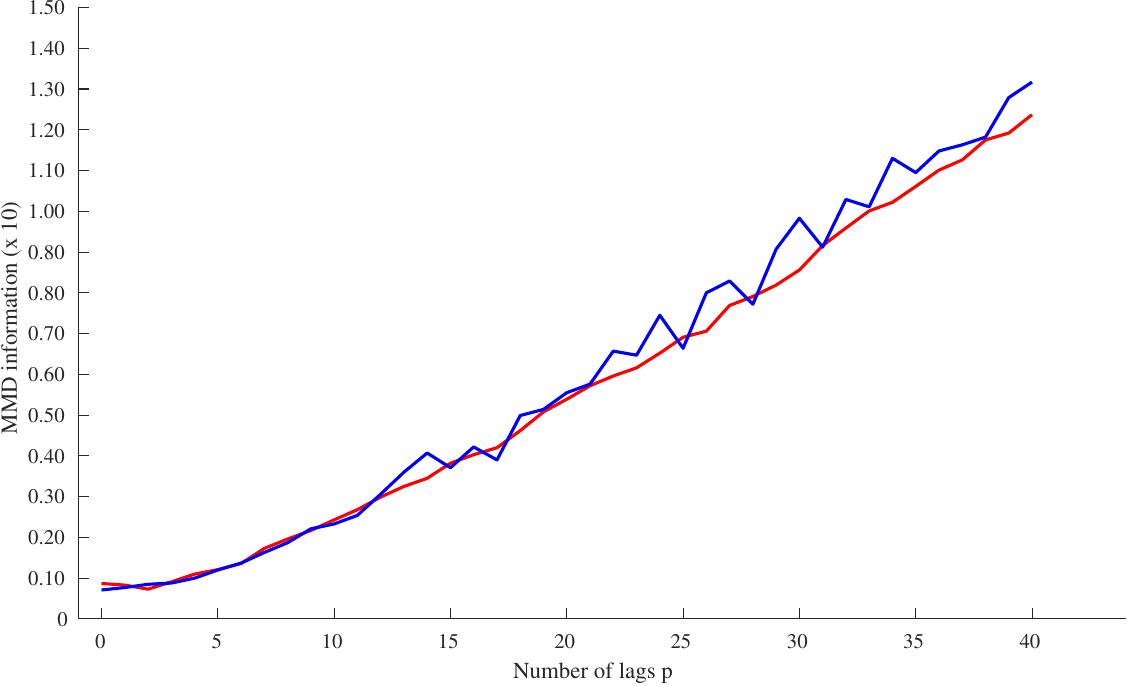}
    \subcaption{MMD criterion. Non-linear MA    model.}\label{MMD_validation_NLts}
  \end{minipage}
  \begin{minipage}[b]{0.45\textwidth}
    \centering
    \includegraphics[width=0.92\linewidth]{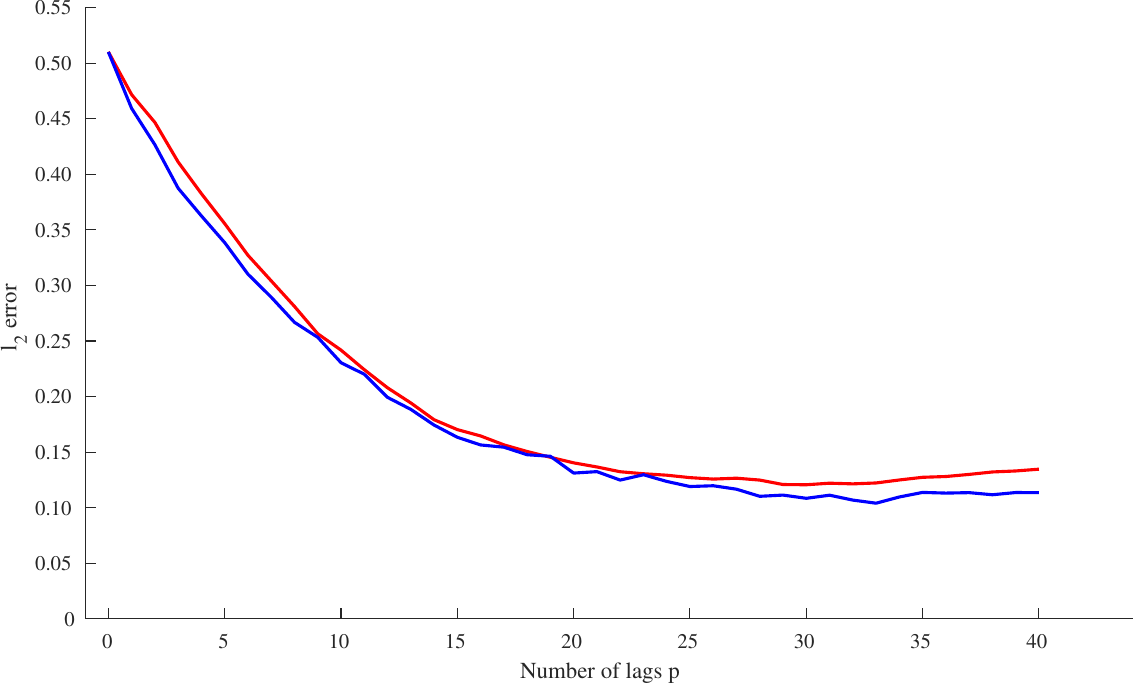}
    \subcaption{$\ell_2$-error. Non-linear MA model.}\label{NLts_error}
  \end{minipage}
  
  \caption{MMD information criterion. Estimation under misspecification.}
\end{figure}

First, let us illustrate this selection method with 
an ARMA model. Its  parameter is $\theta = (\phi,\psi,\sigma^2_u) $, where the observations are generated according to $x_t=\phi x_{t-1} + v_t+\psi v_{t-1}$, $v_t = \sqrt{\sigma^2_u} u_{t}$, $t=1,\ldots,1000$. The true parameter is set as $\theta^*=(0.8,0.15,0.05) $, and \textcolor{black}{$u_t$ is set as in case 2, Subsection \ref{subsec:ARMA}, with $\epsilon=0.1$ and $\kappa = 25$.}
We estimate $\tilde{\theta}^{(k)}_{N,T}$, $k\in\{1,2\}$, only once for $p=0,\ldots,40$. 
Figure \ref{MMD_validation_ARMA} displays the pattern of $p\mapsto \mathcal{D}_{p}(\hat{P}_{\tilde{\theta}^{(k)}_{N,T}},P^{\text{test}}_T)$ (\textcolor{black}{or $p\mapsto \mathcal{D}_p$ to be short}). Figure \ref{ARMA_error} displays the $\ell_2$-error between the estimated and the true parameters as a function of $p$, in the same way as in Subsection \ref{subsec:ARMA}.

The selection procedure is also assessed through 
the GARCH(1,1) model with a parameter $\theta = (\omega,\beta,\alpha) $, where $x_t = \sqrt{h_t}u_t$ with $h_t = \omega+\beta h_{t-1}+\alpha x^2_{t-1}$, $t=1,\ldots,1000$, $\theta^*=(0.05,0.92,0.05) $, and $u_t=\nu_t/\sqrt{3}$ with $\nu_t\sim t(3)$. Figure \ref{MMD_validation_GARCH} displays $ \mathcal{D}_p$ and Figure \ref{GARCH_error} displays the $\ell_2$-errors as functions of $p$, as in Subsection \ref{subsec:GARCH}.

Finally, we apply the procedure to the non-linear MA model. The observations are generated according to $x_t=u_t + \psi u^2_{t-1}, t = 1,\ldots,1000$, and $u_t=\nu_t/\sqrt{3}$ with $\nu_t\sim t(3)$. The model parameter value is $\theta=\psi$. The true parameter is $\theta^* = 0.9$. 
Figure \ref{MMD_validation_NLts} displays the corresponding values of $\mathcal{D}_p$. Figure \ref{NLts_error} displays the $\ell_2$-error values, obtained as in Subsection \ref{subsec:NL_MA} .

For the ARMA model, the selection procedure identifies small values of $p$ as optimal, which is consistent with the estimation errors reported in Figure \ref{ARMA_error}. 
\textcolor{black}{For the ISMMD estimator, $\hat{p}=0$  provides the lowest value of $\mathcal{D}_p$. In the PSMMD case, $\hat{p}=0$ is selected as optimal.} 
\textcolor{black}{Although such values are close to the values of $p$ that minimize the corresponding $\ell_2$-error, the MMD-based information criterion tends to select $\hat p=0$, suggesting that contamination in the ARMA process prevents the selection of larger values of $p$, even though additional lags of $y_t$ may contain useful information for accurately estimating $\theta^*$.}

\textcolor{black}{For the GARCH model, larger values of $p$ are preferred, but such choices remain in line with the errors displayed in Figure \ref{GARCH_error}. For ISMMD, the procedure selects $\hat{p}=14$. In the PSMMD case, $\hat{p}=25$ is selected. Choosing a lag order $p \leq 13$ results in a loss of information, as some part of the dynamics that is relevant for parameter estimation is omitted. Conversely, excessively large values of $p$ may introduce noise, thereby reducing the precision of the MMD-based estimation.}

The results for the non-linear MA model are more intriguing. While our selection procedure favors small values of $p$ \textcolor{black}{(ISMMD selects $\hat p =2$; PSMMD selects $\hat p = 0$)}, the $\ell_2$-errors shown in Figure \ref{NLts_error} indicate that larger values of $p$ are preferable for accurate parameter estimation.
This observation should encourage us to exercise caution. A data-driven relevant choice of $p$ based on the similarity between laws does not guarantee that it will always prove to be the best choice for estimating the underlying parameter, particularly in models with a high degree of dependence between observations.

\begin{rem}
    Similar ideas can be developed when we assume $y_t=(x_t,x_{t-\tau_1},\ldots, x_{t-\tau_p})$, for some lag $p$ and time shifts $\tau_1,\ldots,\tau_p$. In this more general setting, it is possible to consider a criterion based on a splitting scheme as above. Such a criterion would be optimized w.r.t. $(p,\tau_1,\ldots,\tau_p)$, under the constraints $\tau_j<\tau_{j+1}$ for any $j\in \{1,\ldots,p-1\}$, in addition to an arbitrarily chosen upper bound for $\tau_p$. Unfortunately, the computational cost of this extended approach would rapidly grow with $p$, without additional constraints on the $\tau_j$.
\end{rem}

\section{Conclusion}

We propose an estimation method for discrete time series models based on MMD and show that the resulting estimators are consistent and asymptotically normally distributed. Our numerical experiments suggest that such MMD-based estimators are robust to model misspecification.

The proposed method strongly relies on the specification of the vectors $y_t$ (and then $\tilde{y}_i$) from the observed initial path $(x_t)$.  Many different constructions of $y_t$, and hence of the MMD loss, could be considered. For example, instead of working with the ``full vectors'' $y_t$ containing $p$ successive lagged variables from $(x_t)$, one could evaluate a sum of MMD distances based on lower-dimensional vectors constructed from pairs $(x_{t-k},x_{t-l})$, for some pairs of indices $(k,l)$, $0\leq k<l$, in the same spirit as the composite likelihood method.

\textcolor{black}{Furthermore, while the stochastic gradient-based algorithm gives good performances in practice, its theoretical properties are difficult to establish in most practical cases of interest. More theoretically grounded optimization methods were recently proposed by~\cite{kang2026gradient}. It is definitely a direction of interest to study such procedures in the setting of time series models.
}

\bigskip
\bigskip

\noindent \textbf{Acknowledgments}

\medskip

\noindent Benjamin Poignard was supported by JSPS KAKENHI Grant (25K16617).

\newpage

\printbibliography

\appendix

\section{Proofs of consistency results}
\label{supp:section_proofs}

\begin{proof}[Proof of Proposition \ref{prop_rho_beta}]
To clarify the notations, assume that the process $(y_t)_{t\in\mathbb{Z}}$ is defined on some space $(\Omega,\mathcal{A},\mathbb{P})$. The notation used in the paper for the marginal distribution of $y_t$ is $P_0$. Thus, we can write for any Borel subset $A\subset\mathbb{R}^q$: $P_0(A) = \mathbb{P}(y_t \in A) $, which does not depend on $t$ as the process $(y_t)_{t\in\mathbb{Z}}$ is stationary. In this proof, we will also need a notation for the distribution of the pair $(y_0,y_t)$: $P_{0:t}(B) := \mathbb{P}\big((y_0,y_t) \in B\big) $ for any measurable set $B\subset \mathbb{R}^q \times \mathbb{R}^q$.
\\
We now consider another measurable space: $\big(\Omega\times\Omega,\sigma(y_0)\otimes \sigma(y_t)\big)$, and we equip it with the probability $\mathbb{P}^{0 \otimes t}$ defined as the image measure of $\mathbb{P}$ through the map $i(\omega)=(\omega,\omega)$. In particular, observe that, for $\mathcal{A}\in \sigma(y_0)$ and $\mathcal{B}\in\sigma(y_t)$, we have $\mathbb{P}^{0 \otimes t}(\mathcal{A}\times \mathcal{B})=\mathbb{P}(\mathcal{A}\cap \mathcal{B})$. Then, for any $\mathcal{A}\in \sigma(y_0)$, we put $\mathbb{P}^0(\mathcal{A}) := \mathbb{P}(\mathcal{A})$ and for any $\mathcal{B}\in \sigma(y_t)$, we put $\mathbb{P}^t(\mathcal{B}) := \mathbb{P}(\mathcal{B})$. In Definition 2.1 of Section 1.6 in~\cite{rio2017}, the $\beta$-mixing coefficient are defined as
\begin{equation}
\label{equation:beta:rio}
\beta\big(\sigma(y_0),\sigma(y_t)\big) = \sup_{\mathcal{C}\in \sigma(y_0)\otimes \sigma(y_t)} | \mathbb{P}^{0 \otimes t}(\mathcal{C}) - \mathbb{P}^{0}\otimes \mathbb{P}^{t} (\mathcal{C})|.
\end{equation}
Other definitions of the latter coefficients can be found in the literature; they are proven to be equivalent to~\eqref{equation:beta:rio} in Section 1.6 of~\cite{rio2017}. Note that
$ \mathbb{P}^t(\{ y_t\in A\}) =  \mathbb{P}(\{ y_t\in A\}) = P_0(A) =  \mathbb{P}(\{ y_0\in A\})  = \mathbb{P}^0(\{ y_0\in A\}) $ for any measurable set $A\subset\mathbb{R}^d$.
\\
Let us now consider, for a fixed $u>0$, the event $\mathcal{C}(u):=\big\{(\omega,\omega') ; u\geq \|y_0(\omega)-y_t(\omega')\|\big\}$. We are going to prove that $\mathcal{C}(u)\in \sigma(y_0)\otimes \sigma(y_t)$. In order to do so, we will prove that $\mathcal{D}(u)\in \sigma(y_0)\otimes \sigma(y_t)$, where $\mathcal{D}(u)=\big\{(\omega,\omega') ; u< \|y_0(\omega)-y_t(\omega')\|\big\}$ is the complement of $\mathcal{C}(u)$. For any $(x,y)\in \mathbb{Q}^q \times \mathbb{Q}^q $ such that $ \|x-y\|> u$, we put $\delta_u(x,y) = (\|x-y\|-u)/2$, and $D_u(x,y) = \{\omega\in\Omega:  y_0(\omega)\in[x-\delta_u(x,y),x+\delta_u(x,y)] \} \times \{\omega'\in\Omega:  y_t(\omega)\in[y-\delta_u(x,y),y+\delta_u(x,y)] \} \in \sigma(y_0)\times \sigma(y_t) \subset  \sigma(y_0)\otimes \sigma(y_t)$. As $\mathcal{D}(u)$ can be written as the countable union
$$
\mathcal{D}(u) = \bigcup_{ \{ (x,y)\in \mathbb{Q}^q \times \mathbb{Q}^q: \|x-y\|> u \} } D_u(x,y),
$$
then $\mathcal{D}(u)$ also belongs to $\sigma(y_0)\otimes \sigma(y_t)$. Thus, so does its complement $\mathcal{C}(u)$.
\\
By~\eqref{equation:beta:rio}, for any $u>0$,
\begin{align*}
\beta\big(\sigma(y_0),\sigma(y_t)\big)
& \geq  | \mathbb{P}^{0 \otimes t}\big(\mathcal{C}(u)\big) - \mathbb{P}^{0}\otimes \mathbb{P}^{t} \big(\mathcal{C}(u)\big)|
\\
& = | \mathbb{P}^{0 \otimes t}\big(\{ u\geq \|y_0-y_t\| \}\big) - \mathbb{P}^{0}\otimes \mathbb{P}^{t} \big( \{ u\geq \|y_0-y_t\| \} \big)|
\\
& = | P_{0:t}\big(\{(x,y)\in\mathbb{R}^q\times\mathbb{R}^q : u\geq \|x-y\| \}\big) - P_0\otimes P_0\big( \{(x,y)\in\mathbb{R}^q\times\mathbb{R}^q : u\geq \|x-y\| \}\big)|.
\end{align*}
Thanks to our assumptions on the kernel, we have
 \begin{align*}
 \varrho_t
 & = \left| \mathbb{E}\left< \Phi(y_t) - \int \Phi(y) P_0(dy),\Phi(y_0) - \int \Phi(y) P_0(dy) \right>_{\mathcal{H}} \right|
 \\
 & = \left| \int  k(x,y) P_{0:t}\big(d(x,y)\big) - \iint k(x,y) P_0(dx)  P_0(dy)\right|
 \\
 & =  \left| \int  \left(\int \mathbf{1}_{\{u\geq \|x-y\|\}} f(u) du\right) P_{0:t}\big(d(x,y)\big) - \iint  \left(\int \mathbf{1}_{\{u > \|x-y\|\}}\right) f(u) du \, P_0(dx)  P_0(dy)\right|
 \\
 & = \left| \int_{0}^\infty \Big( \int   \mathbf{1}_{\{u \geq  \|x-y\|\}} P_{0:t}\big(d(x,y)\big) - \iint \mathbf{1}_{\{u \geq \|x-y\|\}} \ P_0(dx)  P_0(dy)  \Big) f(u) du \right|
 \\
 & = \left| \int_{0}^\infty \Big(  P_{0:t}\big(\{(x,y): u\geq \|x-y\| \}\big) - P_0\otimes P_0( \{ (x,y): u\geq \|x-y\| \}\big) \Big) f(u) \, du \right|
 \\
 & \leq  \int_{0}^\infty \left|   \mathbb{P}^{0 \otimes t}\big(\mathcal{C}(u)\big) - \mathbb{P}^{0}\otimes \mathbb{P}^{t} \big(\mathcal{C}(u)\big) \right| f(u) \, du
 \\
 & \leq \int_{0}^\infty \beta\big(\sigma(y_0),\sigma(y_t)\big) f(u) du
 \\
 & = \beta\big(\sigma(y_0),\sigma(y_t)\big) \leq \beta\big(\sigma(y_0,y_{-1},\dots),\sigma(y_t,y_{t+1},\dots)\big) = \beta(t),
\end{align*}
which concludes the proof.
\end{proof}

\begin{proof}[Proof of Proposition \ref{prop:pierre:nonasymptotic}]
The arguments are essentially the same as for Theorem 3.1 of \cite{cherief2022finite}: 
\begin{align*}
\mathcal{D}(P_{\hat{\theta}_T},P_0)
& \leq \mathcal{D}(P_{\hat{\theta}_T},P_T) + \mathcal{D}(P_T,P_0) \text{ (triangle inequality)}
\\
& \leq  \min_{\theta\in\Theta } \mathcal{D}(P_{\theta},P_T) + \mathcal{D}(P_T,P_0) \text{ (definition of $\hat{\theta}_T$)}
\\
& \leq \min_{\theta\in\Theta } \mathcal{D}(P_{\theta},P_0) + 2 \mathcal{D}(P_T,P_0) \text{ (triangle inequality again).}
\end{align*}
By taking the expectation on both sides, we get
$$
\mathbb{E}\left[ \mathcal{D}(P_{\hat{\theta}_T},P_0) \right]
\leq \min_{\theta\in\Theta } \mathcal{D}(P_{\theta},P_0) + 2 \mathbb{E}\left[ \mathcal{D}(P_T,P_0) \right] \leq \min_{\theta\in\Theta } \mathcal{D}(P_{\theta},P_0) + 2 \sqrt{\frac{1 + 2\Sigma_T}{T}},
$$
where we used~(\ref{expectation:badrandpierre}) of the main text in the last inequality.
\end{proof}

\begin{proof}[Proof of Proposition \ref{prop:pierre:consistency:1}]
Define $M(\theta) := - \mathcal{D}(P_{\theta},P_0)$ and $M_T(\theta) := - \mathcal{D}(P_{\theta},P_T)$. With these notations, $\hat{\theta}_T = \arg\max_\theta M_T(\theta)$. Note that
$$|M_T(\theta)-M(\theta)|
 = | \mathcal{D}(P_{\theta},P_0) - \mathcal{D}(P_{\theta},P_T) |
 \leq \mathcal{D}(P_0, P_T).$$
Taking the supremum w.r.t. $\theta$ yields
\begin{equation}
\label{proof:prop:pierre:consistency:1:1}
\sup_{\theta\in\Theta} |M_T(\theta)-M(\theta)| \leq \mathcal{D}(P_0, P_T).
\end{equation}
On the other hand, an application of Markov's inequality on~(\ref{expectation:badrandpierre}) gives
\begin{equation}
\label{proof:prop:pierre:consistency:1:2}
\mathbb{P}\left(  \mathcal{D}(P_T,P_0)  \geq \epsilon \right) \leq \frac{\sqrt{1 + 2\Sigma_T} }{\epsilon \sqrt{T}},
\end{equation}
for any $\epsilon>0$. Plugging~\eqref{proof:prop:pierre:consistency:1:1} into~\eqref{proof:prop:pierre:consistency:1:2} brings
$$
\mathbb{P}\left(\sup_{\theta\in\Theta} |M_T(\theta)-M(\theta)| \geq \epsilon \right) \leq \frac{\sqrt{1 + 2\Sigma_T} }{\epsilon \sqrt{T}} \cdot
$$
This shows that $\sup_{\theta\in\Theta} |M_T(\theta)-M_\theta| \rightarrow 0$ in probability. Together with  \textit{(ii)}, this shows that the assumptions of Theorem 5.7 in~\cite{van2000asymptotic} are satisfied. As a consequence, $ \hat{\theta}_T \rightarrow \theta^*$ in probability.
\end{proof}

\begin{proof}[Proof of Proposition \ref{prop:pierre:consistency:2}]
Let $\hat{M}_{N,T}(\theta) := - \mathcal{D}(\hat{P}_\theta,P_T) $, and observe that $\tilde{\theta}_{N,T} = \arg\max \hat{M}_{N,T}(\theta) $. Still using $M(\theta) := - \mathcal{D}(P_{\theta},P_0)$, we have for any $\theta\in\Theta$
\begin{eqnarray*}
\lefteqn{|\hat{M}_{N,T}(\theta)-M(\theta)|
 = | D(\hat{P}_\theta,P_T) - \mathcal{D}(P_{\theta},P_0) |
}\\
& \leq & | D(\hat{P}_\theta,P_T) - \mathcal{D}(P_\theta,P_T) | + | \mathcal{D}(P_\theta,P_T) - \mathcal{D}(P_{\theta},P_0) |
 \leq D(\hat{P}_\theta,P_\theta) + \mathcal{D}(P_T,P_0).
\end{eqnarray*}
By Markov's inequality, this yields
\begin{align}
\mathbb{P}\left(\sup_{\theta\in\Theta} |\hat{M}_{N,T}(\theta)-M(\theta)| \geq \epsilon \right)
&
\leq \mathbb{P}\left( \sup_{\theta\in\Theta} D(\hat{P}_\theta,P_\theta) \geq \frac{\epsilon}{2} \right)
+ \mathbb{P}\left( \mathcal{D}(P_T,P_0) \geq \frac{\epsilon}{2} \right)
\nonumber \\
& \leq \mathbb{P}\left(\sup_{\theta\in\Theta}D( \hat{P}_\theta,P_\theta) \geq \frac{\epsilon}{2} \right) +  \frac{2 \sqrt{1 + 2\Sigma_T} }{\epsilon \sqrt{T}},
\label{equa:proof:pierre:0}
\end{align}
from Proposition~\ref{prop:pierre:nonasymptotic}.
Thus, it mainly remains to bound the first term in the right-hand side.
Observe that the MMD distance with the kernel $k$ is a special case of an integral probability semimetric (IPS), since
$$
\mathcal{D}(P,Q) = \sup_{f\in \Fc} | \mathbb{E}_{Y\sim P}[f(Y)] - \mathbb{E}_{Y\sim Q}[f(Y)]  |,
$$
where $\Fc=\{g\in\mathcal{H}: \|g\|_{\mathcal{H}} \leq 1 \}$ and we recall that $\mathcal{H}$ is the RKHS associated with $k$. From Example 3.6 in~\cite{legramanti2025concentration}, we know  that for any $f\in \mathcal{H}$, $ \|f\|_{\infty} \leq \|f\|_{\mathcal{H}} \sup_{y} k(y,y)$. As $\sup_{y}k(y,y)\leq 1$ by assumption and $\|f\|_{\mathcal{H}}\leq 1$ for $f\in \Fc$, we obtain that $ \|f\|_{\infty} \leq 1$ for $f \in \Fc$.
Moreover, from the same reference, the Rademacher complexity of $\Fc$ (as defined in Definition 2.5 in~\cite{legramanti2025concentration}) satisfies $\mathcal{R}_{\mu,n}(\Fc)\leq 1/\sqrt{n}$ for any $n$ and any probabiliy measure $\mu$, in particular when $\mu=P_\theta$. Thus, we can apply Lemma C1 in~\cite{legramanti2025concentration}: for a fixed $\theta\in\Theta$ and a fixed $\delta>0$,
$$
\mathbb{P}\left( \mathcal{D}(\hat{P}_\theta,P_\theta) \geq \frac{2}{\sqrt{s_N}} + \frac{4}{\sqrt{N}} +  \delta \right) \leq 2 \exp\left[  -\frac{s_N \delta^2}{2} \right] + 2 s_N \beta( \lfloor \sqrt{N} \rfloor; \theta),
$$
where $s_N = \lfloor N / (2 \lfloor \sqrt{N} \rfloor) \rfloor $.
We keep our fixed $\delta>0$ but we now consider a finite grid $\mathcal{G}=\{\theta_1,\dots,\theta_M\} \subset \Theta$. A union bound gives
\begin{equation}
\label{equa:proof:pierre:1}
\mathbb{P}\left( \sup_{\theta\in\mathcal{G}} \mathcal{D}(\hat{P}_\theta,P_\theta) \geq \frac{2}{\sqrt{s_N}} + \frac{4}{\sqrt{N}} +  \delta \right) \leq 2 M \exp\left[  -\frac{s_N \delta^2}{2} \right] + 2 M s_N \sup_{\theta\in\mathcal{G}}\beta( \lfloor \sqrt{N} \rfloor;\theta).
\end{equation}
Now, let us assume that $\mathcal{G}=\{\theta_1,\dots,\theta_M\} \subset \Theta$ is an $\alpha$-cover of $\Theta$ for a given $\alpha>0$, that is: for any $\theta\in\Theta$, there is $\theta'\in\mathcal{G}$ such that $\|\theta-\theta\|' \leq \alpha$. By definition of the covering number $\Nc(\alpha,\Theta,\|\cdot\|)$, the grid $\mathcal{G}$ can be chosen such that $M:=\Nc(\alpha,\Theta,\|\cdot\|)$ the covering number of $\Theta$. Assume for now there is a $L>0$ (possibly random) such that $|\mathcal{D}(\hat{P}_\theta,P_\theta)-\mathcal{D}(\hat{P}_{\theta'},P_{\theta})| \leq L \|\theta-\theta'\|^{\gamma} $ for any couple $(\theta,\theta')\in \Theta\times \Gc$. We will check later that this is actually a consequence of \textit{(iii)} and make $L$ explicit at the same time. For any $\theta\in\Theta$ and any $\theta'\in\mathcal{G}$, we have
$$
\mathcal{D}(\hat{P}_\theta,P_\theta) \leq \mathcal{D}(\hat{P}_{\theta'},P_{\theta'}) + L \|\theta-\theta ' \|^{\gamma},
$$
and we know that $\theta'$ can be chosen such that $ \|\theta-\theta ' \| \leq \alpha$. Therefore,
$$
\sup_{\theta\in\Theta} \mathcal{D}(\hat{P}_\theta,P_\theta) \leq \sup_{\theta'\in\mathcal{G}} \mathcal{D}(\hat{P}_{\theta'},P_{\theta'}) + \alpha^\gamma L.
$$
By plugging the latter inequality into~\eqref{equa:proof:pierre:1}, we obtain
\begin{equation}
\mathbb{P}\left( \sup_{\theta\in\Theta} \mathcal{D}(\hat{P}_\theta,P_\theta) \geq
\alpha^\gamma L + \frac{2}{\sqrt{s_N}} + \frac{4}{\sqrt{N}}  + \delta \right) \leq 2 M \exp\left[  -\frac{s_N \delta^2}{2} \right] + 2 M s_N \sup_{\theta\in\mathcal{G}}\beta( \lfloor \sqrt{N} \rfloor;\theta).
\end{equation}
There is $N_0$ large enough so that, for any $N\geq N_0$, $2/\sqrt{s_N} + 4/\sqrt{N} \leq \delta$. For such an $N$, the above equation can be rewritten as
\begin{equation*}
\mathbb{P}\left( \sup_{\theta\in\Theta} \mathcal{D}(\hat{P}_\theta,P_\theta) \geq
\alpha^\gamma L + 2 \delta \right) \leq 2 M \exp\left[  -\frac{s_N \delta^2}{2} \right] + 2 M s_N \sup_{\theta\in\mathcal{G}}\beta( \lfloor \sqrt{N} \rfloor;\theta).
\end{equation*}
Then, we get
\begin{align}
\mathbb{P}\left( \sup_{\theta\in\Theta} \mathcal{D}(\hat{P}_\theta,P_\theta) \geq 3 \delta\right)
& =
\mathbb{P}\left( \sup_{\theta\in\Theta} \mathcal{D}(\hat{P}_\theta,P_\theta) \geq 3 \delta , \alpha^\gamma L < \delta \right) + \mathbb{P}\left( \sup_{\theta\in\Theta} \mathcal{D}(\hat{P}_\theta,P_\theta) \geq 3 \delta , \alpha^\gamma L \geq \delta \right) \nonumber \\
& \leq 
\mathbb{P}\left( \sup_{\theta\in\Theta} \mathcal{D}(\hat{P}_\theta,P_\theta) \geq 2 \delta+ \alpha^\gamma L , \alpha^\gamma L < \delta \right) + \mathbb{P}( \alpha^\gamma L \geq \delta ) 
\nonumber \\
& \leq  2 M \exp\left[  -\frac{s_N \delta^2}{2} \right] + 2 M s_N \sup_{\theta\in\mathcal{G}}\beta( \lfloor \sqrt{N} \rfloor;\theta) + \frac{\alpha^\gamma}{\delta} \mathbb{E}[L] .
\label{equa:proof:pierre:2}
\end{align}
    It is now time to specify $L$. Let us do this task in the PSMMD case, that includes ISMMD particularly. 
The CSMMD case can be managed similarly. Due to~(\ref{dec_MMD}) of the main text and with our notations (recall~(\ref{not_simul_y})), we have
{\small{\begin{align*}
\mathcal{D}(\hat{P}_\theta,P_\theta )
- \mathcal{D}(\hat{P}_{\theta'},P_{\theta'})
&
=
\left\| \frac{1}{N}\sum_{i=1}^N \Phi\big(\psi(\theta,\underline{u}^{(i)})\big) - \mathbb{E} \Phi\big(\psi(\theta,\underline{u}^{(i)})\big) \right\|_{\mathcal{H}}
- \left\| \frac{1}{N}\sum_{i=1}^N \Phi\big(\psi(\theta',\underline{u}^{(i)})\big) - \mathbb{E} \Phi\big(\psi(\theta',\underline{u}^{(i)})\big) \right\|_{\mathcal{H}}
\\
&
\leq \left\| \frac{1}{N}\sum_{i=1}^N \big\{ \Phi\big(\psi(\theta,\underline{u}^{(i)})\big)- \Phi\big(\psi(\theta',\underline{u}^{(i)})\big)\big\} - \mathbb{E}\Big[\Phi\big(\psi(\theta,\underline{u}^{(i)})\big)-\Phi\big(\psi(\theta',\underline{u}^{(i)})\big)\Big] \right\|_{\mathcal{H}}
\\
&
\leq
\underbrace{
\left\{ \frac{1}{N}\sum_{i=1}^N m(\underline{u}^{(i)})  + \mathbb{E}\big[m(\underline{u}^{(i)})\big] \right\}
}_{=: L} \|\theta-\theta'\|^\gamma,
\end{align*}}}
where we used \textit{(iv)} for the last inequality. The same upper bound obviously applies after swapping $\theta$ and $\theta'$. This defines $L$, and we have
$  \mathbb{E}[L] \leq 2 \mathbb{E}[m(\underline{u})] < +\infty.  $
Coming back to~\eqref{equa:proof:pierre:2}, this yields
\begin{eqnarray*}
\lefteqn{ \mathbb{P}\left( \sup_{\theta\in\Theta} \mathcal{D}(\hat{P}_\theta,P_\theta) \geq 3 \delta\right)
 \leq
2 \Nc(\alpha,\Theta,\|\cdot\|) \exp\left[  -\frac{s_N \delta^2}{2} \right] + 2 \Nc(\alpha,\Theta,\|\cdot\|)  s_N \sup_{\theta\in\mathcal{G}}\beta( \lfloor \sqrt{N} \rfloor;\theta) }\\
&+& 2 \frac{\alpha^\gamma}{\delta}  \mathbb{E}[m(\underline{u})]
 \leq
\mathcal{O}(\alpha^{-\pi}) \exp\left[  -\frac{s_N \delta^2}{2} \right] + \mathcal{O}(\alpha^{-\pi})  s_N \sup_{\theta\in\mathcal{G}}\beta( \lfloor \sqrt{N} \rfloor;\theta)  + 2 \frac{\alpha^\gamma}{\delta}  \mathbb{E}[m(\underline{u})],
\end{eqnarray*}
where we used \textit{(iii)} to upper bound $\Nc(\alpha,\Theta,\|\cdot\|)$.
Choose $\delta = \varepsilon/6$, so that $3\delta = \varepsilon/2$, and
$\alpha = 1/N $ to get
$$
\mathbb{P}\left( \sup_{\theta\in\Theta} \mathcal{D}(\hat{P}_\theta,P_\theta) \geq \frac{\varepsilon}{2} \right)
 \leq 
\mathcal{O}(N^\pi) \exp\left[  -\frac{s_N \varepsilon^2}{72} \right] + \mathcal{O}(N^\pi)  s_N \sup_{\theta\in\mathcal{G}}\beta( \lfloor \sqrt{N} \rfloor;\theta)  + \frac{12 \mathbb{E}[m(\underline{u})] }{\varepsilon N^\gamma}\cdot
$$
Observe that $s_N = \mathcal{O}(N^{1/2})$ and we assumed that $\sup_{\theta\in\mathcal{G}}\beta(n;\theta) = o(1/n^{2\pi+1}) $, which implies 
$$\sup_{\theta\in\mathcal{G}} \beta(\lfloor \sqrt{N} \rfloor;\theta) = o(1/N^{\pi+1/2}) .$$ 
Thus, we obtain
$$
\mathbb{P}\left( \sup_{\theta\in\Theta} \mathcal{D}(\hat{P}_\theta,P_\theta) \geq \frac{\varepsilon}{2} \right)
 \leq 
\mathcal{O}(N^\pi) \exp\left[  -\frac{s_N \varepsilon^2}{72} \right] + o(1)  + \frac{12 \mathbb{E}[m(\underline{u})] }{\varepsilon N^\gamma}
 \xrightarrow[N\rightarrow\infty]{}  0.
$$
Recalling~\eqref{equa:proof:pierre:0}, this shows that
$$ \sup_{\theta\in\Theta} |\hat{M}_{N,T}(\theta)-M(\theta)| \xrightarrow[T,N\rightarrow\infty]{ \text{prob.} } 0. $$
Together with \textit{(ii)}, this means the assumptions of Theorem 5.7 in~\cite{van2000asymptotic} are satisfied. Therefore, we have obtained
$$
\tilde{\theta}_{N,T} \xrightarrow[T,N\rightarrow\infty]{ \text{prob.} } \theta^*,
$$
which ends the proof.
\end{proof}

\section{Proofs of asymptotic normality results}
\label{supp:section_proofs_AN}

We recall our ``ideal'' estimator of $\theta^*$: $\hat\theta_T=\arg\min L_T(\theta)$, with 
$$  L_T(\theta):=  \Eb_{(Y,Y')\sim P_\theta\otimes P_\theta}\big[k(Y,Y')\big]
- \frac{2}{T}\sum_{t=1}^T  \Eb_{Y\sim P_\theta} \big[k(y_{t},Y)\big]   
+ \frac{1}{T^2}\sum_{t,t'=1}^T k(y_t,y_{t'}).$$
Our second estimator of $\theta^*$ is the simulation-based $\tilde\theta_{N,T}=\arg\min L_{N,T}(\theta)$, with 
$$  L_{N,T}(\theta):= \frac{1}{N^2}\sum_{i,i'=1}^{N} k(\tilde y_i,\tilde y_{i'}) 
- \frac{2}{TN}\sum_{t=1}^T \sum_{i=1}^{N} k(y_{t},\tilde y_i)   
+ \frac{1}{T^2}\sum_{t,t'=1}^T k(y_t,y_{t'}).$$

\begin{proof}[Proof of Proposition \ref{AN_theta_T}]
The proof is rather standard because $\hat\theta_T$ is a M-estimator and $\nabla_\theta L_T(\theta)$ appears as an average.
Indeed, due to our model assumptions, the loss can be rewritten
$$  L_T(\theta)= \int  k\big(\psi(\theta,u),\psi(\theta,u')\big)\, P_\Ub(du)\, P_\Ub(du')
- \frac{2}{T}\sum_{t=1}^T \int k\big(y_{t},\psi(\theta,u)\big)\, P_\Ub (du)   
+ \frac{1}{T^2}\sum_{t,t'=1}^T k(y_t,y_{t'}).$$
By a limited expansion, we have
\begin{equation}
 0= \nabla_\theta L_T(\hat \theta_T)= \nabla_\theta L_T(\theta^*) + \nabla^2_{\theta,\theta^\top} L_T(\bar\theta)\cdot (\hat \theta_T - \theta^*), 
\label{Taylor_theta_T}    
\end{equation}
for some (random) parameter $\bar\theta\in \Theta$ s.t. $\|  \bar\theta - \theta^* \|\leq \|\hat\theta_T -\theta^*\|$.
The result follows if we prove the asymptotic normality of $\sqrt{T}\nabla_\theta L_T(\theta^*)$ and the weak consistency of 
$\nabla^2_{\theta,\theta^\top} L_T(\theta)$ uniformly on a neighborhood of $\theta^*$.

\mds 

By differentiating the loss map and since $k$ is symmetrical, we get
\begin{eqnarray*}
\lefteqn{  \nabla_\theta L_T(\theta^*) = 2\int \nabla_\theta \psi^\top(\theta^*,u) \nabla_1 k\big(\psi(\theta^*,u),\psi(\theta^*,u')\big) \, P_\Ub(du)\, P_\Ub(du')   }\\
&-&\frac{2}{T}\sum_{t=1}^T \int \nabla_\theta \psi^\top(\theta^*,u) \nabla_1 k\big(\psi(\theta^*,u),y_{t}\big)\, P_\Ub (du).    
\end{eqnarray*}
Note that 
$$ \Eb\big[ \nabla_\theta L_T(\theta^*) \big]= \nabla_\theta \Eb\big[  L_T(\theta) \big]_{|\theta=\theta^*}      
= \nabla_\theta \Dc^2(P_\theta, P_0)_{|\theta=\theta^*} =0,
$$
by the definition of $\theta^*$.
By the Cramer-Wold device and a CLT for strongly stationary $\beta$-mixing sequences
(Theorem 10.7 of \cite{bradley2007introduction} in our case), we obtain that
 $\sqrt{T}\nabla_\theta  L_T(\theta^*)$ weakly tends to a $\Nc\big(0,V_0(\theta^*)\big)$ due to (a).

\mds 

Moreover, simple algebra yields
{\small{\begin{eqnarray*}
    \lefteqn{ \nabla^2_{\theta,\theta^\top} L_T(\theta) =2\int \Big\{\nabla_{\theta} \psi^\top(\theta,u) \nabla^2_{1,1} k\big(\psi(\theta,u),\psi(\theta,u')\big)  \nabla_{\theta^\top} \psi(\theta,u) }\\
    &+& \nabla_{\theta} \psi^\top(\theta,u) \nabla^2_{1,2} k\big(\psi(\theta,u),\psi(\theta,u')\big)  \nabla_{\theta^\top} \psi(\theta,u')  + \sum_{l=1}^q \partial_{l} k\big(\psi(\theta,u),\psi(\theta,u')\big)  \nabla^2_{\theta,\theta^\top} \psi_l(\theta,u) \Big\}
    \, P_\Ub(du)\, P_\Ub(du') \\   
&-&\frac{2}{T}\sum_{t=1}^T \int \Big\{ \nabla_{\theta} \psi^\top(\theta,u) \nabla^2_{1,1} k\big(\psi(\theta,u),y_{t}\big)\nabla_{\theta^\top} \psi(\theta,u) + \sum_{l=1}^q \partial_{l} k\big(\psi(\theta,u),y_{t}\big)\nabla^2_{\theta,\theta^\top} \psi_l(\theta,u)\Big\}\, P_\Ub (du) \\
&=& I(\theta) - \int f_\theta(y) \, P_T(dy). 
\end{eqnarray*}}}
Note that the expectation of the latter term is $V_1(\theta)$.
Deduce that $\nabla^2_{\theta,\theta^\top} L_T(\bar\theta)$ tends to $V_1(\theta^*)$ in probability from {\it (c)} and the continuity of $V_1(\cdot)$ at $\theta^*$. Indeed, 
$$ \nabla^2_{\theta,\theta^\top} L_T(\bar\theta) - V_1(\theta^*)=
\nabla^2_{\theta,\theta^\top} L_T(\bar\theta) - V_1(\bar\theta) + o_P(1)
= \int f_{\theta^*} d(P_T - P_0) + o_P(1)=o_P(1),$$
because the sequence $\big(f_{\theta^*}(y_t)\big)$ is beta-mixing and satisfies a weak LLN.
Then,~(\ref{Taylor_theta_T}) provides the asymptotic normality of $\sqrt{T}(\hat \theta_T-\theta^*)$;
its asymptotic variance is obviously 
$V_1(\theta^*)^{-1}V_0(\theta^*) V_1(\theta^*)^{-1}$.
\end{proof}

\begin{proof}[Proof of Proposition \ref{AN_theta_NT}]
As usual with M-estimation, the estimated parameter $\tilde \theta_{N,T}$ satisfies the first-order equation, i.e., 
$$ 0= \nabla_\theta L_{N,T}(\tilde \theta_{N,T})= \nabla_\theta L_{N,T}(\theta^*) + \nabla^2_{\theta,\theta^\top} L_{N,T}(\bar\theta)\cdot (\tilde \theta_{N,T} - \theta^*), $$
for some parameter $\bar\theta\in \Theta$ s.t. $\| \bar\theta - \theta^* \|\leq \|\tilde\theta_{N,T} -\theta^*\|$.
By assumption, $\tilde\theta_{N,T}$ tends to $\theta^*$ in probability. This implies $\bar\theta - \theta^*=o_P(1)$.
Let us prove that, when $T$ and $N$ tends to $+\infty$ so that $N/(N+T)\rightarrow \lambda$, we have
\begin{enumerate}
\item[(a)] $\sqrt{T+N} \nabla_\theta L_{N,T}(\theta^*)$ tends in law towards a $\Nc(0,\Sigma)$, and
\item[(b)] $\nabla^2_{\theta,\theta^\top} L_{N,T}(\bar\theta)$ tends to $\tilde \Sigma$ in probability.
\end{enumerate}
We will straightforwardly deduce from (a) and (b) that
$$ \sqrt{N+T}(\tilde \theta_{N,T} - \theta^*) 
\xrightarrow[N,T\rightarrow\infty]{ \text{ law}}
\Nc\big( 0, \tilde\Sigma^{-1}\Sigma \tilde \Sigma^{-1} \big) ,$$
under our conditions of regularity.

\mds 

{\it Proof of (a):}
By differentiating the loss map, we get
$$ \nabla_\theta L_{N,T}(\theta)_{|\theta=\theta^*} = \frac{1}{N^2}\sum_{i,i'=1}^{N} \big\{ z_i \nabla_1 k(\tilde y_i,\tilde y_{i'}) + z_{i'} \nabla_1 k(\tilde y_{i'},\tilde y_i) \big\} 
- \frac{2}{TN}\sum_{t=1}^T \sum_{i=1}^{N}  z_i \nabla_1 k(\tilde y_i, y_{t}) ,$$
with our notations.
Thus, rewrite $ \nabla_\theta L_{N,T}(\theta^*) = U_1 - U_2+r_{N,T} + O_P(N^{-1})$, by setting
$$ U_1:=\frac{1}{N(N-1)} \sum_{i,i'=1, i\neq i'}^{N} h(\zeta_i,\zeta_{i'}) ,$$
$$ U_2:= \frac{1}{TN}\sum_{t=1}^T \sum_{i=1}^{N} g( \zeta_i,y_t),\;\text{and}\;  r_{N}:=\frac{1}{N^2}\sum_{i=1}^N z_i\nabla_1 k(\tilde y_i,\tilde y_{i}) .$$
Note that $r_{N}=O_P(N^{-1})$ by Markov's inequality, and $r_N$ will then be negligible.
Note that the sequences $(\zeta_i)_{i\geq 1}$ and $(y_t)_{t\geq 1}$ in $\Rb^{q(1+d)}$ and $\Rb^{q}$ respectively are independent by the definition of our simulation schemes. 
Moreover, they are $\beta$-mixing by assumption. 

\mds 

By definition of $\theta^*$, we have $\nabla_\theta \Dc^2(P_\theta,P_0)_{|\theta=\theta^*}=0$, which implies
$$ \int h(\zeta_1,\zeta_2) \, P_\zeta (d\zeta_1)\, P_\zeta (d\zeta_2) = \int  g(\zeta_1,y) \, P_\zeta (d\zeta_1)\, P_{0}(dy) =:E.  $$
Indeed, in the definition of the MMD distance, we consider independent realizations under $P_\theta$ and $P_0$, contrary to its estimator $L_{N,T}(\theta)$ 
~\footnote{Note that $E= \Eb[U_2]$, but $E\neq \Eb[U_1]$ in general because the latter expectation has to take into account dependencies between the vectors $\zeta_i$ (except in the ISMMD case). This is a classical feature of U-statistics with dependent data.}.
Rewrite 
\begin{eqnarray*}
 \lefteqn{ \nabla_\theta L_{N,T}(\theta^*) = U_1 - \int h(\zeta_1,\zeta_2) \, P_\zeta (d\zeta_1)\, P_\zeta (d\zeta_2) - \Big( U_2 - \int  g(\zeta_1,y) \, P_\zeta (d\zeta_1)\, P_{0}(dy) \Big) + O_P(\frac{1}{N}) }\\
 &=& (U_1 - E) - (U_2 - E) + O_P(\frac{1}{N}).    \hspace{10cm}
\end{eqnarray*}
$U_1$ is a U-statistic based on the symmetrical kernel $h$ and on dependent observations in general. By Hajek projection, we approximate $U_1 - E$ by 
$$\hat U_1:= \frac{2}{N}\sum_{i=1}^N h_{(1)}(\zeta_i).$$
Similarly, we approximate $U_2-E$ by
$$\hat U_2:= \frac{1}{N}\sum_{i=1}^N g_{(1)}(\zeta_i)+  \frac{1}{T}\sum_{t=1}^T g_{(2)}(y_t).$$
Note that 
$$ \sqrt{N+T}\big( \hat U_1-\hat U_2\big)=\frac{\sqrt{N+T}}{N}\sum_{i=1}^N (2h_{(1)}-g_{(1)})(\zeta_i) -  \frac{\sqrt{N+T}}{T}
   \sum_{t=1}^T g_{(2)}(y_t),$$
that is the difference between two independent centered sequences. Since $N/(N+T)\rightarrow \lambda $, apply the Cramer-Wold device and a CLT for strongly stationary $\beta$-mixing sequences: due to Theorem 10.7 in~\cite{bradley2007introduction} and invoking Assumption \ref{AN_theta_NT_beta_zeta} of the main text, we have 
\begin{equation}
\frac{\sqrt{N+T}}{N}\sum_{i=1}^N (2h_{(1)}-g_{(1)})(\zeta_i) 
\xrightarrow[N,T\rightarrow\infty]{ \text{ law}} \Nc(0,\Sigma_1/\lambda).
\label{AN_h1_g1}    
\end{equation} 
Similarly, applying the same CLT for the $\beta$-mixing sequence $(y_t)$, we get the asymptotic normality of $T^{-1}\sum_{t=1}^T g_{(2)}(y_t)$: due to Assumption \ref{AN_theta_NT_beta} of the main text, we have
\begin{equation}
 \frac{\sqrt{N+T}}{T}\sum_{t=1}^T g_{(2)}(y_t) 
 \xrightarrow[N,T\rightarrow\infty]{ \text{ law}}
 \Nc\big(0,\Sigma_2/(1-\lambda)\big).
 \label{AN_g2}
\end{equation}
By independence, this yields the asymptotic normality of $\sqrt{N+T}\big( \hat U_1-\hat U_2\big)$.
The result (a) follows if we state that $U_k-E-\hat U_k$, $k\in \{1,2\}$, are $o_P(1/\sqrt{N+T})$.    
Since $U_1$ is a U-statistic with dependent observations, Lemma 2 of~\cite{yoshihara1976limiting} (or Proposition 2 of~\cite{denker1983u}) yields
$U_1- E-\hat U_1=o_P(1/\sqrt{N})$ (recall Assumption \ref{AN_theta_NT_beta_zeta}).
Concerning $U_2$, note that
$$ U_2- E-\hat U_2=\frac{1}{NT}\sum_{i=1}^N \sum_{t=1}^T 
\Big\{ g(\zeta_i,y_t) - \Eb_{Y\sim P_0}[g(\zeta_i,Y)] - \Eb_{\zeta\sim P_{\zeta}}[g(\zeta,y_t)]+ E\Big\}. $$
It is sufficient to prove that the variance of any component of 
$U_2- E-\hat U_2$ is $o(1/(N+T))$. Let us focus on an arbitrarily chosen component of the vector-valued map $g$, say $g_l$ for some $l\in \{1,\ldots,q\}$.
Moreover, denote $E_l:= \int  g_l(\zeta,y) \, dP_\zeta (\zeta)\, dP_{0}(y)$ its expectation.
Thus, we have just to state that
$$ \Delta_l := \Eb\Big[  \Big( \frac{1}{NT}\sum_{i=1}^N \sum_{t=1}^T \big\{ 
 g_l(\zeta_i,y_t) - \Eb_{Y\sim P_0}[g_l(\zeta_i,Y)] - \Eb_{\zeta\sim P_{\zeta}}[g_l(\zeta,y_t)]+ E_l\big\} \Big)^2 \Big] = o\Big( \frac{1}{N+T} \Big).$$
Setting $\Delta(i_1,t_1,i_2,t_2):=\Eb\big[ H_l(\zeta_{i_1},\zeta_{i_2},y_{t_1},y_{t_2})\big]$, this means proving 
\begin{equation}
     \Delta_l=\frac{1}{N^2 T^2}\sum_{i_1,i_2=1}^N \sum_{t_1,t_2=1}^T \Delta(i_1,t_1,i_2,t_2) = o\Big( \frac{1}{N+T}\Big) .
    \label{order_residual_term}
\end{equation}

To this aim, let us state two technical lemmas that are direct extensions of Lemma 1 of~\cite{yoshihara1976limiting} and then given without any proof. 
Let $(i_1,i_2,t_1,t_2)$ be arbitrary positive integers. Denote $\iota :=(i_1,i_2,t_1,t_2)$.  
Let $A_1$ and $A_2$ (resp. $B_1$ and $B_2$) be arbitrary borel subsets in $\Rb^{q(1+d)}$ (resp. $\Rb^{q}$). 
For a given $\iota$, define the probability measures $Q^\iota$ and $Q_\zeta^\iota$ on $\Rb^{2q(1+d)}\times \Rb^{2q}$ by
$$ Q^\iota(A_1\times A_2\times B_1 \times B_2):= P\big( (\zeta_{i_1},\zeta_{i_2})\in A_1\times A_2\big)
P\big((y_{t_1},y_{t_2})\in B_1\times B_2\big),\; \text{and} $$
$$ Q_\zeta^\iota(A_1\times A_2\times B_1 \times B_2):= P( \zeta_{i_1} \in A_1) P( \zeta_{i_2}\in A_2)
P\big((y_{t_1},y_{t_2})\in B_1\times B_2\big). $$
\begin{lem}
Let $J(\zeta,\zeta',y,y')$ be a Borel function from $\Rb^{2q(1+d)}\times \Rb^{2q}$ to $\Rb$ such that
$$  \max\Big( \int | J(\zeta,\zeta',y,y')|^{1+\delta_1} \, dP_{\zeta}(\zeta)\, dP_{\zeta}(\zeta') ;  \int | J(\zeta,\zeta',y,y')|^{1+\delta_1} \, 
d P_{(\zeta_{i_1},\zeta_{i_2})}(\zeta,\zeta') \Big) \leq M(y,y')<\infty ,$$
for any couple $(y,y')\in \Rb^{2q}$ and any couple of indices $(i_1,i_2)$, $i_1<i_2$. Moreover,
$ \sup_{t,t'}\Eb\Big[  M^{1/(1+\delta_1)}(y_{t},y_{t'}) \Big]=: M^* <\infty .$
Then, 
$$ \Big| \int J(\zeta,\zeta',y,y') \, \big(dQ^\iota - dQ_\zeta^\iota\big)(\zeta,\zeta',y,y') \Big| \leq 4 M^* \beta_\zeta^{\delta_1/(2+\delta_1)}(i_2-i_1) .$$
\label{tech_lem_1}
\end{lem}

Similarly, the roles of the variables $\zeta_i$ and $y_t$ can be interchanged, with new mixing coefficients: 
given $\iota$, define the probability measure $Q_y^\iota$ on $\Rb^{2q(1+d)}\times \Rb^{2q}$ by
$$ Q_y^\iota(A_1\times A_2\times B_1 \times B_2):= P_{\theta^*}\big( (\zeta_{i_1},\zeta_{i_2})\in A_1\times A_2\big) 
P_0(y_{t_1}\in B_1) P_0(y_{t_2}\in B_2). $$
\begin{lem}
Let $J(\zeta,\zeta',y,y')$ be a Borel function from $\Rb^{2q(1+d)}\times \Rb^{2q}$ to $\Rb$ such that
$$  \max\Big( \int | J(\zeta,\zeta',y,y')|^{1+\delta_2} \, dP_0(y)\, dP_0(y') ;  \int | J(\zeta,\zeta',y,y')|^{1+\delta_2} \, 
d P_{(y_{t_1},y_{t_2})}(y,y') \Big) \leq \bar M(\zeta,\zeta')<\infty ,$$
for any couple $(\zeta,\zeta')\in \Rb^{2q(1+d)}$ and any couple of indices $(t_1,t_2)$, $t_1<t_2$. Moreover,
$ \sup_{i,j}\Eb\Big[  \bar M^{1/(1+\delta_2)}(\zeta_{i},\zeta_{j}) \Big]=:\bar M^* <\infty .$
Then, 
$$ \Big| \int J(\zeta,\zeta',y,y') \, \big(dQ^\iota - dQ_y^\iota\big)(\zeta,\zeta',y,y') \Big| \leq 4 \bar M^* \beta^{\delta_2/(2+\delta_2)}(t_2-t_1) .$$
\label{tech_lem_2}
\end{lem}

To prove~(\ref{order_residual_term}), let us mimic the same reasoning as in Lemma 2 of~\cite{yoshihara1976limiting}; invoking our Lemma~\ref{tech_lem_1} and our Lemma~\ref{tech_lem_2}, replace their map $J$ with $H_l$.
To this aim, it is important to note that 
$$\Delta(i_1,t_1,i_2,t_2):=\int H_l(\zeta,\zeta',y,y')\, dQ^\iota (\zeta,\zeta',y,y'),$$ 
$$\int H_l(\zeta,\zeta',y,y')\, dQ_\zeta^\iota (\zeta,\zeta',y,y')=0,\; \text{and}\; 
\int H_l(\zeta,\zeta',y,y')\, dQ_y^\iota (\zeta,\zeta',y,y')=0.$$
To fix the ideas, assume $1\leq i_1 < i_2 \leq N$ and $1\leq t_1\leq t_2 \leq T$, with $t_2-t_1\leq i_2-i_1$  (case 1). Then, Lemma~\ref{tech_lem_1} implies
$$ |\Delta(i_1,t_1,i_2,t_2)| \leq 4 M^* \beta_\zeta^{\delta_1/(2+\delta_1)}(i_2-i_1).$$
By summing over all t-uplets $\iota$ in case 1, we get 
$$ |\sum_{\{(i_1,i_2);i_1< i_2\}} \sum_{\{(t_1,t_2);0\leq t_2-t_1\leq i_2-i_1\}}\Delta(i_1,t_1,i_2,t_2)| \leq 4 M^* NT\sum_{k=1}^N (k+1)\beta_\zeta^{\delta_1/(2+\delta_1)}(k) .$$
Other similar cases are 
\begin{itemize}
\item 
$1\leq i_2 \leq  i_1 \leq N$ and $1\leq t_1\leq t_2 \leq T$, with $t_2-t_1\leq i_1-i_2$;
\item 
$1\leq i_1 < i_2 \leq N$ and $1\leq t_2<t_1 \leq T$, with $t_1-t_2 \leq  i_2-i_1$;
\item 
$1\leq i_2\leq i_1 \leq N$ and $1\leq t_2 < t_1\leq T$, with $t_1-t_2\leq i_1-i_2  $.
\end{itemize}
All these cases are treated in the same way as case 1 and provide the same upper-bound.

Now, assume $1\leq t_1 < t_2 \leq N$ and $1\leq i_1\leq i_2 \leq T$, with $i_2-i_1\leq t_2-t_1$  (case 2). Then, applying Lemma~\ref{tech_lem_2} and by the same reasoning as for case 1, we get
$$ |\sum_{t_1< t_2} \sum_{i_1\leq i_2}\Delta(i_1,t_1,i_2,t_2)| \leq 4 \bar M^* NT\sum_{t=1}^T (t+1)\beta^{\delta_2/(2+\delta_2)}(t) .$$
And similarly for the three similar situations, as above. Globally, we have obtained
$$ |\frac{1}{N^2 T^2}\sum_{i_1,i_2=1}^N \sum_{t_1,t_2=1}^T \Delta(i_1,t_1,i_2,t_2)| = 
O\Big(\frac{1}{NT} \sum_{k=1}^{N} (k+1)\beta_\zeta^{\delta_1/(2+\delta_1)} (k) + \frac{1}{NT} \sum_{t=1}^{T} (t+1)\beta^{\delta_2/(2+\delta_2)} (t)    \Big).
$$
Deduce from Assumption \ref{AN_theta_NT_beta_zeta} that
$\sum_{k=1}^{N} (k+1)\beta_\zeta^{\delta_1/(2+\delta_1)} (k) \sim N^{2-(2+\delta_1')\delta_1/(\delta_1'(2+\delta_1))}.$
Similarly, invoking Assumption \ref{AN_theta_NT_beta}, we have
$\sum_{t=1}^{T} (t+1)\beta^{\delta_2/(2+\delta_2)} (t) \sim T^{2-(2+\delta_2')\delta_2/(\delta_2'(2+\delta_2))}.$
This yields
$$ \frac{1}{N^2 T^2}\sum_{i_1,i_2=1}^N \sum_{t_1,t_2=1}^T \Delta(i_1,t_1,i_2,t_2) = 
O\Big(\frac{1}{N T^{(2+\delta_2')\delta_2/(\delta_2'(2+\delta_2))-1}}+ \frac{1}{T N^{(2+\delta_1')\delta_1/(\delta_1'(2+\delta_1))-1}} \Big)=
o(\frac{1}{N}+ \frac{1}{T}),
$$
since $(2+\delta_k')\delta_k/\big(\delta_k'(2+\delta_k)\big)>1$ when $\delta_k'<\delta_k$, $k\in \{1,2\}$.
This proves (a).

\mds 

{\it Proof of (b):}
For any $\theta$ in a neighborhood of $\theta^*$, simple calculations yield the matrix
\begin{eqnarray}
    \lefteqn{ \nabla^2_{\theta,\theta^\top} L_{N,T}(\theta) =\frac{2}{N^2} \sum_{i,i'=1}^N \Big\{\nabla_{\theta} \psi^\top(\theta,\underline{u}^{(i)}) \nabla^2_{1,1} k\big(\psi(\theta,\underline{u}^{(i)}),\psi(\theta,\underline{u}^{(i')})\big)  \nabla_{\theta^\top} \psi(\theta,\underline{u}^{(i)}) } \nonumber\\
    &+& \nabla_{\theta} \psi^\top(\theta,\underline{u}^{(i)}) \nabla^2_{1,2} k\big(\psi(\theta,\underline{u}^{(i)}),\psi(\theta,\underline{u}^{(i')})\big)  \nabla_{\theta^\top} \psi(\theta,\underline{u}^{(i')})  \nonumber \\
    &+& \sum_{l=1}^q \partial_{l} k\big(\psi(\theta,\underline{u}^{(i)}),\psi(\theta,\underline{u}^{(i')})\big)  \nabla^2_{\theta,\theta^\top} \psi_l(\theta,\underline{u}^{(i)}) \Big\} \nonumber \\   
&-&\frac{2}{NT}\sum_{t=1}^T \sum_{i=1}^N\Big\{ \nabla_{\theta} \psi^\top(\theta,\underline{u}^{(i)}) \nabla^2_{1,1} k\big(\psi(\theta,\underline{u}^{(i)}),y_{t}\big)\nabla_{\theta^\top} \psi(\theta,\underline{u}^{(i)}) \nonumber \\
&+& \sum_{l=1}^q \partial_{l} k\big(\psi(\theta,\underline{u}^{(i)}),y_{t}\big)\nabla^2_{\theta,\theta^\top} \psi_l(\theta,\underline{u}^{(i)})\Big\}   \nonumber \\ 
&=& \frac{1}{N(N-1)}\sum_{i,i'=1, i\neq i'}^{N} \bar h_\theta(\underline{u}^{(i)},\underline{u}^{(i')}) - 
\frac{2}{TN}\sum_{t=1}^T \sum_{i=1}^{N} \bar g_\theta(\underline{u}^{(i)},y_t) + O_P(\frac{1}{N}). \hspace{3cm}
\label{dec_Hessian_thetaNT}
\end{eqnarray}
Due to Assumption \ref{lipschitz_ULLN} of the main text, 
$$   \| \nabla^2_{\theta,\theta^\top} L_{N,T}(\bar\theta) - \nabla^2_{\theta,\theta^\top} L_{N,T}(\theta^*) \| = o_P(1).  $$
Thus, we can focus on the case $\theta=\theta^*$ and it remains to state the weak consistency of $\nabla^2_{\theta,\theta^\top} L_{N,T}(\theta^*)$.

\mds 

First, concerning the U-statistic with the kernel $\bar h_{\theta^*}$, we have
\begin{equation}
 \frac{1}{N(N-1)}\sum_{i,i'=1, i\neq i'}^{N} \big\{ \bar h_{\theta^*}(\underline{u}^{(i)},\underline{u}^{(i')})   - \Eb[ \bar h_{\theta^*}(\underline{u}^{(i)},\underline{u}^{(i')})] \big\} 
 \xrightarrow[N\rightarrow\infty]{ \text{ prob.}} 0.
\label{first_LLN}     
\end{equation} 
Indeed, $\bar h_\theta(\underline{u}^{(i)},\underline{u}^{(i')})$ is a symmetrical measurable map of $(\xi_i,\xi_{i'})$.
Thus, (\ref{first_LLN}) is only a LLN for a U-statistic based on the beta-mixing process $(\xi_i)$.
Applying Theorem 1 of~\cite{arcones1998}, this is guaranteed under the moment condition~(\ref{cond_LLN_barh}) of the main text. The latter limit $ \Eb\big[ \bar h_{\theta^*}(\underline{u}^{(i)},\underline{u}^{(i')})\big]$ can be replaced by 
$\int \bar h_{\theta^*}(u,v) \, dP_\Ub(u) \, dP_\Ub(v)$, i.e., under an independence assumption. 
Indeed, noting that the sequence $\big(\beta_\xi(n)\big)$ is bounded by some power of $n$, we have
$$\frac{1}{N(N-1)}\sum_{i,i'=1, i\neq i'}^{N} 
\beta^{\nu/(1+\nu)}_\xi ( |i-i'|) = O\big(
N^{-1}\sum_{k=1}^N \beta^{\nu/(1+\nu)}_\xi ( k ) \big)= o(1).$$
Applying Lemma 1 of~\cite{yoshihara1976limiting}, we get 
$$ \lim_{N\rightarrow \infty}  \frac{1}{N(N-1)}\sum_{i,i'=1, i\neq i'}^{N} \Eb[ \bar h_{\theta^*}(\underline{u}^{(i)},\underline{u}^{(i')})] = \int \bar h_{\theta^*}(u,v) \, P_\Ub(du) \, P_\Ub(dv),$$
yielding
\begin{equation}
 \frac{1}{N(N-1)}\sum_{i,i'=1, i\neq i'}^{N} \bar h_{\theta^*}(\underline{u}^{(i)},\underline{u}^{(i')})   
 \xrightarrow[N\rightarrow\infty]{ \text{ prob.}}
 \int \bar h_{\theta^*}(u,v) \, P_\Ub(du) \, P_\Ub(dv). 
\label{first_LLN_V2}     
\end{equation} 

Second, let us prove that
\begin{equation}
 D_{N,T}(\theta^*):= \frac{1}{NT}\sum_{i=1}^{N} \sum_{t=1}^{T} \Big\{ \bar g_{\theta^*}(\underline{u}^{(i)},y_t) - \Eb\big[ \bar g_{\theta^*}(\underline{u}^{(1)},y_t) \big]\Big\}
 \xrightarrow[N\rightarrow\infty]{ \text{ prob.}} 0.
\label{second_LLN}     
\end{equation} 
It is sufficient to prove that $\Eb\big[ \| D_{N,T}(\theta^*) \|^2\big]\rightarrow 0$.
Set $\tilde g_{\theta^*}(\underline{u}^{(i)},y_t):=\bar g_{\theta^*}(\underline{u}^{(i)},y_t)   - \Eb\big[\bar g_{\theta^*}(\underline{u}^{(i)},y_t) \big]$.
Note that 
$\bar g_\theta(\underline{u}^{(i)},y_t)$ is a measurable map of 
$(\xi_i,y_t)$. 
Reasoning for any component of $\tilde g_{\theta^*}(\underline{u}^{(i)},y_t)$, say $(\ell,\ell') \in \{1,\ldots,d\}^2$, Davydov's inequality and the independence between the processes $(y_t)$ and $(\xi_i)$ imply
\begin{eqnarray*}
    \lefteqn{ \Eb\big[\| D_{N,T}(\theta^*) \|^2\big] \leq  \frac{1}{(NT)^2} \sup_{(\ell,\ell')} \sum_{i,i'} \sum_{t,t'} 
    | \text{Cov}\big(\tilde g_{\theta^*,(\ell,\ell')}(\underline{u}^{(i)},y_t), 
    \tilde g_{\theta^*,(\ell,\ell')}(\underline{u}^{(i')},y_{t'}) \big) | }\\
    &\leq & \frac{C_0}{(NT)^2} \sum_{i,i'} \sum_{t,t'}  \alpha \big( \sigma(\xi_i,y_t), \sigma(\xi_{i'},y_{t'})  \big)^{1-2/r} \| \tilde g_{\theta^*} (\underline{u}^{(i)},y_{t}) \|^2_r  \\
    &\leq & \frac{C_0}{(NT)^2} \sum_{i,i'} \sum_{t,t'}  \big\{ \beta_\xi \big( |i-i'| )^{1-2/r} + \beta (|t-t'|)^{1-2/r}\big\} \| \tilde g_{\theta^*} (\underline{u}^{(i)},y_{t}) \|^2_r  \\
    &\leq &  \frac{C_1}{N^2} \sum_{i,i'}   \beta_\xi ( |i-i'|  )^{1-2/r}  
    +\frac{C_1}{T^2} \sum_{t,t'} \beta ( |t-t'|)^{1-2/r} \\
        &\leq &  \frac{C_2}{N} \sum_{k=1}^N   \beta_\xi ( k )^{1-2/r}  
    +\frac{C_2}{T} \sum_{t=1}^T \beta ( t)^{1-2/r},
\end{eqnarray*}
for some positive constants $C_0,C_1,C_2$ (recall equation (\ref{cond_LLN_barh}) of the main text). Since the sequence $\big(\beta_\xi(n)\big)$ (resp. $\big(\beta(t)\big)$) is bounded by some power of $n$ (resp. $t$) and $1-2/r>0$, check that 
$\sum_{k=1}^N  \beta_\xi ( k )^{1-2/r} = o(N)$ (resp. $\sum_{t=1}^T \beta ( t)^{1-2/r}=o(T)$).
This provides $\Eb\big[\|D_{N,T}(\theta^*)\|^2\big]= o(1)$, and then~(\ref{second_LLN}). 
Thanks to~(\ref{first_LLN_V2}),~(\ref{second_LLN}) and~(\ref{dec_Hessian_thetaNT}), deduce 
$\nabla^2_{\theta,\theta^\top} L_{N,T}(\bar\theta) - \tilde\Sigma=o_P(1)$, proving {\it (b)}.
This concludes the proof of the (1).

When $N/T\rightarrow 0$ (case (2)), the arguments are exactly the same, except that (\ref{AN_h1_g1}) and~(\ref{AN_g2}) are replaced by
$$ \frac{\sqrt{N}}{N}\sum_{i=1}^N (2h_{(1)}-g_{(1)})(\zeta_i) 
\xrightarrow[N\rightarrow\infty]{ \text{ law}}
\Nc(0,\Sigma_1), \;\text{and} \; 
 \frac{\sqrt{N}}{T}\sum_{t=1}^T g_{(2)}(y_t) =
 \frac{\sqrt{N}}{\sqrt{T}}\frac{1}{\sqrt{T}}\sum_{t=1}^T g_{(2)}(y_t)=o_P(1).$$

Finally, when $T/N\rightarrow 0$ (case (3)), note that (\ref{AN_h1_g1}) and~(\ref{AN_g2}) are replaced by
$$ \frac{\sqrt{T}}{N}\sum_{i=1}^N (2h_{(1)}-g_{(1)})(\zeta_i)= 
\frac{\sqrt{T}}{\sqrt{N}} \frac{1}{\sqrt{N}}\sum_{i=1}^N (2h_{(1)}-g_{(1)})(\zeta_i) = o_P(1), \;\text{and} \;  
 \frac{\sqrt{T}}{T}\sum_{t=1}^T g_{(2)}(y_t) 
 \xrightarrow[T\rightarrow\infty]{ \text{law}}
 \Nc\big(0,\Sigma_2\big).$$
\end{proof}

\newpage

\section{Additional numerical experiments}\label{supp:sec:additional_experiments}

\subsection{Gradient descent algorithm}\label{supp:subsec:SGD}

Following the discussion on the implementation methods described in Section \ref{sec:implementation} of the main text, two ISMMD estimators $\tilde{\theta}^{(1)}_{N,T}$ may be implemented: ISMMD and ISMMD-SGD, where $N=1000$, $T_0=100$. As for the PSMMD method, two PSMMD estimators $\tilde{\theta}^{(2)}_{N,T}$ may be implemented: PSMMD and PSMMD-SGD, where $N=1000$. 
In their non-stochastic and stochastic versions, the innovations are drawn following the Gaussian distribution (centered, unit variance); in the non-stochastic gradient case, they are drawn only once before the iterations start. The performances of the various algorithms will be compared through the ARMA and the non-linear MA models studied in Subsections \ref{subsec:ARMA} and \ref{subsec:NL_MA} of the main text, respectively. We consider:
\begin{itemize}
    \item the non-linear MA model with parameter $\theta=\psi$, where $\theta^*=0.9$, $T=1000$ and $u_t=t(3)/\sqrt{3}$.
    \item the ARMA process with parameter $\theta=(\phi,\psi,\sigma^2_u)$, where we set $\theta^* =(0.8,0.15,0.05)$, $T=1000$ and $v_t = \sqrt{\sigma^2_u}u_t$, with $u_t=t(3)/\sqrt{3}$. 
\end{itemize}

\begin{figure}[htbp]
\centering
  \begin{minipage}[b]{0.45\textwidth}
    \centering
    \includegraphics[width=0.95\linewidth]{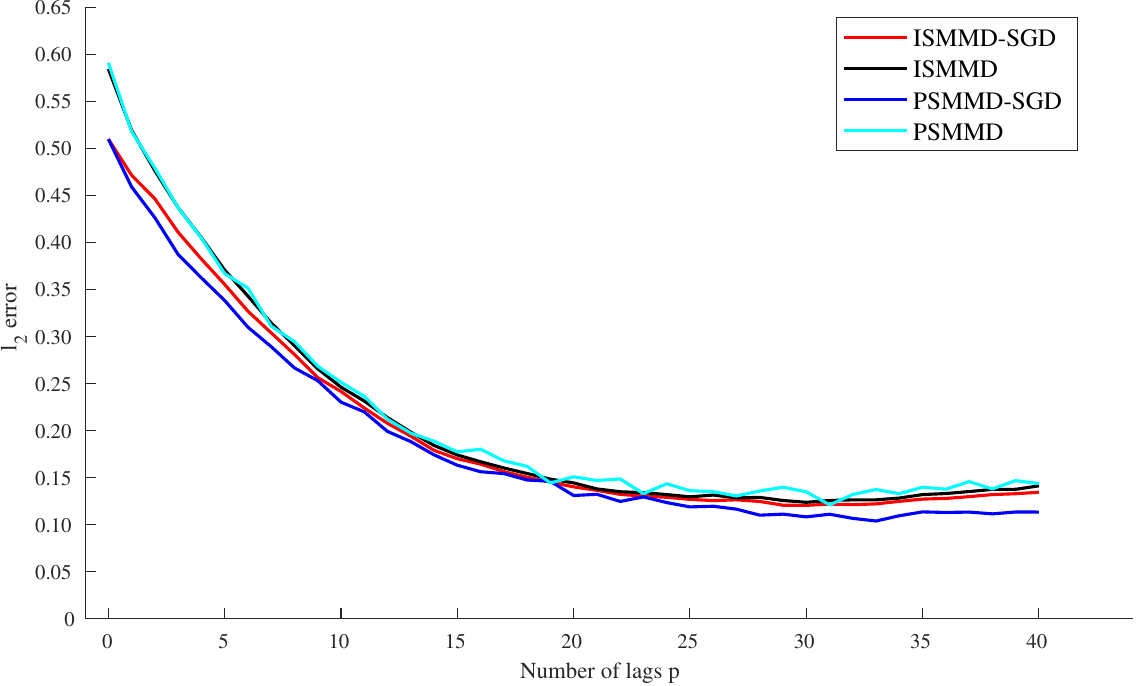}
    \subcaption{Non-linear MA model - $\theta^*=0.9$.}\label{figure_NLS_S_a9_1000_SGD_DET}
  \end{minipage}
  \begin{minipage}[b]{0.45\textwidth}
    \centering
    \includegraphics[width=0.95\linewidth]{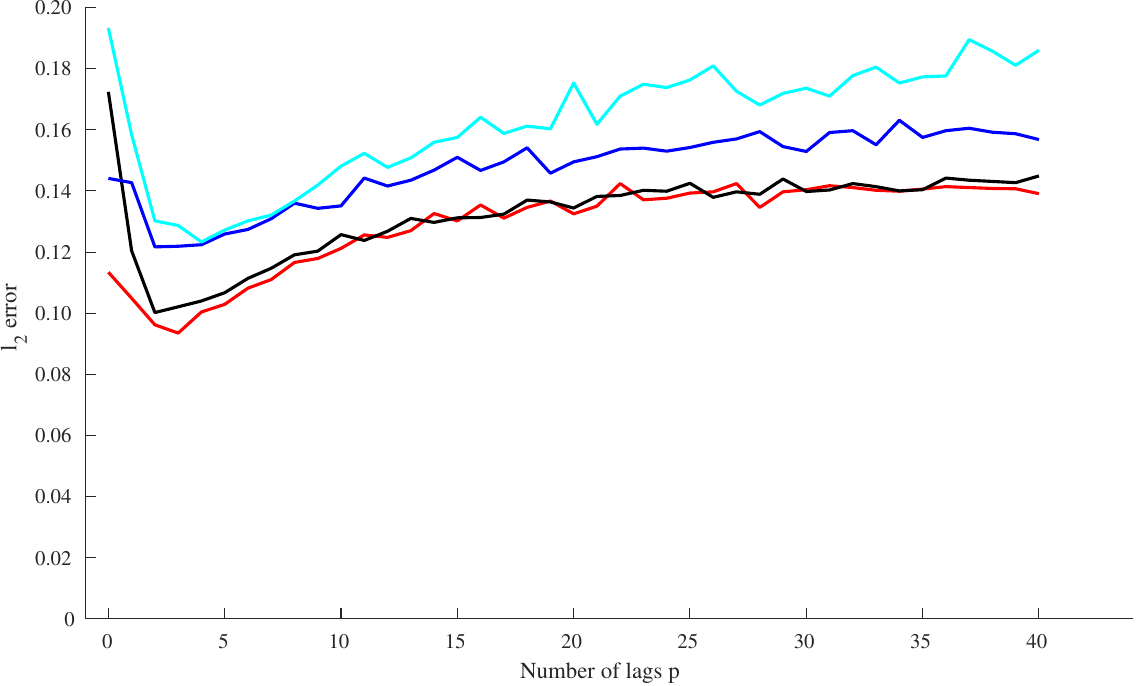}
    \subcaption{ARMA model - $\theta^* =(0.8,0.15,0.05)$.}\label{figure_ARMA_IO_p8_t15_1000_SGD_DET}
  \end{minipage}
  \caption{ISMMD/ISMMD-SGD and PSMMD/PSMMD-SGD errors. Each point shows the average of 100 batches.}
\end{figure}

Figures \ref{figure_NLS_S_a9_1000_SGD_DET} and \ref{figure_ARMA_IO_p8_t15_1000_SGD_DET} show the $\ell_2$ errors under misspecification (as in Case 2 of the main text) of the different estimators, averaged over 100 independent batches, for the latter two models, respectively. In all cases, ISMMD-SGD and PSMMD-SGD perform slightly better than ISMMD and PSMMD. The differences in terms of $\ell_2$ error are substantially more pronounced for the ARMA model, where a larger number of parameters must be estimated.

\newpage

\subsection{Kernel specification in MMD: a sensitivity analysis}\label{supp:subsec:kernel_comp}

\begin{figure}[htbp]
\centering
  \begin{minipage}[b]{0.45\textwidth}
    \centering
    \includegraphics[width=0.92\linewidth]{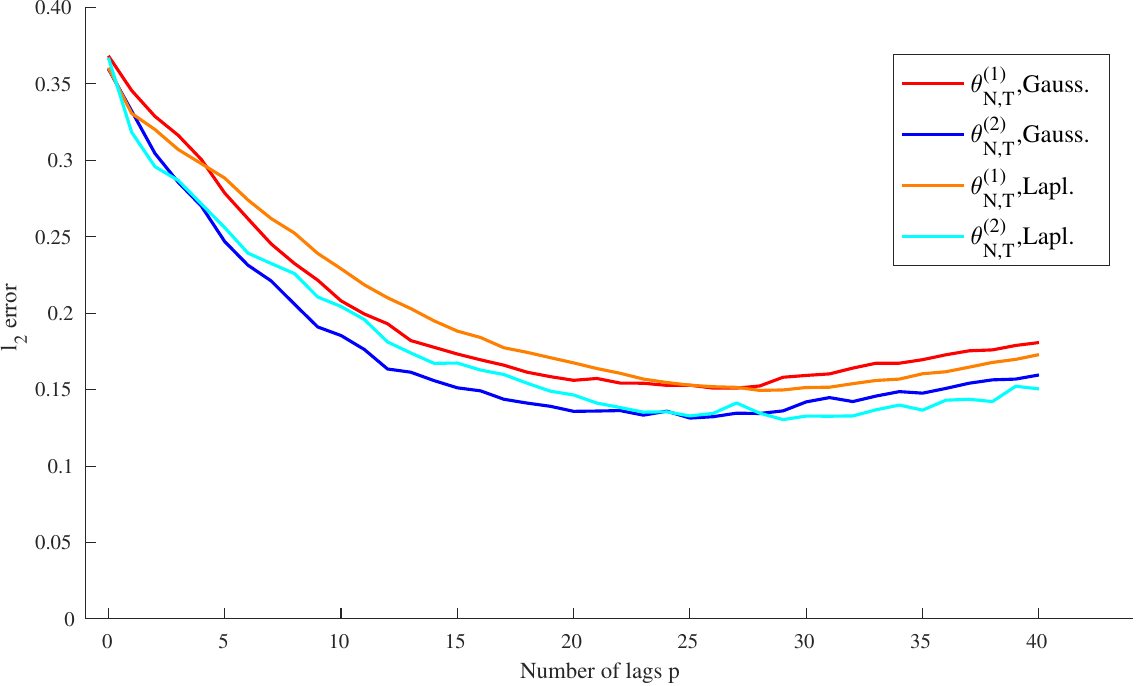}
    \subcaption{Non-linear MA. $\theta^*=0.7$, $\eta_t = t(3)/\sqrt{3}$. $T=300$. \\ Estimation under misspecification.}\label{figure_NLS_S_a7_kernel}
  \end{minipage}
   \begin{minipage}[b]{0.45\textwidth}
    \centering
    \includegraphics[width=0.92\linewidth]{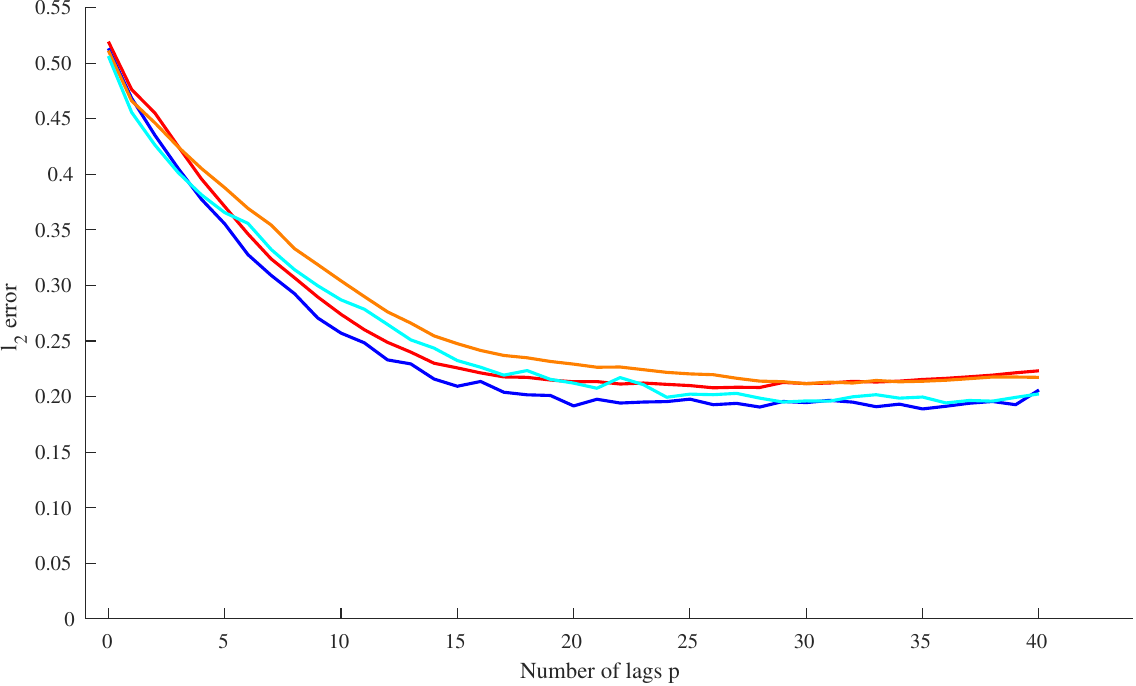}
    \subcaption{Non-linear MA. $\theta^*=0.9$, $\eta_t = t(3)/\sqrt{3}$. $T=300$. \\ Estimation under misspecification.}\label{figure_NLS_S_a9_kernel}
  \end{minipage}

  \vspace*{0.8cm}
  
  \begin{minipage}[b]{0.45\textwidth}
    \centering
    \includegraphics[width=0.92\linewidth]{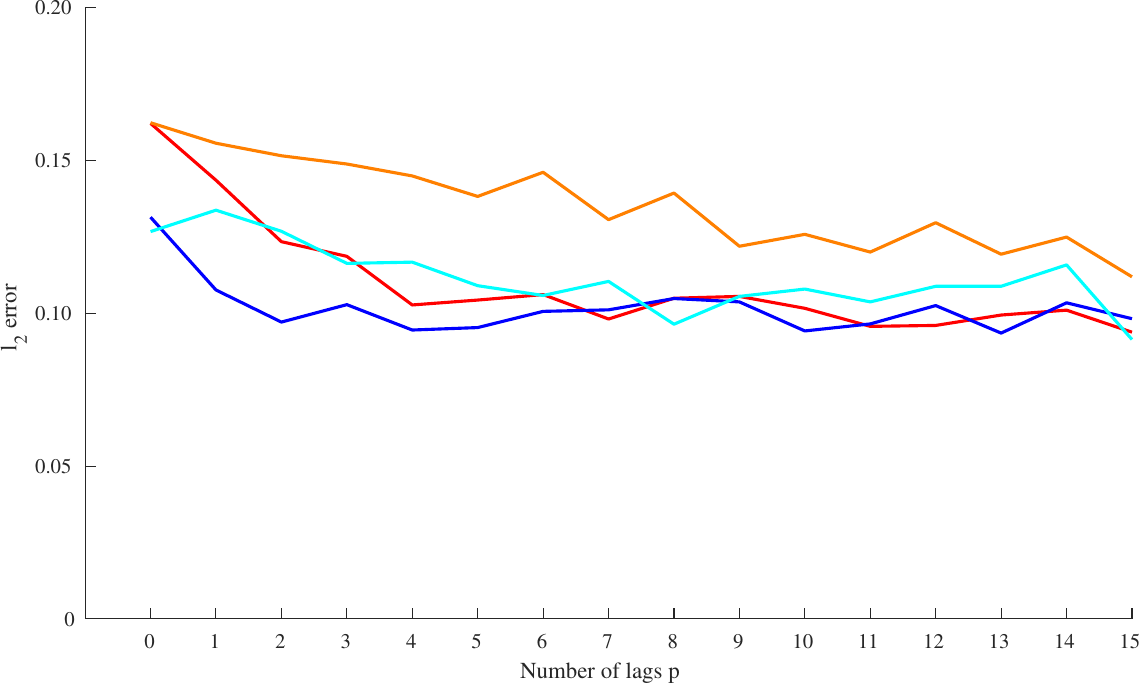}
    \subcaption{GARCH. $\theta^*=(0.05,0.92,0.05)$, $\eta_t = t(3)/\sqrt{3}$. $T=300$. Estimation under misspecification.}\label{figure_GARCH_S_b92_kernel}
  \end{minipage}
  \begin{minipage}[b]{0.45\textwidth}
    \centering
    \includegraphics[width=0.92\linewidth]{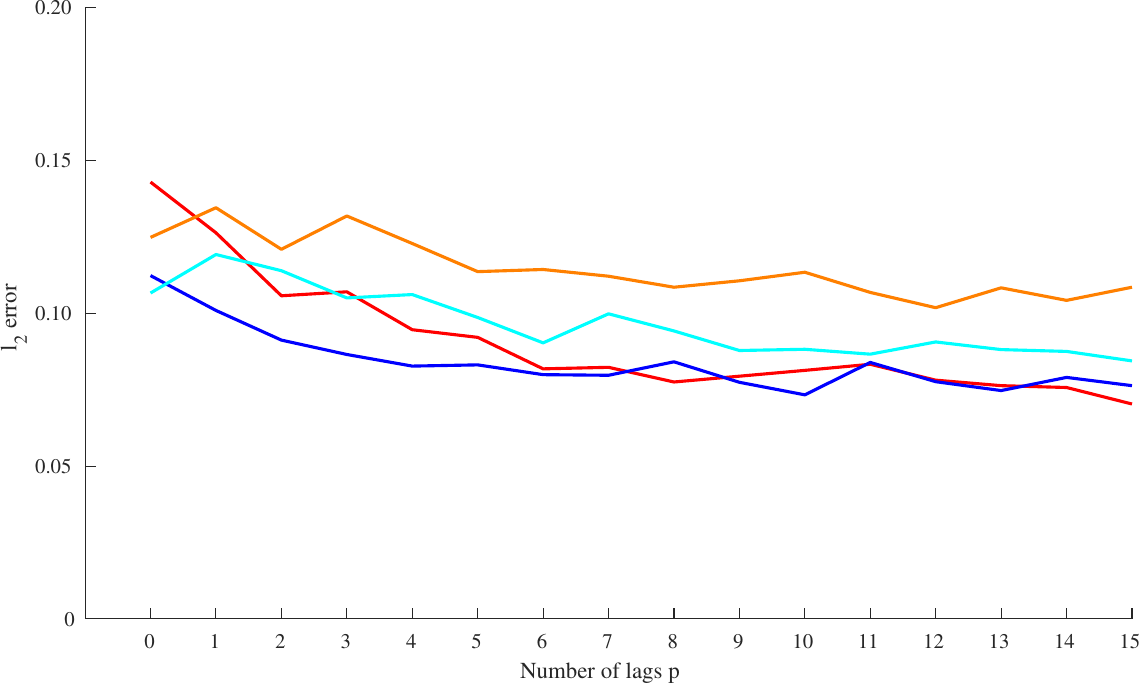}
    \subcaption{GARCH. $\theta^*=(0.05,0.85,0.1)$, $\eta_t = t(3)/\sqrt{3}$. $T=300$. Estimation under misspecification.}\label{figure_GARCH_S_b85_kernel}
  \end{minipage}
  \caption{$\ell_2$-errors for Gaussian and Laplace kernels. Each point shows the average of 100 batches.}
\end{figure}

The MMD-based procedure requires choosing a kernel function $k(y,y')$. In this subsection, we assess the sensitivity of the estimation accuracy with respect to the choice of $k(\cdot,\cdot)$, here the Gaussian kernel $k(y,y')=\exp(-\|y-y'\|^2/(2\sigma^2))$ and the Laplace kernel $k(y,y')=\exp(-\|y-y'\|/\sigma)$. In both cases, we will set $\sigma$ using the median heuristic applied to the sample of observations $y_t$. The performances of the procedure depending on $k(\cdot,\cdot)$ will be compared through the GARCH and the non-linear MA models studied in Subsections \ref{subsec:GARCH} and \ref{subsec:NL_MA} of the main text, respectively. We consider
\begin{itemize}
    \item the non-linear MA model with parameter $\theta=\psi$, where $\theta^*=0.7$ or $\theta^*=0.9$, $T=300$ and $u_t=t(3)/\sqrt{3}$;
    \item the GARCH process with parameter $\theta=(\phi,\psi,\sigma^2_u)$, where we set $\theta^*=(0.05,0.92,0.05)$ or $\theta^* =(0.8,0.15,0.05)$, $T=300$ and $u_t=t(3)/\sqrt{3}$. 
\end{itemize}
Figures \ref{figure_NLS_S_a7_kernel}-\ref{figure_NLS_S_a9_kernel} and Figures \ref{figure_GARCH_S_b92_kernel}-\ref{figure_GARCH_S_b85_kernel} display the $\ell_2$ errors under misspecification of the different estimators, averaged over 100 independent batches, for the non-linear MA and GARCH models, respectively. Both kernels result in similar patterns with respect to $p$, where their corresponding errors are decreasing with $p$. However, the Gaussian kernel-based ISMMD and PSMMD perform better than the Laplace kernel in most cases. 

\newpage

\subsection{SV model}

\begin{figure}[htbp]
\centering
  \begin{minipage}[b]{0.45\textwidth}
    \centering
    \includegraphics[width=0.92\linewidth]{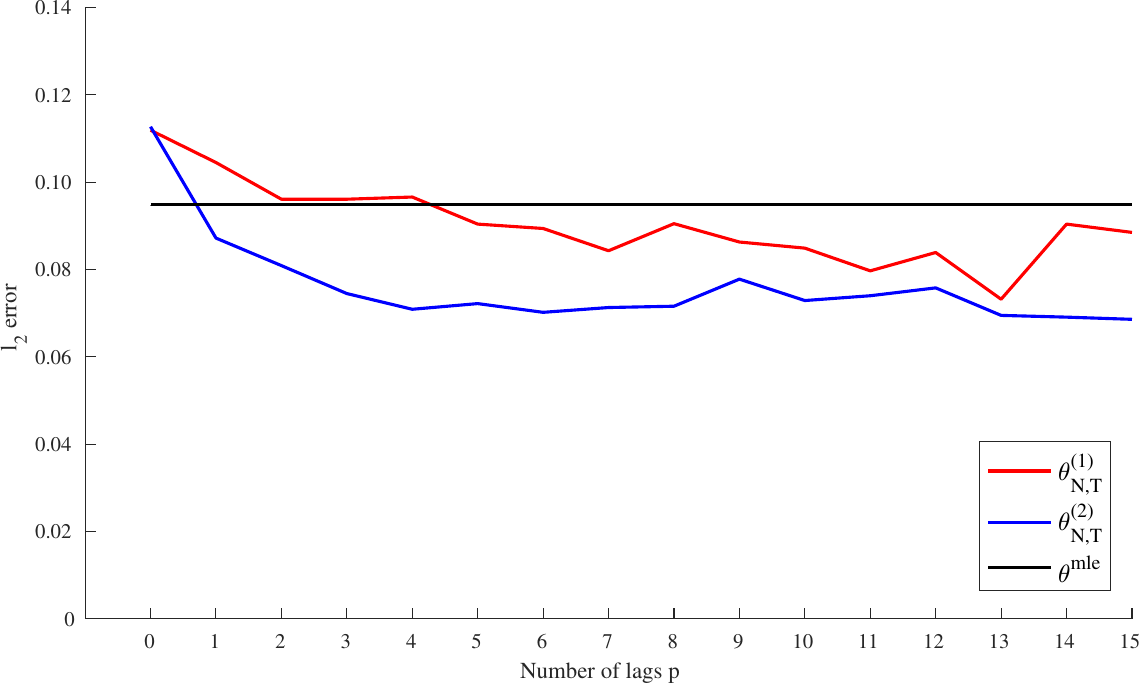}
    \subcaption{Case 1: $\eta_t \sim \mathcal{N}(0,1)$. $T=300$. \\ Estimation under correct specification.}\label{figure_SV_G_b8_300}
  \end{minipage}
  \begin{minipage}[b]{0.45\textwidth}
    \centering
    \includegraphics[width=0.92\linewidth]{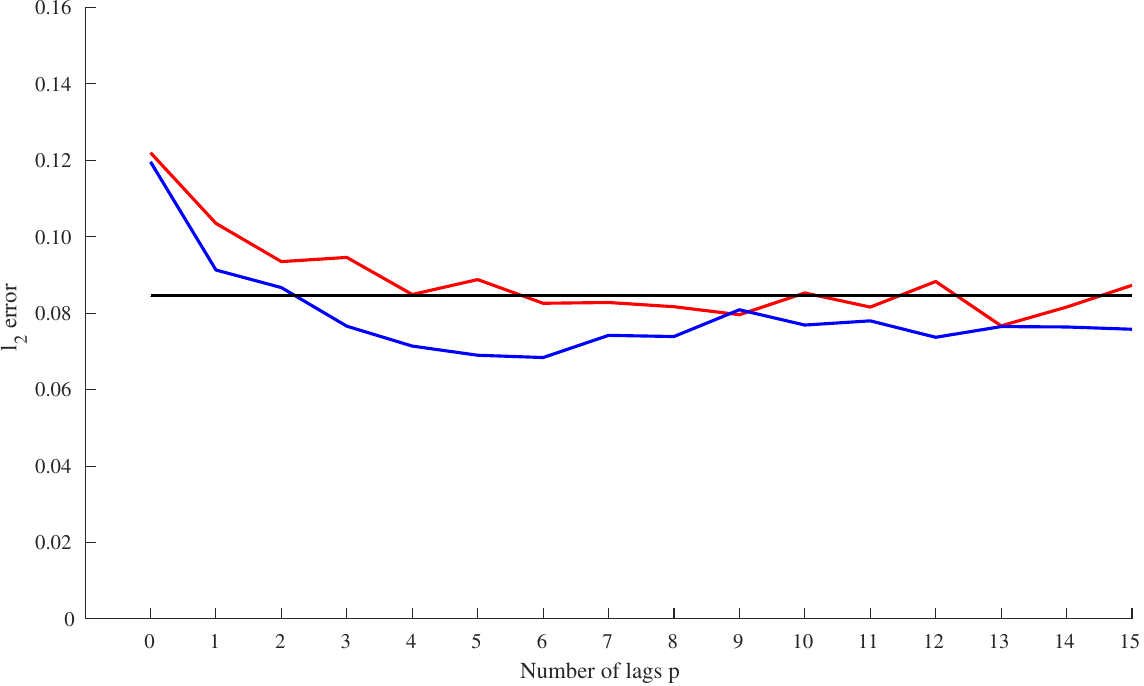}
    \subcaption{Case 2:  $\eta_t = t(3)/\sqrt{3}$. $T=300$. \\ Estimation under misspecification.}\label{figure_SV_S_b8_300}
  \end{minipage}

  \vspace*{0.8cm}
  
  \begin{minipage}[b]{0.45\textwidth}
    \centering
    \includegraphics[width=0.92\linewidth]{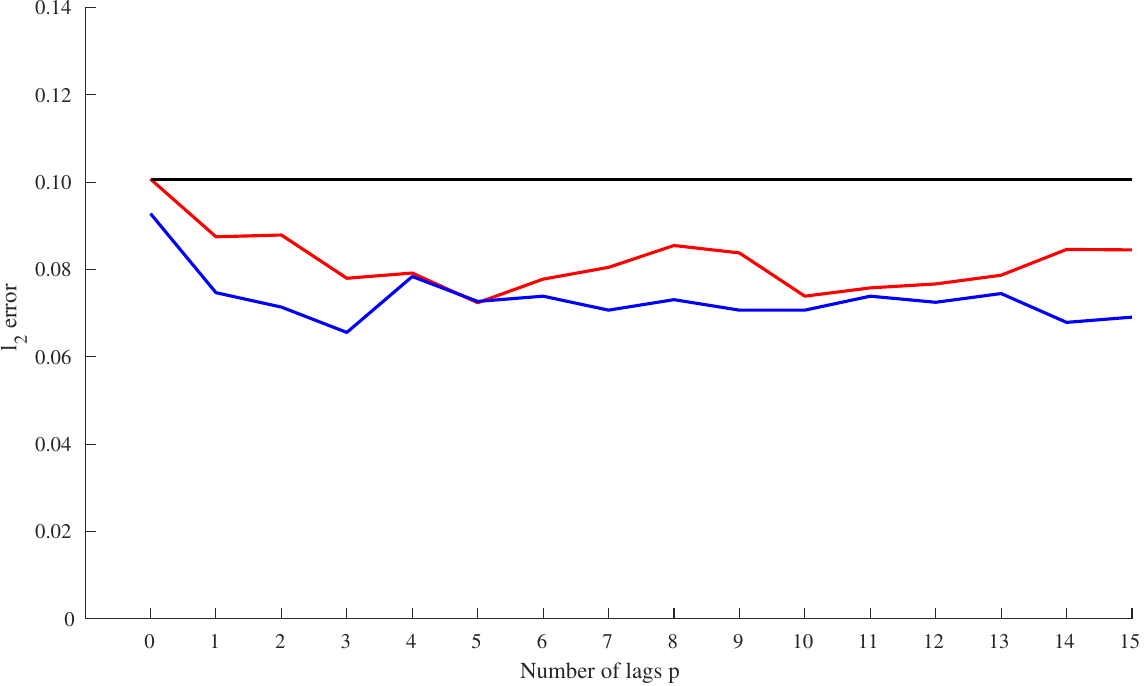}
    \subcaption{Case 1: $\eta_t \sim \mathcal{N}(0,1)$. $T=1000$. \\ Estimation under correct specification.}\label{figure_SV_G_b8_1000}
  \end{minipage}
  \begin{minipage}[b]{0.45\textwidth}
    \centering
    \includegraphics[width=0.92\linewidth]{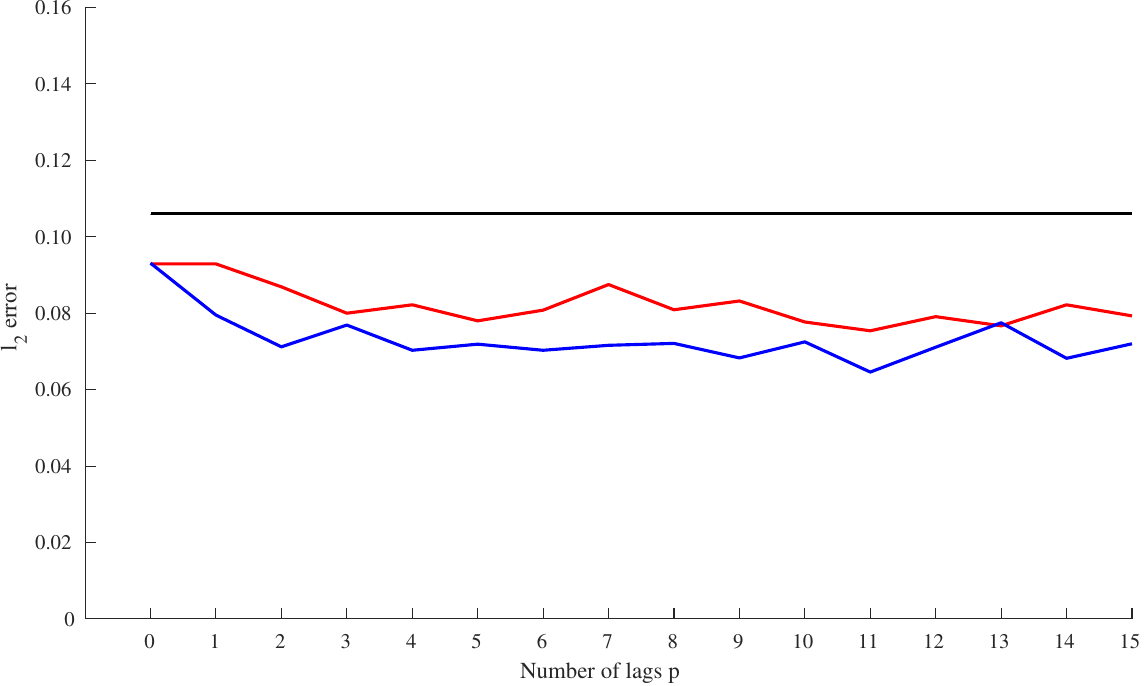}
    \subcaption{Case 2:  $\eta_t = t(3)/\sqrt{3}$. $T=1000$. \\ Estimation under misspecification.}\label{figure_SV_S_b8_1000}
  \end{minipage}
  \caption{SV model - $\theta^*=(0.8,0.05,0.15)$. Each point shows the average of 100 batches.}
\end{figure}

We consider the same setting as in Subsection \ref{subsec:SV} of the main text, where the SV model parameter is $\theta=(\phi,\sigma_{\eta},\sigma_x)$, except the true parameter is set as $\theta^* = (0.8,0.05,0.15)$ here.
The $\ell_2$ distances for the different estimators are displayed in Figures \ref{figure_SV_G_b8_300}-\ref{figure_SV_S_b8_1000}. Unlike our findings in Subsection \ref{subsec:SV} of the main text, a lower value of the autoregressive parameter $\phi$ results in better performances for PSMMD. Both ISMMD and PSMMD outperform the particle filter-based estimator, at least for large values of $p$ in the misspecified case (Figure \ref{figure_SV_S_b8_300}).

\newpage 

\subsection{GARCH model}

\begin{figure}[htbp]
\centering
  \begin{minipage}[b]{0.45\textwidth}
    \centering
    \includegraphics[width=0.92\linewidth]{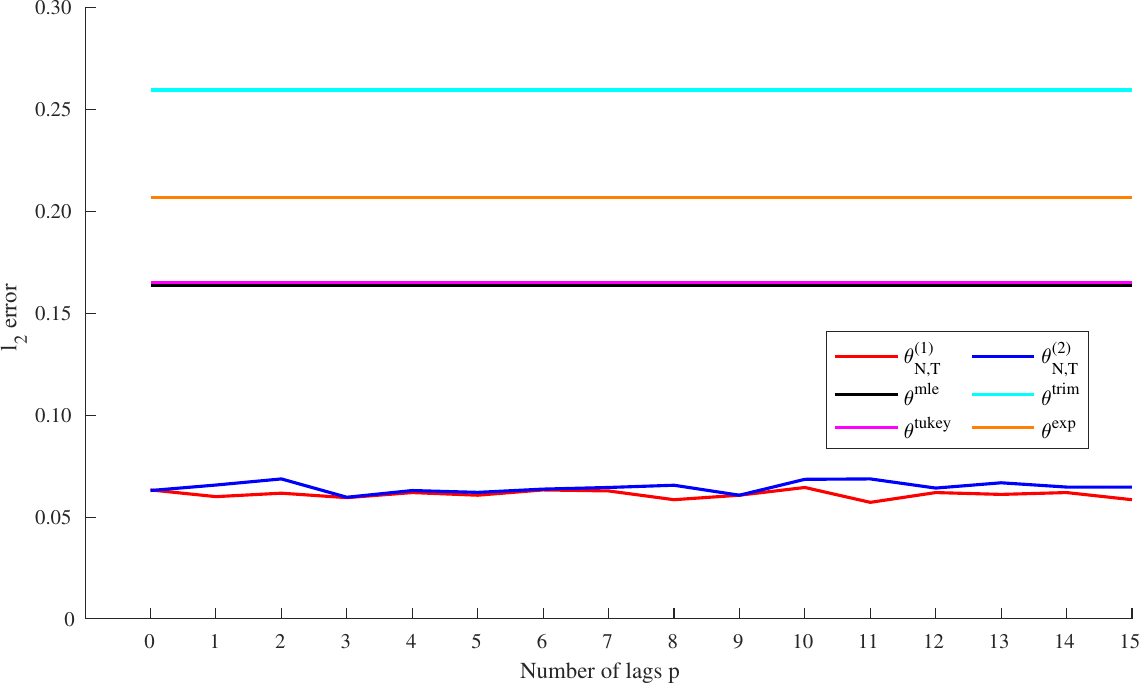}
        \subcaption{Case 1: $u_t \sim \mathcal{N}(0,1)$. $T=300$. \\ Estimation under correct specification.}\label{figures_GARCH_G_b85_c1_300}
  \end{minipage}
  \begin{minipage}[b]{0.45\textwidth}
    \centering
    \includegraphics[width=0.92\linewidth]{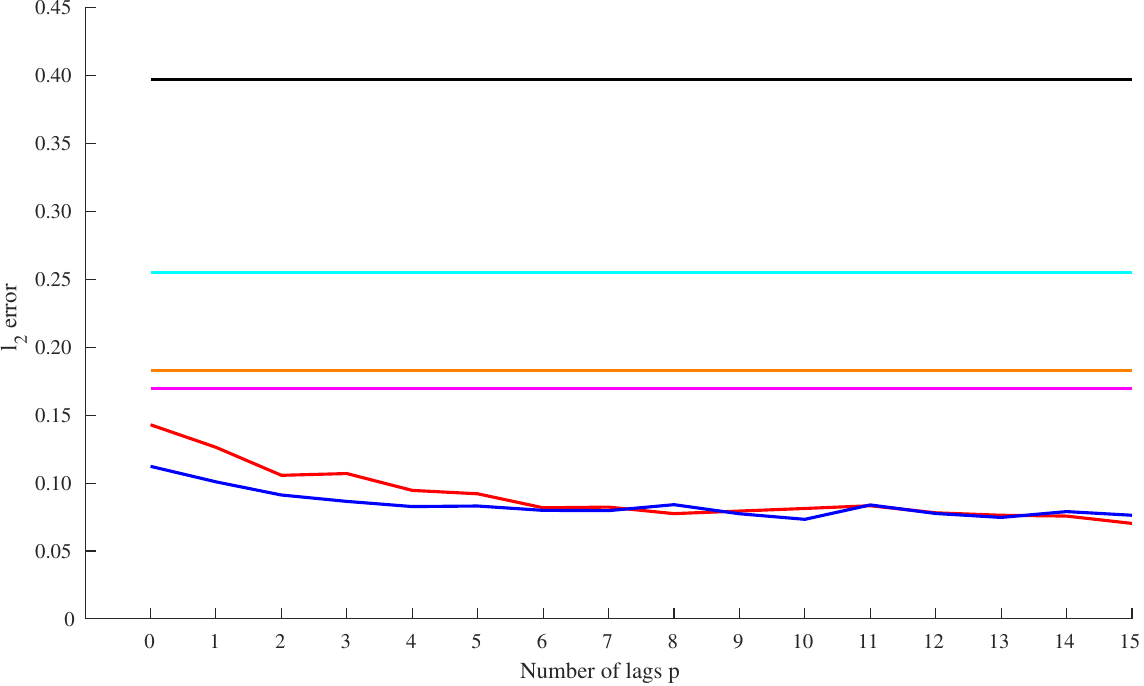}
        \subcaption{Case 2: $u_t = t(3)/\sqrt{3}$. $T=300$. \\ Estimation under misspecification.}\label{figures_GARCH_S_b85_c1_300}
  \end{minipage}

  \vspace*{0.8cm}
  
  \begin{minipage}[b]{0.45\textwidth}
    \centering
    \includegraphics[width=0.92\linewidth]{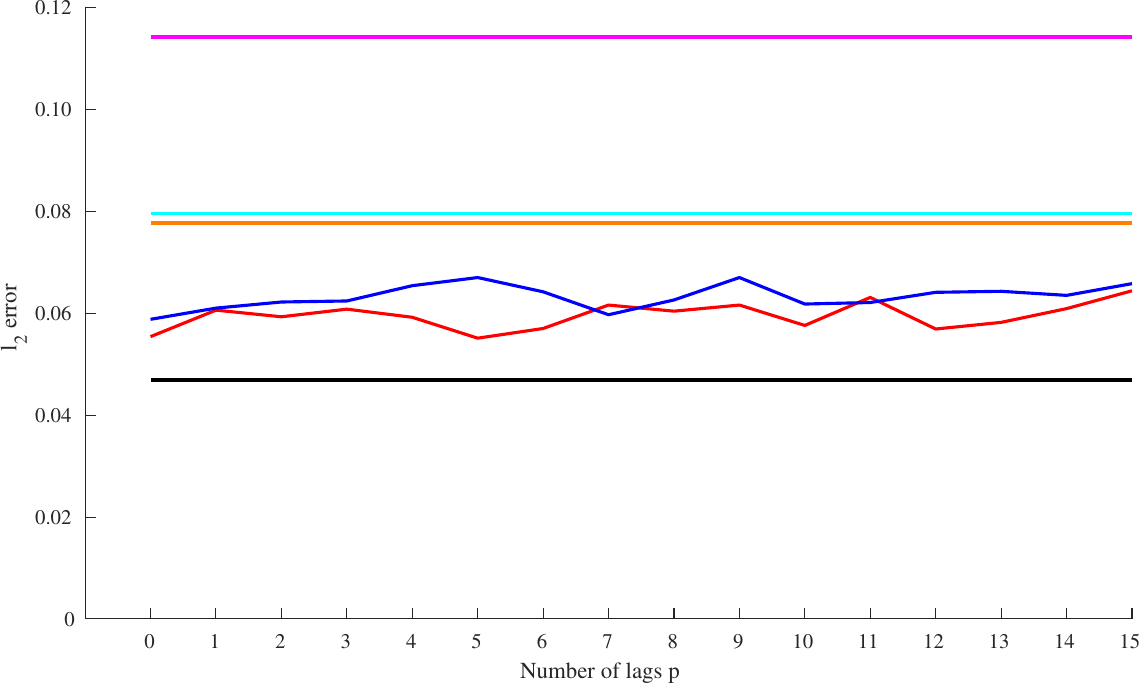}
        \subcaption{Case 1: $u_t \sim \mathcal{N}(0,1)$. $T=1000$. \\ Estimation under correct specification.}\label{figures_GARCH_G_b85_c1_1000}
  \end{minipage}
  \begin{minipage}[b]{0.45\textwidth}
    \centering
    \includegraphics[width=0.92\linewidth]{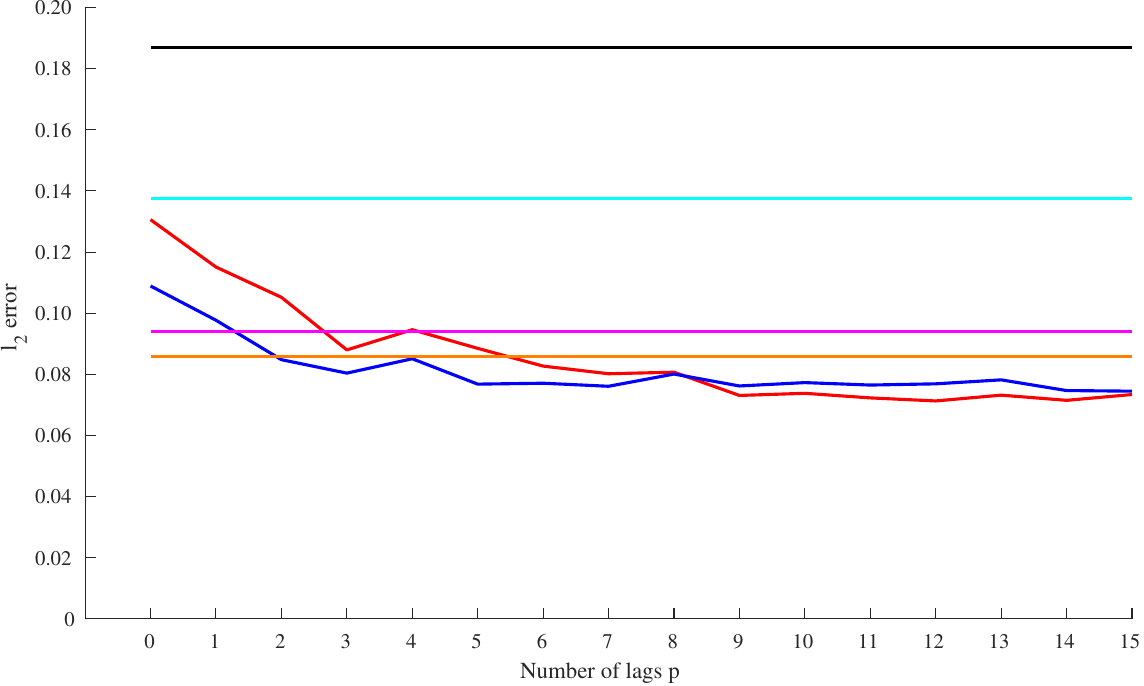}
        \subcaption{Case 2: $u_t = t(3)/\sqrt{3}$. $T=1000$. \\ Estimation under misspecification.}\label{figures_GARCH_S_b85_c1_1000}
  \end{minipage}
  \caption{GARCH - $\theta^*=(0.05,0.85,0.1)$. Each point shows the average of 100 batches.}
\end{figure}

Define the GARCH model parameters $\theta=(\omega,\beta,\alpha)$. The same experiments as in Subsection \ref{subsec:GARCH} of the main text are conducted but with $\theta^* = (0.05,0.85,0.1)$ now. The results are displayed in Figures \ref{figures_GARCH_G_b85_c1_300}-\ref{figures_GARCH_S_b85_c1_1000}, and align with our findings in Subsection \ref{subsec:GARCH} of the main text, except for the large sample case under correct specification, where the QML method performs better. \textcolor{black}{In the large sample case, and under misspecification, MMD yields better accuracy than MNWM.}
    
\newpage 

\subsection{ARMA model}

\begin{figure}[htbp]
\centering
  \begin{minipage}[b]{0.45\textwidth}
    \centering
    \includegraphics[width=0.92\linewidth]{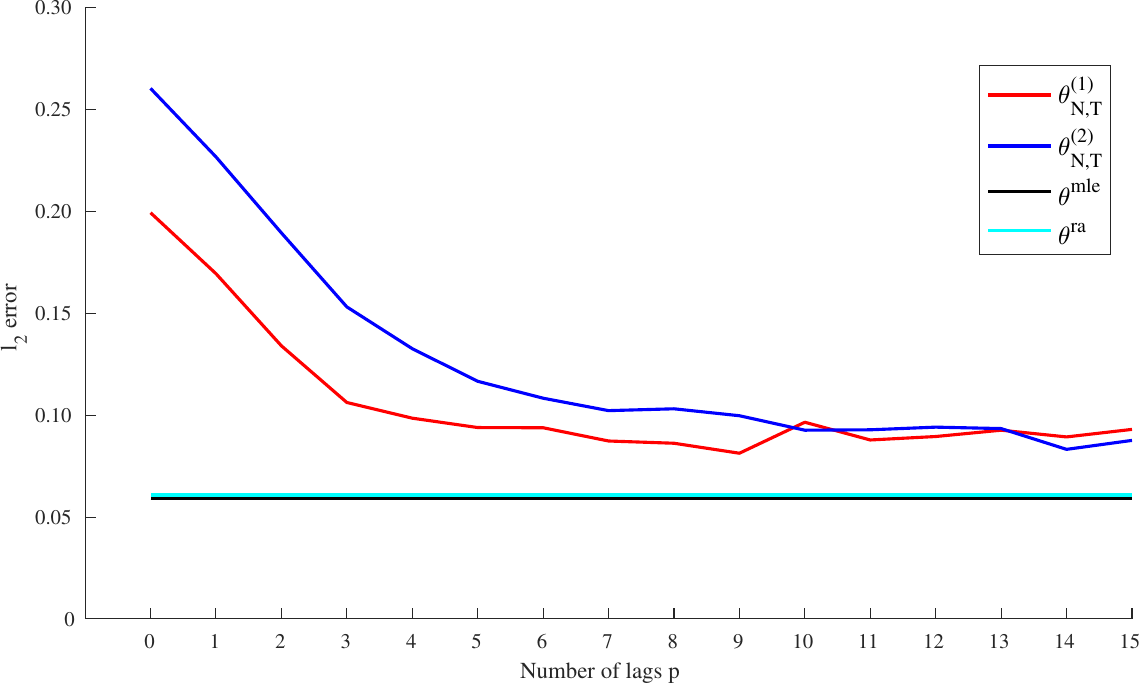}
    \subcaption{Case 1: $u_t \sim \mathcal{N}(0,1)$. $T=300$. \\ Estimation under correct specification.}\label{figures_ARMA_G_p9_t08_300}
  \end{minipage}
   \begin{minipage}[b]{0.45\textwidth}
     \centering
     \includegraphics[width=0.92\linewidth]{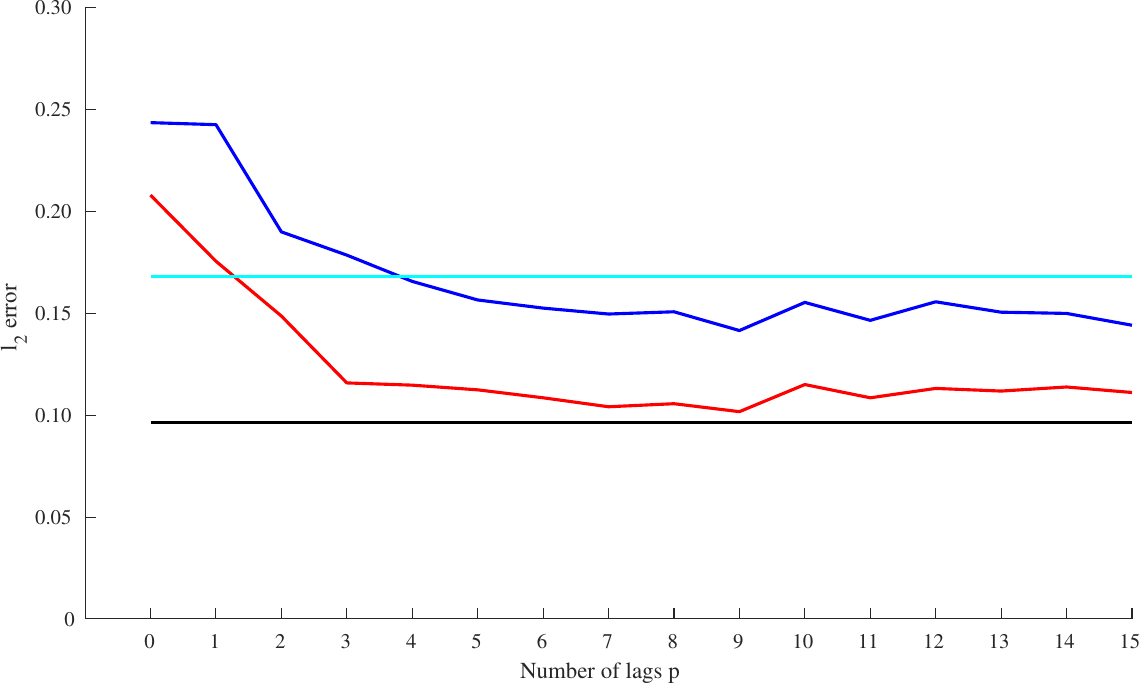}
     \subcaption{Case 2: $u_t \sim (1-\epsilon)\mathcal{N}(0,1)+\epsilon \mathcal{N}(0,\kappa)$. $T=300$. \\ Estimation under contamination.}\label{figures_ARMA_IO_p9_t08_300}
   \end{minipage}

   \vspace*{0.8cm}
   
   \begin{minipage}[b]{0.45\textwidth}
    \centering
    \includegraphics[width=0.92\linewidth]{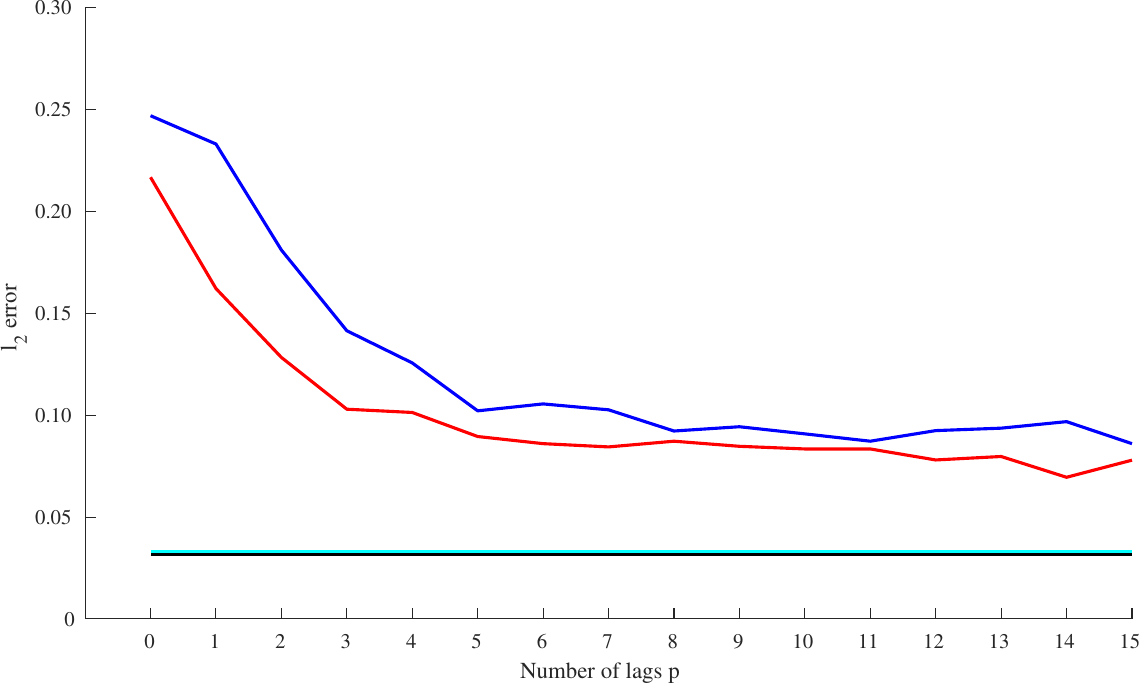}
    \subcaption{Case 1: $u_t \sim \mathcal{N}(0,1)$. $T=1000$. \\ Estimation under correct specification.}\label{figures_ARMA_G_p9_t08_1000}
  \end{minipage}
   \begin{minipage}[b]{0.45\textwidth}
     \centering
     \includegraphics[width=0.92\linewidth]{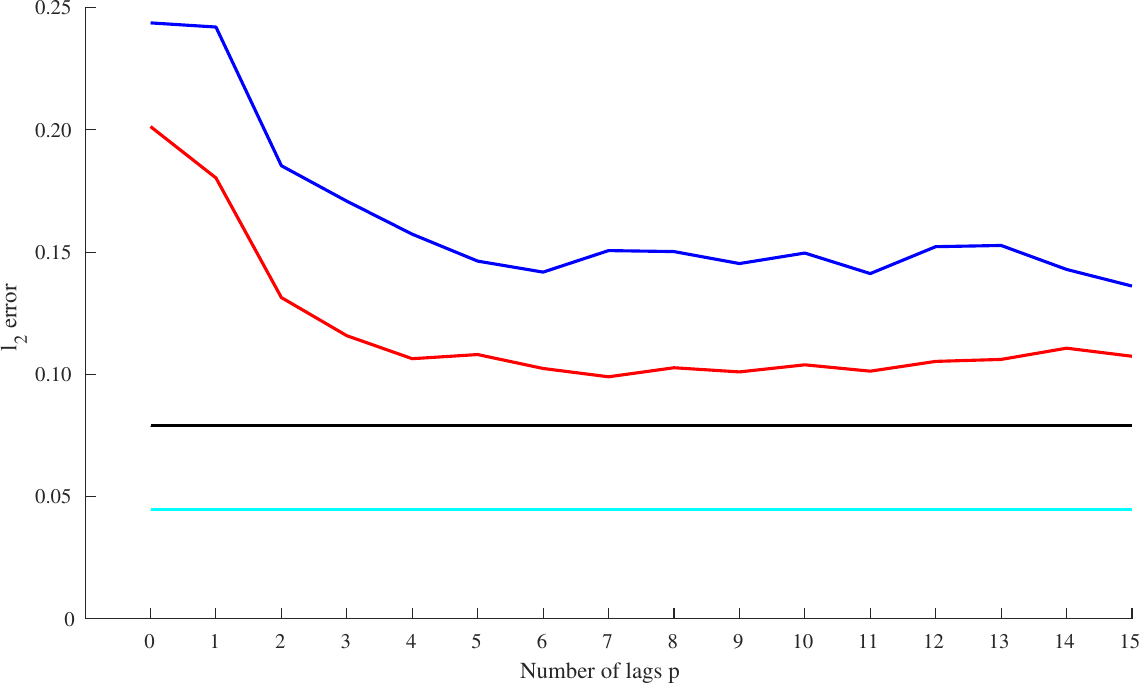}
     \subcaption{Case 2: $u_t \sim (1-\epsilon)\mathcal{N}(0,1)+\epsilon \mathcal{N}(0,\kappa)$. $T=1000$. \\ Estimation under contamination.}\label{figures_ARMA_IO_p9_t08_1000}
   \end{minipage}
   
  \caption{ARMA - $\theta^*=(0.9,0.08,0.03)$. Each point shows the average of 100 batches.}
\end{figure}

Define the ARMA model parameters $\theta=(\phi,\psi,\sigma^2_u)$. We conduct the same experiments as in Subsection \ref{subsec:ARMA} of the main text when $\theta^* = (0.9,0.08,0.03)$. The results displayed in Figures \ref{figures_ARMA_G_p9_t08_300}-\ref{figures_ARMA_IO_p9_t08_1000} show that the Gaussian QML-based method outperforms our method with this choice of $\theta^*$. \textcolor{black}{However, the RA estimator is outperformed by the MMD procedure in the small sample regime.} This suggests that the results in Subsection \ref{subsec:ARMA} should be interpreted with caution.

\newpage 

\subsection{Non-linear MA model}

\begin{figure}[htbp]
\centering
  \begin{minipage}[b]{0.45\textwidth}
    \centering
    \includegraphics[width=0.92\linewidth]{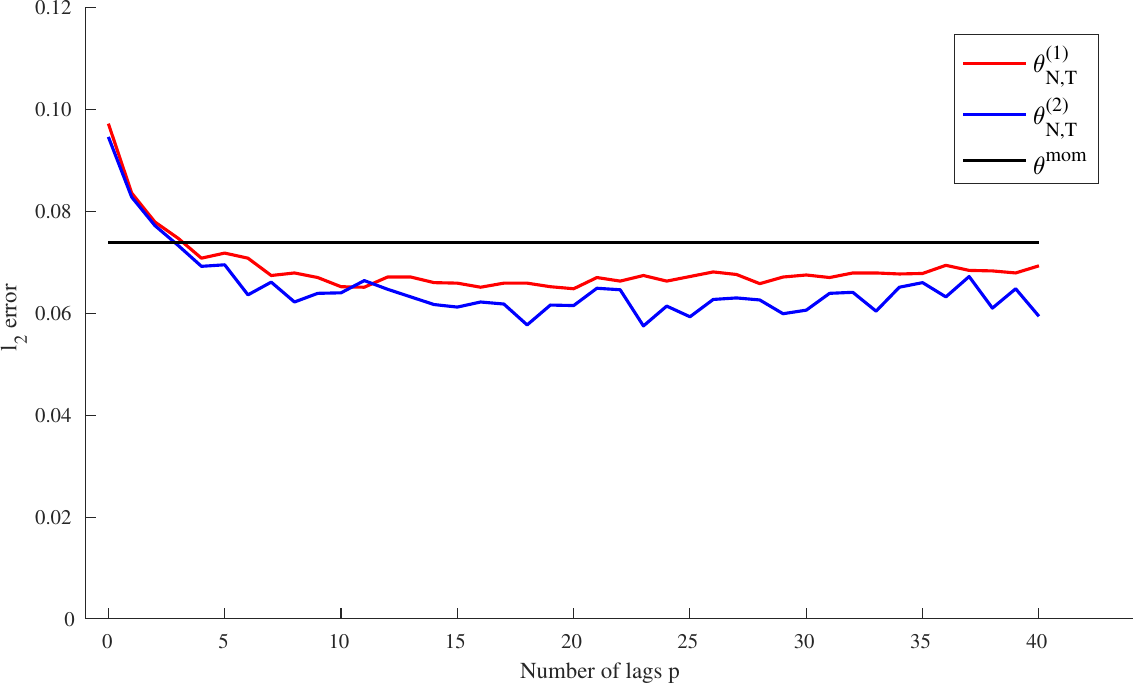}
    \subcaption{Case 1: $u_t \sim \mathcal{N}(0,1)$. $T=300$.\\ Estimation under correct specification.}\label{figures_NLS_G_a7_300}
  \end{minipage}
  \begin{minipage}[b]{0.45\textwidth}
    \centering
    \includegraphics[width=0.92\linewidth]{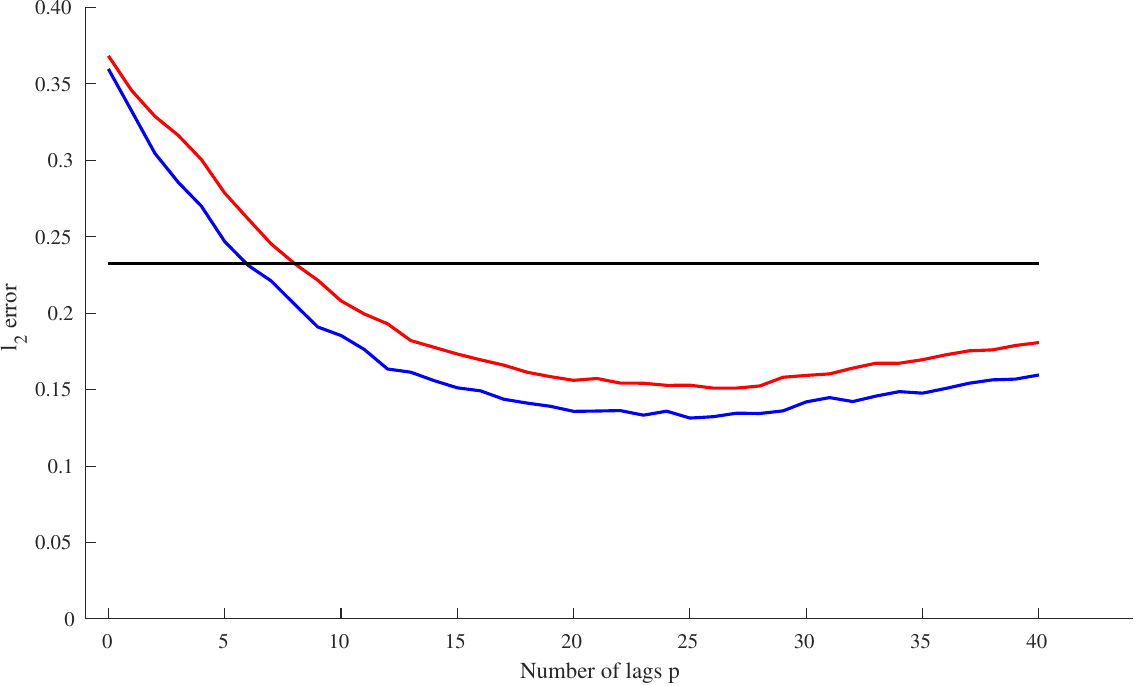}
    \subcaption{Case 2: $u_t = t(3)/\sqrt{3}$. $T=300$. \\ Estimation under misspecification.}\label{figures_NLS_S_a7_300}
  \end{minipage}

  \vspace*{0.8cm}
  
  \begin{minipage}[b]{0.45\textwidth}
    \centering
    \includegraphics[width=0.92\linewidth]{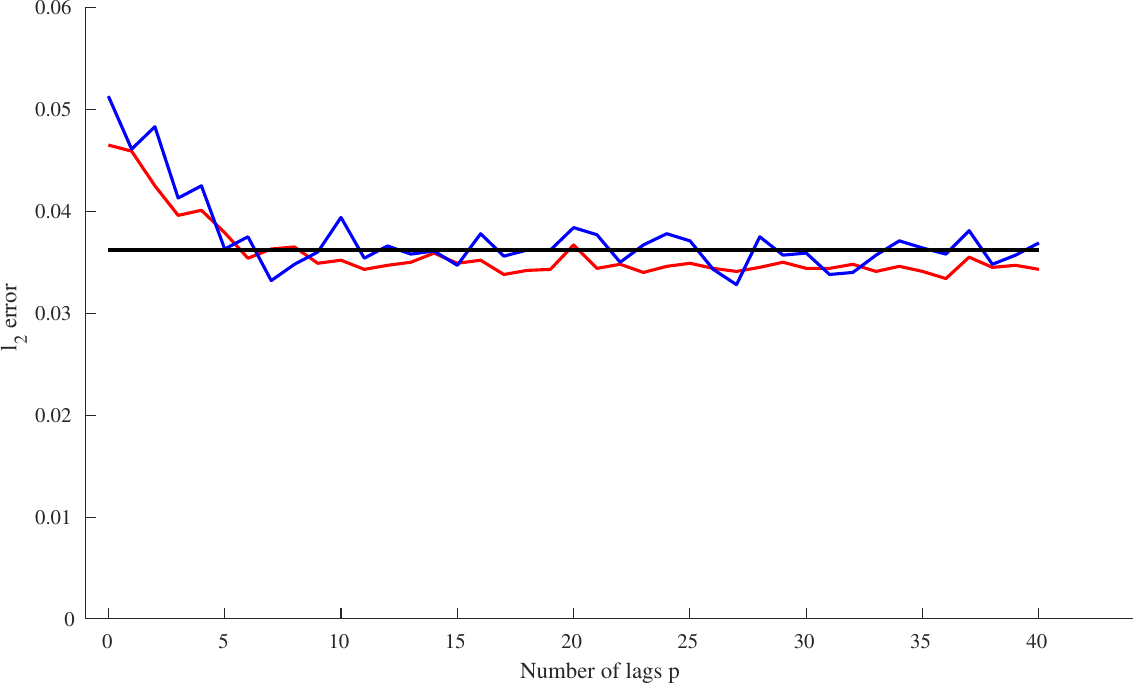}
    \subcaption{Case 1: $u_t \sim \mathcal{N}(0,1)$. $T=1000$. \\ Estimation under correct specification.}\label{figures_NLS_G_a7_1000}
  \end{minipage}
  \begin{minipage}[b]{0.45\textwidth}
    \centering
    \includegraphics[width=0.92\linewidth]{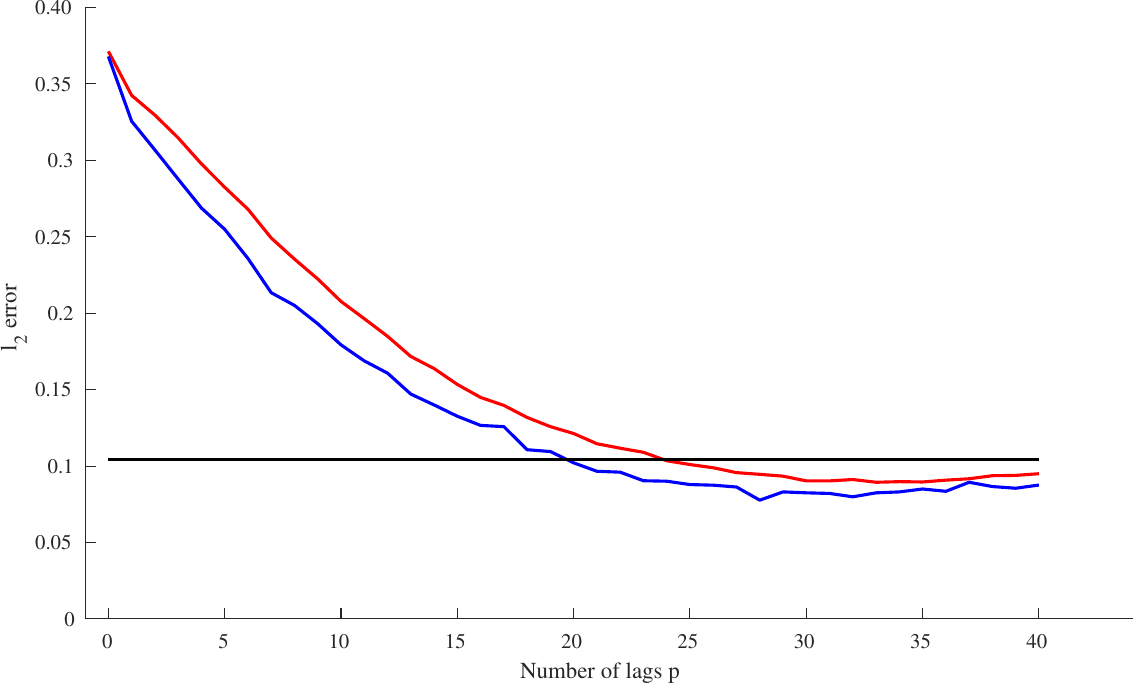}
    \subcaption{Case 2: $u_t = t(3)/\sqrt{3}$. $T=1000$. \\ Estimation under misspecification.}\label{figures_NLS_S_a7_1000}
  \end{minipage}
  \caption{Non-linear MA model - $\theta^*=0.7$. Each point shows the average of 100 batches.}
\end{figure}

The same experiments as in Subsection \ref{subsec:NL_MA} of the main text are carried out. The non-linear MA(1) parameter is $\theta=\psi$. We set the true parameter as $\theta^*=0.7$. The errors are displayed in Figures \ref{figures_NLS_G_a7_300}-\ref{figures_NLS_S_a7_1000}. Under correct specification, ISMMD and PSMMD outperform the moment estimator for lags $p\geq 5$. Under misspecification and small sample size, both ISMMD (resp. PSMMD) outperforms the moment estimator for $p\geq 8$ (resp. $p\geq 6$). ISMMD (resp. PSMMD) outperforms the moment estimator when $p \geq 24$ (resp. $p\geq 20$) under the large sample regime. 

\newpage 

\subsection{Ricker model}

\begin{figure}[htbp]
\centering
  \begin{minipage}[b]{0.45\textwidth}
    \centering
    \includegraphics[width=0.92\linewidth]{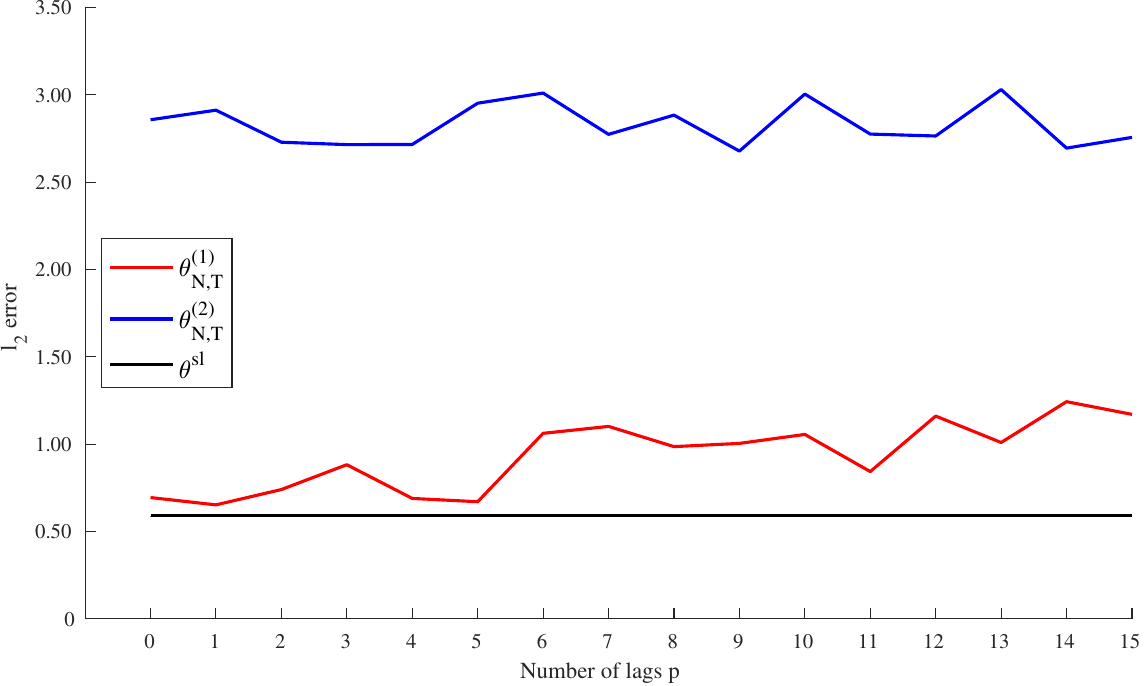}
    \subcaption{Case 1: $u_t \sim \mathcal{N}(0,1)$. $T=300$. \\ Estimation under correct specification.}\label{figures_Ricker_G_l5_p5_300}
  \end{minipage}
 \begin{minipage}[b]{0.45\textwidth}
    \centering
    \includegraphics[width=0.92\linewidth]{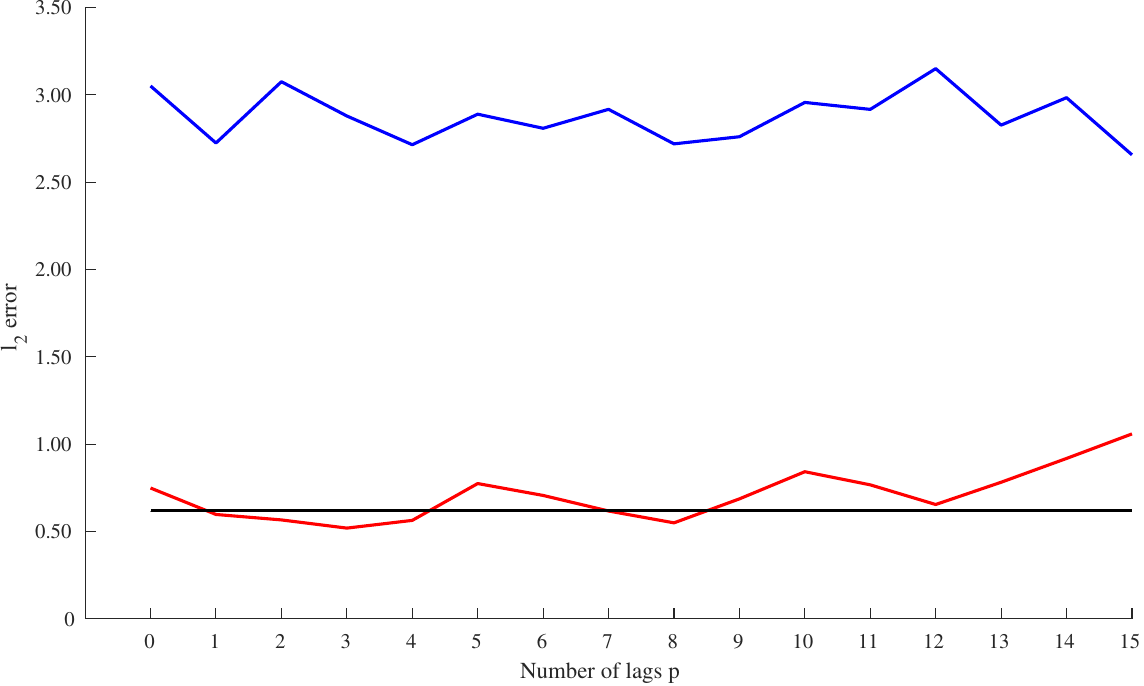}
    \subcaption{Case 2: $u_t = t(3)/\sqrt{3}$. $T=300$. \\ Estimation under misspecification.}\label{figures_Ricker_S_l5_p5_300}
  \end{minipage}

  \vspace*{0.8cm}
  
  \begin{minipage}[b]{0.45\textwidth}
    \centering
    \includegraphics[width=0.92\linewidth]{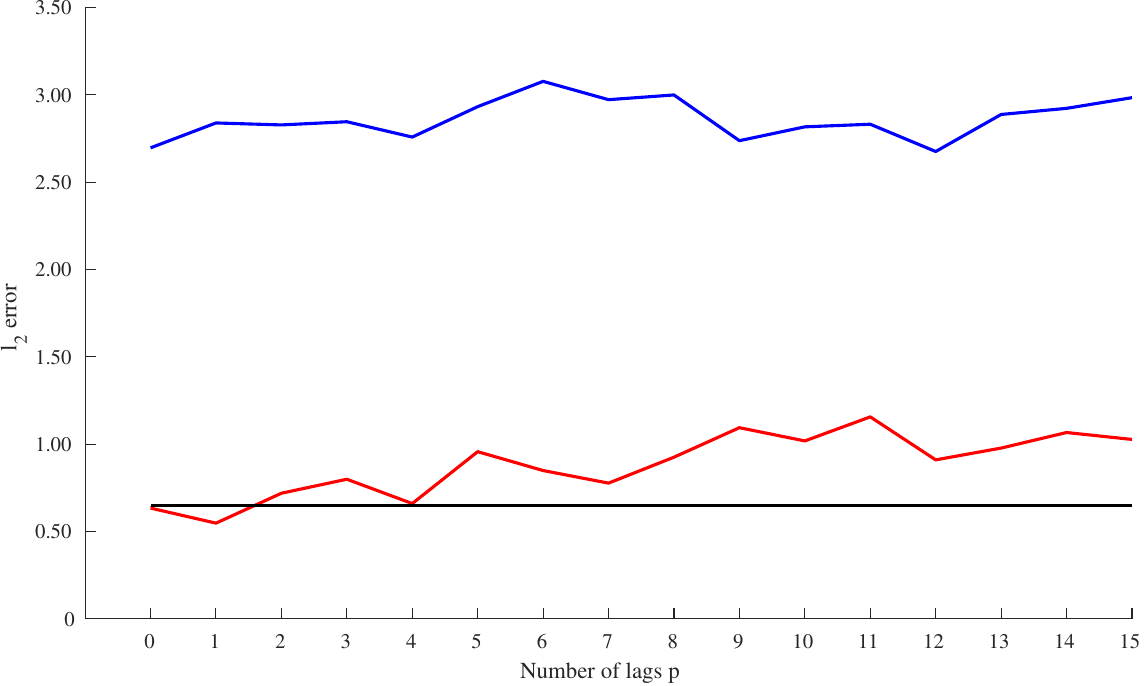}
    \subcaption{Case 1: $u_t \sim \mathcal{N}(0,1)$. $T=1000$. \\ Estimation under correct specification.}\label{figures_Ricker_G_l5_p5_1000}
  \end{minipage}
  \begin{minipage}[b]{0.45\textwidth}
    \centering
    \includegraphics[width=0.92\linewidth]{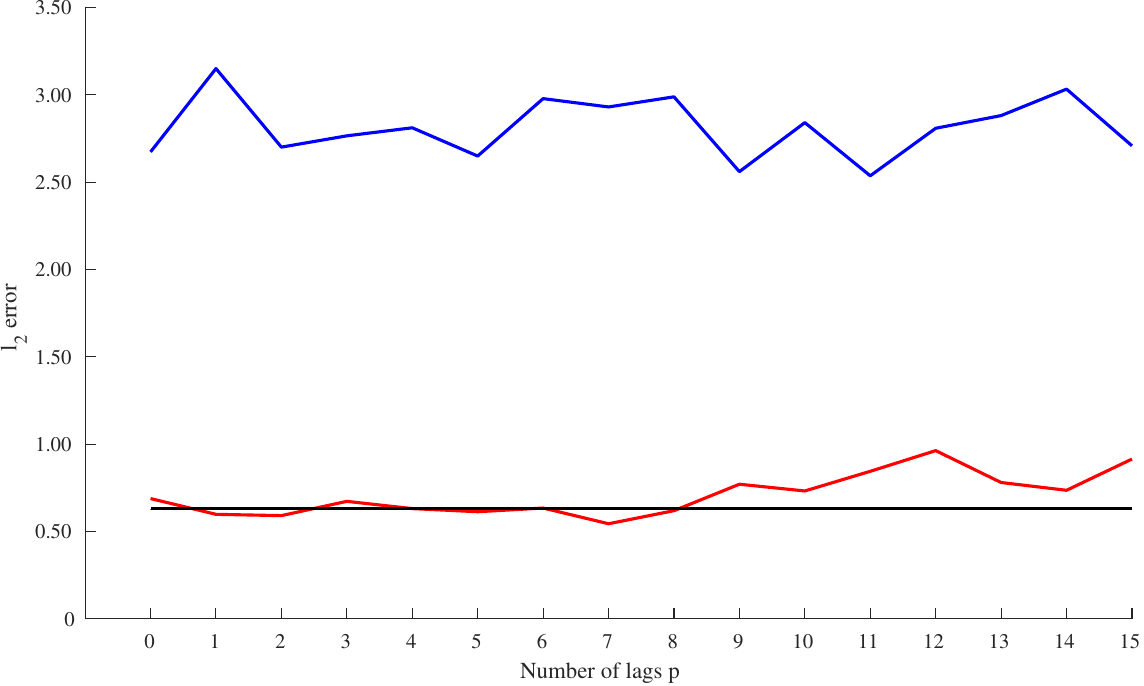}
    \subcaption{Case 2: $u_t = t(3)/\sqrt{3}$. $T=1000$. \\ Estimation under misspecification.}\label{figures_Ricker_S_l5_p5_1000}
  \end{minipage}
  \caption{Ricker model - $\theta^* = (\log(5),0.03,5)$. Each point shows the average of 100 batches.}
\end{figure}

Define the Ricker model parameters $\theta=(\log(r),\sigma_u,\phi)$. The same experiments are in Subsection \ref{subsec:Ricker} of the main text are conducted when $\theta^* = (\log(5),0.03,5)$. The errors are displayed in Figures \ref{figures_Ricker_G_l5_p5_300}-\ref{figures_Ricker_S_l5_p5_1000}. ISMMD performs in the same way as the SL method, and may yield better results for some $p$ lags. On the other side, PSMMD does not give convenient results in this experiment.

\end{document}